\documentclass[12pt,document,nofootinbib,superscriptaddress,onecolumn,preprintnumbers,balancelastpage]{article}

\pdfoutput=1

\usepackage{jheppub_1}

\usepackage{subfig}
\usepackage[countmax]{subfloat}
\usepackage{framed}
\usepackage{mdframed}

\usepackage{multirow}

\usepackage{slashed}

\usepackage{booktabs} 

\usepackage{array}

\makeatletter  
\newcommand{\thickhline}{%
    \noalign {\ifnum 0=`}\fi \hrule height 1pt
    \futurelet \reserved@a \@xhline
}

\newcolumntype{"}{@{\hskip\tabcolsep\vrule width 1pt\hskip\tabcolsep}}

\makeatother

\usepackage{latexsym}
\usepackage{amssymb}
\usepackage{epsfig,amsmath,graphics}
\usepackage{listings}
\usepackage{color}
\usepackage{dcolumn}
\usepackage{slashed}
\usepackage{cancel}
\usepackage{comment,latexsym} 
\usepackage{bm}
\usepackage{verbatim}
\usepackage{tabularx}
\usepackage{dcolumn}
\definecolor{Mahogany}{rgb}{0.62,0.24,0.15}
\definecolor{colorLink}{rgb}{0.7,0,0}
\definecolor{colorCite}{rgb}{0,.7,0}
\definecolor{colorURL}{rgb}{0,0,0.7}
\usepackage[colorlinks=true,linktocpage=true,linkcolor=black,citecolor=colorCite,urlcolor=colorURL]{hyperref}
\usepackage{setspace}
\usepackage{xspace}
\usepackage{multirow}
\usepackage{titlesec}
\usepackage[bbgreekl]{mathbbol}
\usepackage{bm}
\usepackage{float}
\usepackage{placeins}
\usepackage{afterpage}
\usepackage{footmisc}
\usepackage{breakcites}

\usepackage{empheq}
\usepackage[most]{tcolorbox}
 \tcbset{myformula/.style={colback=white!40,colframe=black,
every box/.style={highlight math style={colback=LightBlue!50!white,colframe=Navy}}
}}

\usepackage[toc,page]{appendix}

\usepackage{etoolbox}
\appto\appendix{\addtocontents{toc}{\protect\setcounter{tocdepth}{1}}}

\usepackage{scalerel}

\usepackage{arydshln}

\allowdisplaybreaks

\expandafter\def\expandafter\normalsize\expandafter{%
    \normalsize
    \setlength\abovedisplayskip{8pt}
    \setlength\belowdisplayskip{8pt}
    \setlength\abovedisplayshortskip{8pt}
    \setlength\belowdisplayshortskip{8pt}
}

\newcommand{\rpii}{{\kappa_\text{I}}}
\newcommand{\rpiii}{{\kappa_\text{II}}}
\newcommand{\rpiiii}{{\kappa_\text{III}}}
\newcommand{\rpiiC}{{\kappa^*_\text{I}}}
\newcommand{\rpiiiC}{{\kappa^*_\text{II}}}

\newcommand{\full}{\text{full}}

\newcommand{\s}{\hspace{0.8pt}}

\definecolor{darkblue}{rgb}{0,0,0.5}

\definecolor{colorTC}{rgb}{.2,.7,.2}

\newcommand{\qq}{\mathcal{Q}} 

\newcommand{\PP}{\mathbb{d}}

\newcommand{\RPIi}{\,\,\xrightarrow[\hspace{5pt}\text{RPI-I}\hspace{5pt}]{}\,\,}
\newcommand{\RPIii}{\,\,\xrightarrow[\hspace{5pt}\text{RPI-II}\hspace{5pt}]{}\,\,}
\newcommand{\RPIiii}{\,\,\xrightarrow[\hspace{5pt}\text{RPI-III}\hspace{5pt}]{}\,\,}
\newcommand{\gauge}{\,\,\xrightarrow[\hspace{5pt}\text{Gauge}\hspace{5pt}]{}\,\,}

\newcommand{\colleta}{\,\,\xrightarrow[\hspace{5pt}\tilde{\eta} = 0 \hspace{5pt}]{}\,\,}

\DeclareRobustCommand{\Sec}[1]{Sec.~\ref{#1}}
\DeclareRobustCommand{\Secs}[2]{Secs.~\ref{#1} and \ref{#2}}
\DeclareRobustCommand{\App}[1]{App.~\ref{#1}}
\DeclareRobustCommand{\Tab}[1]{Table~\ref{#1}}
\DeclareRobustCommand{\Tabs}[2]{Tables~\ref{#1} and \ref{#2}}

\DeclareRobustCommand{\Eq}[1]{Eq.~(\ref{#1})}
\DeclareRobustCommand{\Eqs}[2]{Eqs.~(\ref{#1}) and (\ref{#2})}
\DeclareRobustCommand{\Ref}[1]{Ref.~\cite{#1}}
\DeclareRobustCommand{\Refs}[1]{Refs.~\cite{#1}}

\newcommand{\tr}{\text{tr}}

\newcommand{\bPhi}{ \boldsymbol \Phi}
\newcommand{\bS}{ \boldsymbol S}
\newcommand{\bM}{ \boldsymbol M}

\newcommand{\bPhiA}{ \boldsymbol{\Phi}_\alc}
\newcommand{\bPhiD}{ \boldsymbol{\Phi}_D}
\newcommand{\bPhiF}{ \boldsymbol{\Phi}_F}

\newcommand{\bC}{ \boldsymbol C}
\newcommand{\bV}{ \boldsymbol V}
\newcommand{\bW}{ \boldsymbol W}

\newcommand{\bD}{ \boldsymbol{V}_{n \cdot A}}
\newcommand{\bPhialc}{ \boldsymbol{\Phi}_\alc}
\newcommand{\bU}{ \tilde{\boldsymbol U}}

\newcommand{\D}{\mathbb{D}}
\newcommand{\Q}{\mathbb{Q}}
\newcommand{\RCA}{\boldsymbol{\Omega}}

\newcommand{\be}{\begin{equation}}
\newcommand{\ee}{\end{equation}}
\newcommand{\bea}{\begin{eqnarray}}
\newcommand{\eea}{\end{eqnarray}}

\newcommand{\alc}{\mathcal{A}}

\newcommand{\uu}{\tilde u}

\newcommand{\nocontentsline}[3]{}
\newcommand{\tocless}[2]{\bgroup\let\addcontentsline=\nocontentsline#1{#2}\egroup}

\preprint{MIT--CTP 5143}

\title{
Circumnavigating Collinear Superspace
}

\author[a]{Timothy Cohen,}
\author[b]{Gilly Elor,}
\author[c]{Andrew J.~Larkoski,}
\author[d,e]{and Jesse Thaler\,}

\affiliation[a]{\footnotesize Institute of Theoretical Science, University of Oregon, 120 Willamette Hall, 1371 E 13th Avenue, Eugene, OR 97403 U.S.A}
\affiliation[b]{\footnotesize Department of Physics, University of Washington, Seattle, Box 1560, WA 98195, U.S.A.}
\affiliation[c]{\footnotesize Department of Physics , Reed College, 3203 SE Woodstock Boulevard, Portland, OR 97202, U.S.A.}
\affiliation[d]{\footnotesize Center for Theoretical Physics, Massachusetts Institute of Technology, Cambridge, MA 02139, U.S.A.}
\affiliation[e]{\footnotesize Department of Physics, Harvard University, 17 Oxford Street, Cambridge, MA 02138, U.S.A.}

\emailAdd{tcohen@uoregon.edu}
\emailAdd{gelor@uw.edu}
\emailAdd{larkoski@reed.edu}
\emailAdd{jthaler@mit.edu}

\abstract{
In this paper, we extend the collinear superspace formalism to include the full range of $\mathcal{N} = 1$ supersymmetric interactions.
Building on the effective field theory rules developed in a companion paper --- \emph{Navigating Collinear Superspace}~\cite{Cohen:2018qvn} --- we construct collinear superspace Lagrangians for theories with non-trivial $F$- and $D$-term auxiliary fields.
For (massless) Wess-Zumino models, the key ingredient is a novel type of Grassmann-valued supermultiplet whose lowest component is a (non-propagating) fermionic degree of freedom.
For gauge theories coupled to charged chiral matter, the key ingredient is a novel type of vector superfield whose lowest component is a non-propagating gauge potential.
This unique vector superfield is used to construct a gauge-covariant derivative; while such an object does not appear in the standard full superspace formalism, it is crucial for modeling gauge interactions when the theory is expressed on a collinear slice.
This brings us full circle, by showing that all types of $\mathcal{N} = 1$ theories in four dimensions can be constructed in collinear superspace from purely infrared considerations.
We speculate that supersymmetric theories with $\mathcal{N} > 1$ could also be implemented using similar collinear superspace constructions.
}

\begin{document} 
\maketitle

\setcounter{page}{2}
\begin{spacing}{1.2}

\pagebreak

\clearpage
\setcounter{page}{3}
\addtocontents{toc}{\protect\vspace{-20pt}}
\section{Casting Off}
\label{sec:intro}
The study of supersymmetric (SUSY) models has deepened our understanding of the nature of quantum field theory~\cite{Seiberg:1994aj,Seiberg:1994rs, Seiberg:1994pq, Maldacena:1997re, Bianchi:2008pu, Komargodski:2009rz,Festuccia:2011ws,Kahn:2015mla,Ferrara:2015tyn,Dall'Agata:2016yof,Delacretaz:2016nhw,Pestun:2007rz, Erickson:2000af}.
In particular, significant insights into $\mathcal{N} = 4$ SUSY Yang-Mills were exposed by realizing that it could be formulated in superspace~\cite{Salam:1974yz,Ferrara:1974ac} using light-cone coordinates~\cite{Mandelstam:1982cb, Brink:1982pd, Brink:1982wv}, see also \emph{e.g.}~\cite{Green:1996um,Maldacena:1997re,Belitsky:2004yg,Kallosh:2009db,Hearin:2010dw,Ramond:2009hb}.
To this end, our goal has been to understand the effective field theory rules for expressing $\mathcal{N} = 1$ SUSY theories on the light cone in the language of collinear superspace~\cite{Cohen:2016jzp,Cohen:2016dcl}.
This provides a setting where states are packaged into superfields, thereby making SUSY manifest.
In our companion paper~\cite{Cohen:2018qvn}, we focused on theories that could be constructed exclusively in terms of propagating degrees of freedom.
In this paper, we introduce building blocks for collinear superspace that allow us to construct models whose interactions require non-trivial auxiliary fields.
This brings us full circle, by demonstrating how all known $\mathcal{N} = 1$ SUSY Lagrangians can be written in collinear superspace.%
\footnote{We choose to call the setting for these theories ``collinear superspace'' as opposed to the more traditional nomenclature ``light-cone superspace'' since our focus here is on the bottom-up effective theory rules.
For some related references on light-cone superspace, see~\Refs{Siegel:1981ec, Brink:1981nb, Hassoun:1982mr, Mandelstam:1982cb, Brink:1982pd, Brink:1982wv}.}

Expressing a model in collinear superspace requires picking a light cone, as specified by a pair of light-like vectors $n^\mu$ and $\bar{n}^\mu$.
Using a spinor-helicity decomposition, is it convenient to write these vectors in terms of a pair of orthogonal bosonic spinors, $\xi_\alpha$ and $\tilde{\xi}_\alpha$:
\begin{align}
n^\mu = \tilde{\xi}\, \sigma^\mu\, \tilde{\xi}^\dagger\,, \qquad \bar{n}^\mu = \xi\, \sigma^\mu\, \xi^{\dagger}\,,
\label{eq:nAndnBar}
\end{align} 
where we follow the standard convention of suppressing spinor indices when no confusions would arise, see \emph{e.g.}~\cite{Dreiner:2008tw}.
To achieve properly normalized light-cone coordinates with $n^2 = 0 = \bar{n}^2$ and $n\cdot\bar{n} = 2$, the light-cone spinors must have mass dimension $0$ and satisfy $\xi^\alpha\s \tilde{\xi}_\alpha= 1 = -  \tilde{\xi}^\alpha\s \xi_\alpha$.
Despite choosing explicit light-cone directions, we can restore Lorentz invariance of the $S$-matrix by imposing the three types of reparametrization invariance (RPI) \cite{Manohar:2002fd, Becher:2014oda, Cohen:2019wxr,Kogut:1969xa} on the Lagrangian.

To reduce from the full $\mathcal{N}=1$ superspace to collinear superspace, we rewrite the standard superspace coordinates $\big(\theta^\alpha, \theta^{\dag\dot\alpha}\big)$ in terms of the light-cone spinors and two Grassmann numbers $\bigl\{ \eta, \tilde{\eta} \bigr\} = 0$ with mass dimension $-1/2$: 
\begin{align}
&\ \theta^\alpha = \xi^\alpha\s \eta + \tilde{\xi}^{\alpha}\s \tilde{\eta}\, , \qquad \theta^{\dagger\dot{\alpha}} = \eta^\dagger \s \xi^{\dagger \dot{\alpha}} + \tilde{\eta}^\dagger \s \tilde{\xi}^{\dagger \dot{\alpha}} \, .
\label{eq:theta}
\end{align}
Collinear superspace~\cite{Cohen:2016jzp,Cohen:2016dcl} then corresponds to the following restriction~\cite{Cohen:2018qvn}:
\begin{align}
\tilde{\eta} = 0 \qquad \Longrightarrow \qquad \text{``collinear superspace''}\,.
\label{eq:etaEq0}
\end{align}
One implication is that the $\theta_\alpha$ expansion of a standard superfield no longer includes $F$- and $D$-terms as its highest component, since $\theta^\alpha\s \theta_\alpha = 2\, \eta \s \tilde{\eta}\, \Longrightarrow\, 0$.
As was shown in our companion paper~\cite{Cohen:2018qvn}, one can then express $\mathcal{N} = 1$ SUSY Yang-Mills on the light cone using only propagating degrees of freedom.\footnote{Formulating the theory with this approach has similarities with on-shell superspace~\cite{Gates:1982an, Brink:1980cb, Elvang:2013cua}, with the important difference that we do not require fields to be exactly on the mass shell.}

Our goal here is to understand how to re-introduce the non-propagating degrees of freedom within the collinear superspace framework.
This is required in order to implement generic $\mathcal{N} = 1$ theories --- like Wess-Zumino models and gauge theories with charged chiral matter --- where the $F$- and $D$-terms have non-trivial equations of motion. 
Remarkably, this can be accomplished in collinear superspace even though $\tilde{\eta} = 0$.  
The key is to introduce novel superfields with exotic quantum numbers, constraints, and RPI transformation properties.

In the case of generic Wess-Zumino models with chiral multiplets $\bPhi$, we will introduce fermionic superfields $\bU$ and $\bC$ with mass dimension $+3/2$ (instead of the standard bosonic superfields with mass dimension $+1$).
These satisfy the following constraints: 
\begin{equation}
\bar{\D} \bU = 0\,, \qquad \bar{\D} \bC = -\sqrt{2}\s i\, \PP_\perp^* \bPhi\,.
\end{equation}
Here, $\D$ is a covariant derivative in collinear superspace and $\PP_\perp$ is a space-time derivative perpendicular to the light cone.
While $\bU$ (despite being fermionic) satisfies the standard constraint for a collinear chiral multiplet, $\bC$ satisfies an exotic  ``almost chiral'' constraint that is only possible because Lorentz invariance is obscured.
In the case of massless theories, $\bU$ and $\bC$ can be regarded as auxiliary superfields whose primary role is to make RPI manifest.
This will provide us with a new way to deal with helicity-flipping interactions such Yukawa couplings, without having to introduce sources as in~\Ref{Cohen:2016dcl}.
For massive theories, the fermionic degrees of freedom in $\bU$ and $\bC$ become physical so they must remain in the spectrum.
These exotic multiplets also make it straightforward to incorporate spontaneous SUSY breaking.

In the case of (non-)Abelian gauge theories with charged chiral matter, we will introduce a real superfield $\bD$ with mass dimension $+1$ (instead of the standard real superfield with mass dimension $0$).
Similar to above, one can regard $\bD$ as an auxiliary superfield whose primary role is to make RPI (and gauge symmetry) manifest.
Under RPI-I, this field transforms as
\begin{equation}
\bD \RPIi \bD + \rpiiC\, \bPhialc + \rpii\, \bPhialc^\dagger\,,
\end{equation}
where $\bPhialc$ is a chiral multiplet that contains the two propagating polarizations of the gauge field.
To reduce the number of auxiliary degrees of freedom, we work exclusively in Wess-Zumino gauge (WZG).
Ordinarily, this gauge fixing would break SUSY.
Remarkably, if we also fix to light-cone gauge (LCG), then both collinear SUSY and (residual) gauge invariance can be made manifest using just $\bPhialc$ and $\bD$.
So while our collinear superspace construction involves only a subset of the gauge auxiliary fields, we can still maintain manifest collinear SUSY despite working in WZG, in contrast with the standard $\mathcal{N} = 1$ treatment.

As in the companion paper~\cite{Cohen:2018qvn}, we will take a ``bottom-up'' perspective.
After identifying the relevant multiplets and their symmetry properties, we write down all interactions consistent with the rules of effective field theory.
Our Lagrangians will be manifestly invariant under gauge symmetries, collinear SUSY, RPI-I, and RPI-III.
Specifically for gauge theories with charged matter, we use $\bPhialc$ and $\bD$ to define gauge covariant derivatives as well as covariant versions of $\bU$ and $\bC$. 
As argued in \Ref{Cohen:2018qvn}, the collinear superspace restriction in \Eq{eq:etaEq0} is incompatible with RPI-II, since this transformation acts to rotate $\tilde{\eta}$.
While it is not manifest in collinear superspace, we can still impose RPI-II on the component action.
In this way, the final Lagrangian has the full Lorentz symmetry, and therefore, full $\mathcal{N} = 1$ SUSY.
In \Ref{Cohen:2018qvn}, RPI-II was enforced ``by hand'' by guessing a form for the RPI-II transformation rules and showing compatibility with the RPI algebra.
Here, the exotic superfields $\bU$, $\bC$, and $\bD$ allow for a straightforward realization of RPI-II, and we indeed recover the transformation rules from \Ref{Cohen:2018qvn}
after integrating out the off-shell degrees of freedom.

In the spirit of coming full circle, we will also provide a ``top-down'' perspective on the origin of the exotic superfields $\bU$, $\bC$, and $\bD$.
In retrospect, the main point is rather elementary.
In full $\mathcal{N} = 1$ superspace, a generic superfield $\bS$ can be written as a function of $\big(\eta, \eta^\dagger, \tilde{\eta}, \tilde{\eta}^\dagger\big)$.
Setting $\tilde{\eta} = 0$ will result in loss of information unless, just as for a Taylor expansion, one also knows all derivatives of $\bS$ with respect to $\tilde{\eta}$ and $\tilde{\eta}^\dagger$ at the origin.
In this way, the exotic superfields are effectively the higher-order coefficients of the $\big(\tilde{\eta},\tilde{\eta}^\dagger\big)$ expansion of a full $\mathcal{N} = 1$ superfield.
We speculate that this construction will be useful in understanding theories with $\mathcal{N} > 1$ SUSY, since one could potentially make (part of) the SUSY transformations manifest in collinear superspace at the expense of making (part of) the Lorentz transformations obscure.

The rest of this paper is organized as follows.
In \Sec{sec:formalism} we review the key results of the companion paper \cite{Cohen:2018qvn}.
In \Secs{sec:Fterm}{sec:WZ}, we introduce the fields needed to construct interacting Wess-Zumino models and give both a bottom-up and top-down derivation of the required collinear superspace Lagrangian.
In \Secs{sec:SailingForTheD}{sec:AbelianGaugeTheory}, we follow the same logic for the case of Abelian gauge theories, leaving a discussion of the non-Abelian case to \App{sec:nonAbelianGauge}.
Putting these ingredients together in \Sec{sec:Landfall}, we construct Abelian gauge theories with charged matter, which require the full set of auxiliary multiplets.
We conclude in \Sec{sec:Outlook} with our reflections on the process of constructing collinear superspace and a sketch of future research directions.

\section{Navigating Collinear Superspace}
\addtocontents{toc}{\protect\vspace{2.5pt}}
\label{sec:formalism}

We begin with an overview of our companion paper \emph{Navigating Collinear Superspace}~\cite{Cohen:2018qvn}, with an emphasis on the notation that will be needed here.
We review how to encode residual Lorentz invariance via RPI transformations, how to construct standard collinear superfields which contain only propagating degrees of freedom, and how to build Lagrangians in collinear superspace.\footnote{Regarding conventions, we work with the mostly minus metric, the two-component spinors formalism of~\Ref{Dreiner:2008tw}, and the SUSY formulation from pages 449--453 of~\Ref{Binetruy:2006ad}.  A review of the standard light-cone conventions is given in \Ref{Leibbrandt:1983pj}, although we do not follow their notation in many cases.}

\subsection{Light-Cone Decomposition}
\label{eq:lightcone}
In \Ref{Cohen:2018qvn}, we identified a subset of the SUSY algebra by projecting the SUSY generators with the bosonic spinor $\xi^\alpha$.
We can obtain the full set of $\mathcal{N} = 1$ supercharges by also projecting with $\tilde{\xi}$:%
\begin{align}
\label{eq:SUSYcharges}
\Q^{\full} \equiv \xi^\alpha \qq_\alpha^\full\,, \qquad \tilde{\Q}^{\full} \equiv \tilde{\xi}^\alpha \qq_\alpha^\full \,. 
\end{align}
Keeping $\tilde{\eta} \neq 0$ for now, the operator expressions for these generators are
\begin{align}
\label{eq:collinearSUSYcharges}
\Q^{\full} =i\s \frac{\partial}{\partial \eta}  - \eta^\dagger\, \PP - \tilde{\eta}^\dagger\, \PP_\perp\,, \qquad \tilde{\Q}^{\full} =     i \s\frac{\partial}{\partial \tilde{\eta}} - \eta^\dagger\,  \PP_\perp^*  - \tilde{\eta}^\dagger\, \tilde{\PP}  \,,
\end{align}
where we have defined the projected derivatives 
\begin{align}
\PP &=  \bar{n} \cdot \partial = \xi^\alpha\, \big(\sigma \cdot \partial\big)_{\alpha \dot{\alpha}} \, \xi^{\dagger \dot{\alpha}}, &
\tilde{\PP} &= n \cdot \partial = \tilde{\xi}^\alpha\, \big(\sigma \cdot \partial\big)_{\alpha \dot{\alpha}} \, \tilde{\xi}^{\dagger \dot{\alpha}} \, ,\notag\\[7pt] 
\PP_\perp &=  \xi^\alpha\, \big(\sigma \cdot \partial\big)_{\alpha \dot{\alpha}} \, \tilde{\xi}^{\dagger \dot{\alpha}},&
\PP_\perp^* &= \tilde{\xi}^\alpha\, \big(\sigma \cdot \partial\big)_{\alpha \dot{\alpha}} \, \xi^{\dagger \dot{\alpha}}\,.
\end{align}
In deriving \Eq{eq:collinearSUSYcharges}, we have used the identity $\sigma^\mu \partial_\mu =  \tilde{\xi}\s \tilde{\xi}^\dagger\, \PP + \xi\s \xi^\dagger\, \tilde{\PP} + \xi\s \tilde{\xi}^\dagger\, \PP_\perp^* + \tilde{\xi}\s \xi^\dagger\, \PP_\perp$, which implies
\begin{align}
\theta\, \sigma \cdot \partial\, \theta^\dagger = \eta\s \eta^\dagger\, \PP + \eta \s\tilde{\eta}^\dagger\, \PP_\perp + \tilde{\eta} \s\eta^\dagger\, \PP_\perp^* + \tilde{\eta}\s \tilde{\eta}^\dagger\, \tilde{\PP}\,.
\end{align}

In collinear superspace with $\tilde{\eta} = 0$, the SUSY algebra reduces to
\begin{align}
\label{eq:collinearSUSYalgebra}
\bigl\{\Q, \Q^\dagger \bigr\} = 2\s i\, \PP \, , \quad \text{where} \quad \Q = \Q^{\full} \bigg|_{\tilde{\eta}=0} = i\s  \frac{\partial}{\partial \eta}  - \eta^\dagger \PP\,. 
\end{align}
Remarkably, this reduced algebra encodes the full structure of $\mathcal{N} = 1$ SUSY, as long as one also imposes RPI, as discussed further in \Sec{subsec:RPIIsuperspace}.

Next, we project the superspace covariant derivative operator along the $\xi$ and $\tilde{\xi}$ directions:
\begin{align}
\mathcal{D}_\alpha^\full = \frac{\partial }{\partial \theta_\alpha} - i\s \big(\sigma \cdot \partial\big)_{\alpha \dot{\alpha}} \,\theta^{\dagger \dot{\alpha}} & =  \tilde{\xi}_\alpha\s \frac{\partial}{\partial \eta} - \xi_\alpha \s\frac{\partial}{\partial \tilde{\eta}} -i\s \big(\sigma \cdot \partial\big)_{\alpha \dot{\alpha}} \big(\eta^\dagger\s \xi^{\dagger \dot{\alpha}} + \tilde{\eta}^\dagger\s \tilde{\xi}^{\dagger \dot{\alpha}}\big) \notag\\[7pt] 
& =  \tilde{\xi}_\alpha\, \D^\full - \xi_\alpha\, \tilde{\D}^\full \, . 
\label{eq:DFullDecomposed}
\end{align}
More explicitly, we have:
\begin{align}
\label{eq:Dfull}
 \nonumber
\D^\full  \equiv \xi^\alpha\s \mathcal{D}_{\alpha}^\full &=    \frac{\partial}{\partial \eta} - i\s   \eta^\dagger\, \PP -i\s \tilde{\eta}^\dagger \s\PP_\perp \,, &
\bar{\D}^\full  \equiv \bar{\mathcal{D}}_{\dot{\alpha}}^\full\s \xi^{\dagger \dot{\alpha}} &=   \frac{\partial}{\partial \eta^\dagger}- i \s \eta\s \PP - i \s \tilde{\eta}\, \PP_\perp^* \,,\\[5pt]   
\tilde{\D}^\full  \equiv \tilde{\xi}^\alpha\s \mathcal{D}_{\alpha}^\full &=     \frac{\partial}{\partial \tilde{\eta}} - i\s \eta^\dagger\, \PP_\perp^* - i\s \tilde{\eta}^\dagger\, \tilde{\PP} \,, &
\tilde{\bar{\D}}^\full   \equiv \bar{\mathcal{D}}_{\dot{\alpha}}^\full\s \tilde{\xi}^{\dagger \dot{\alpha}} & = \frac{\partial}{\partial \tilde{\eta}^\dagger}- i\s  \eta\, \PP_\perp - i\s \tilde{\eta}\, \tilde{\PP}\,.
\end{align}
The non-trivial anti-commutation relations between these operators are:
\begin{align}
\Bigl\{\D^\full, \bar{\D}^\full \Bigr\} &= - 2\s i\, \PP \, ,&
\Bigl\{\tilde{\D}^\full, \tilde{\bar{\D}}^\full \Bigr\} &= - 2\s i\, \tilde{\PP}  \, , \nonumber \\[8pt] 
\Bigl\{\D^\full, \tilde{\bar{\D}}^\full \Bigr\} &= - 2\s i\, \PP_\perp \, ,&
\Bigl\{\tilde{\D}^\full, \bar{\D}^\full \Bigr\} &= - 2\s i\, \PP^*_\perp \, ,
\label{eq:DanticommAll}
\end{align}
which will be useful for performing matching calculations in the following sections.
It is also convenient to express the Lorentz-invariant combinations as
\begin{align}
\bar{\mathcal{D}}^\text{full}_{\dot{\alpha}}\, \bar{\mathcal{D}}^{\text{full}\,\dot{\alpha}}  &= \tilde{\bar{\D}}^\text{full}\s\bar{\D}^\text{full} -  \bar{\D}^\text{full}\s \tilde{\bar{\D}}^\text{full} = - 2\, \bar{\D}^\text{full}\s \tilde{\bar{\D}}^\text{full} \,, \notag\\[5pt]
\mathcal{D}^{\text{full}\,\alpha}\, \mathcal{D}^\text{full}_\alpha &= \D^\text{full}\s \tilde{\D}^\text{full} - \tilde{\D}^\text{full}\s \D^\text{full} = 2\,\D^\text{full}\s \tilde{\D}^\text{full} \,.
\label{eq:DsquaredSimplification}
\end{align}

Going to collinear superspace where $\tilde{\eta} = 0$, we define 
\begin{align}
\D \equiv \D^\full \s\big|_{\tilde{\eta}=0} =  \frac{\partial}{\partial \eta} - i \s  \eta^\dagger\, \PP\,, \notag  \quad \quad \,\,
\bar{\D} \equiv \bar{\D}^\full \s \big|_{\tilde{\eta}=0} =  \frac{\partial}{\partial \eta^\dagger}- i\s  \eta\, \PP\,, 
\end{align}
which satisfy $\bigl\{\D, \bar{\D} \bigr\} = - 2\s i\, \PP$.
These collinear covariant derivatives can be used to enforce constraints in collinear superspace and to construct new collinear superfields.
For example, a collinear chiral multiplet satisfies $\bar{\D}\s \bPhi = 0$; see \Sec{subsec:chiralmultiplet} below.
Because $\D^2 = 0 = \bar{\D}^2$, acting $\bar{\D}$ on a generic superfield always yields a chiral superfield.
Similarly, we can define a chiral projector that acts as the identity operator on chiral multiplets:
\begin{equation}
\label{eq:chiralprojector}
\frac{i\s \bar{\D} \D}{2\, \PP} \s\bPhi = \bPhi\,.
\end{equation}

As discussed in \Ref{Cohen:2018qvn}, standard collinear chiral multiplets do not contain $F$-term auxiliary fields, which (naively) presents an obstruction to constructing superpotential-like terms in collinear superspace.
To overcome this issue, we will rely on tilded operators that contain $\partial/ \partial \tilde{\eta}$.
Note that there is no well-defined meaning to ``$\tilde{\D}$'' or ``$\tilde{\bar{\D}}$'' in collinear superspace; for this reason, $\tilde{\D}^\full$ and $\tilde{\bar{\D}}^\full$ were not defined in~\Ref{Cohen:2018qvn}.  
Nevertheless, we can still invoke $\tilde{\D}^\full$ and $\tilde{\bar{\D}}^\full$ \emph{prior} to imposing $\tilde{\eta} = 0$, and this will enable top-down matching of the full $\mathcal{N} = 1$ theory down to collinear SUSY.
For example, in \Sec{subsec:topdownchiral} we define the exotic multiplet $\bC \equiv \tilde{\D}^\full\s \bPhi^\full \big|_{\tilde{\eta}=0}$, whose highest component is the $F$-term auxiliary field needed to construct superpotential interactions.

To simplify the notation, we drop the ``\full'' label when no confusions would arise.
When an unlabeled operator acts on a field that caries a ``\full'' label, then it should interpreted as the full Lorentz covariant operator, \emph{e.g.}\ $\tilde{\D}\s \bPhi^\full \equiv \tilde{\D}^\full\s \bPhi^\full$.
Otherwise, objects should be interpreted as the collinear superspace operators with $\tilde{\eta} = 0$.

\subsection{Reparametrization Invariance}

By explicitly choosing the bosonic spinors $\xi_\alpha$ and $\tilde{\xi}_\alpha$, we are formulating the theory on the light cone defined by \Eq{eq:nAndnBar}.
The transformations that preserve the orthogonality conditions $\xi\s \tilde{\xi} = 1 = -\tilde{\xi}\s  \xi$ correspond to the Lorentz generators that are preserved on the light cone, \emph{i.e.}, the RPI generators.
Via \Eq{eq:nAndnBar}, RPI also preserves the orthogonality conditions $n^2 = 0 = \bar{n}^2$ and $n\cdot \bar{n} = 2$.

\begin{table}[t]
\renewcommand{\arraystretch}{1.8}
\setlength{\arrayrulewidth}{.3mm}
\centering
\small
\setlength{\tabcolsep}{0.45 em}
\begin{tabular}{ |c || c | c | c|}
\hline
 Object &  RPI-I & RPI-II &   RPI-III    \\
 \hline
 \hline
 $\xi^\alpha  $ &     $\xi^\alpha$    & $\xi^\alpha + \rpiii \,\tilde{\xi}^\alpha$ &   $e^{-\kappa_{\rm III}/2 }\, \xi^\alpha $  \\
 $\tilde{\xi}^\alpha $  & $\tilde{\xi}^\alpha + \rpii \,\xi^\alpha$  & $\tilde{\xi}^\alpha$ &    $  e^{\kappa_{\rm III}/2 }\, \tilde{\xi}^\alpha $ \\
\hline
  $\eta $  & $\eta- \rpii\, \tilde{\eta}$   & $\eta$ &    $ e^{\kappa_{\rm III}/2 }\,\eta $ \\ 
  $\tilde{\eta} $  & $\tilde{\eta} $   & $\tilde{\eta} - \rpiii\, \eta$ &    $ e^{-\kappa_{\rm III}/2 } \,\tilde{\eta}$ \\
  \hline
    $\PP $ &  $\PP$   & $\PP + \rpiiiC\, \PP_\perp + \rpiii \,\PP_\perp^*$ & $e^{-\kappa_{\rm III}}\, \PP $ \\ 
  $\tilde{\PP} $ &  $  \tilde{\PP} + \rpiiC\, \PP_\perp^* + \rpii\, \PP_\perp$  & $\tilde{\PP} $ &  $e^{\kappa_{\rm III}}\, \tilde{\PP} $ \\ 
 $\PP_\perp $ &   $ \PP_\perp +\rpiiC \, \PP$   & $\PP_\perp + \rpiii\, \tilde{\PP}$ & $\PP_\perp $ \\ 
  $\PP_\perp^* $ &   $ \PP_\perp^*  + \rpii \, \PP $   & $\PP_\perp^* + \rpiiiC\, \tilde{\PP}$ & $\PP_\perp^*  $ \\
  \hline
        $\D^\full $  &  $\D^\full $   & $\D^\full + \rpiii\, \tilde{\D}^\full $  &    $ e^{-\kappa_{\rm III}/2 }\, \D^\full$ \\
    $\bar{\D}^\full $  & $\bar{\D}^\full $ & $\bar{\D}^\full + \rpiiiC\, \tilde{\bar{\D}}^\full $ & $  e^{-\kappa_{\rm III}/2 }\,\bar{\D}^\full$ \\
        $\tilde{\D}^\full $  &  $\tilde{\D}^\full + \rpii\, \D^\full$ &  $\tilde{\D}^\full $ &  $e^{\kappa_{\rm III}/2 }\, \tilde{\D}^\full $ \\
    $\tilde{\bar\D}^\full $  &  $\tilde{\bar\D}^\full + \rpiiC\, \bar{\D}^\full$ &  $\tilde{\bar\D}^\full $ &  $e^{\kappa_{\rm III}/2 } \,\tilde{\bar\D}^\full $ \\
    \hline
\end{tabular}
\caption{
The RPI transformations of the superspace coordinates and derivative operators.
Note that prior to setting $\tilde{\eta} = 0$, the RPI-II transformations are linear.  
These transformations make it clear that the chiral projector in \Eq{eq:chiralprojector} is RPI-III invariant.
}
\label{table:RPIobjects}
\end{table}

The five RPI generators are traditionally grouped into three categories:
\begin{align}
 \xi &\RPIi \xi \,, & \tilde{\xi} &\RPIi \tilde{\xi}+ \rpii \,  \xi\,, \label{eq:RPI1}\\[2pt]  
 \xi &\RPIii \xi+ \rpiii \, \tilde{\xi}\,,& \tilde{\xi} &\RPIii \tilde{\xi}\,, \label{eq:RPI2}\\[2pt] 
 \xi &\RPIiii e^{-\rpiiii/2} \, \xi\,, &\tilde{\xi} &\RPIiii e^{\rpiiii/2} \, \tilde{\xi}\,, \label{eq:RPI3}
\end{align}
where the $\kappa_\text{I,II}$ are complex numbers and $\rpiiii$ is real. 
A sixth generator, corresponding to imaginary values of $\rpiiii$, is usually elided since it preserves $n^\mu$ and $\bar{n}^\mu$.
Because the superspace coordinates $\theta^\alpha$ and $\bar{\theta}^{\dot{\alpha}}$ are inert under RPI, these transformation imply analogous transformations for $\eta$ and $\tilde{\eta}$, as summarized in \Tab{table:RPIobjects}.

Transformations of matter and gauge component fields will be given in \Tabs{table:RPIfieldsChiral}{table:RPIfieldsGauge}, respectively, once we have introduced the relevant objects.
Likewise, superfield transformations will be given in \Tabs{table:RPIsuperfieldsChiral}{table:RPIsuperfieldsGauge}.
Since many of the derivations are given in~\Ref{Cohen:2018qvn}, we will not restate them here.

Using \Eq{eq:RPI2}, the action of RPI-II transforms $\tilde{\eta}$~\cite{Cohen:2018qvn}.
In collinear superspace where $\tilde{\eta} = 0$, RPI-II cannot be realized at the level of superfields, and only RPI-I and RPI-III can be made manifest.
Nevertheless, to ensure the underlying Lorentz invariance of the theory, RPI-II must be maintained at the component level. 
The non-trivial role of RPI-II will be discussed further in \Sec{subsec:RPIIsuperspace}.

\subsection{Chiral Superfields}
\label{subsec:chiralmultiplet}

We will be working with collinear superfields, which are functions of the space-time coordinate $x^\mu$ and the superspace coordinate $\eta$.
As in ordinary superspace, one can define various constrained superfields.
Collinear chiral superfields satisfy the constraints $\bar{ \D}\s  \bPhi = 0$, and they take the form \cite{Cohen:2018qvn}:
\begin{align}
 \label{eq:PhiSuperfield}
\bPhi\big(x^\mu, \eta, \eta^\dagger\big) = \phi + \sqrt{2} \, \eta\s u + i\s \eta^\dagger\s \eta \, \PP \phi \,,
 \end{align}
where $u \equiv \xi^\alpha \s u_\alpha$ describes the propagating fermion helicity.

To gain more intuition for \Eq{eq:PhiSuperfield}, we can start from a general $\mathcal{N}=1$ superfield with no external Lorentz indices, and Taylor expand it in terms of $\theta$:
\begin{align}
\boldsymbol{S}\big(x^\mu, \theta^\alpha, \theta^{\dagger \dot{\alpha}}\big) = \phi + \theta^\alpha\s u_\alpha &+ \theta^\dagger_{\dot{\alpha}}\s \chi^{\dagger \dot{\alpha}} + \theta^\alpha\s \theta_\alpha\, F + \theta^\dagger_{\dot{\alpha}}\s \theta^{\dagger \dot{\alpha}} \,N + \theta^\alpha\s \sigma^\mu_{\alpha \dot{\alpha}}\s \theta^{\dagger \dot{\alpha}}\, v_\mu   \nonumber \\[4pt]
& + \theta^\alpha\s \theta_\alpha \s\theta^{\dagger}_{\dot{\alpha}}\s \lambda^{\dagger \dot{\alpha}} + \theta^\dagger_{\dot{\alpha}}\s \theta^{\dagger \dot{\alpha}}\s \theta^\alpha\s M_\alpha + \theta^\alpha\s \theta_\alpha \s \theta^\dagger_{\dot{\alpha}}\s \theta^{\dagger \dot{\alpha}}\, D\,, 
 \end{align}
where $\phi$, $F$, $N$, $D$ are complex scalar fields, $u_\alpha$, $\chi_\alpha$, $\lambda_\alpha$, $M_\alpha$ are Weyl fermions, and $v_\mu$ is a vector field.
A standard $\mathcal{N} = 1$ chiral superfield satisfies the constraint $\mathcal{D}_\alpha^\full\s \bPhi^{\full} = 0$, which is equivalent to imposing both 
$\bar{ \D} \s \bPhi^{\full} = 0$ and $\tilde{\bar{ \D}} \s \bPhi^{\full} = 0$:
 \begin{align}
 \label{eq:PhiFulltheta}
\hspace{-7pt} \bPhi^{\full} = \phi - i\s \theta\s \sigma^\mu\s \theta^\dagger\, \partial_\mu \phi + \frac{1}{4}\, \theta \theta \theta^\dagger \theta^\dagger\, \Box\s \phi + \sqrt{2}\, \theta^\alpha u_\alpha - \frac{i}{\sqrt{2}}\, \theta\s \theta\s \left( \partial_\mu u^\alpha\s (\sigma^\mu)_{\alpha \dot{\alpha}}\s \theta^{\dagger \dot{\alpha}}  \right) + \theta \theta\s F \,. 
\end{align}
Following \Ref{Cohen:2018qvn}, we will working with $x^\mu$ (instead of $y^\mu = x^\mu + i\s \theta^\dag \bar{\sigma}^\mu \theta$) coordinates throughout this paper.

To go from ordinary superspace to collinear superspace, we can use \Eq{eq:theta} to expand \Eq{eq:PhiFulltheta} in $\big(\eta, \eta^\dagger\big)$ coordinates, for instance through the replacements $\theta^\alpha\s \theta_\alpha = 2\,  \eta\s \tilde{\eta}$ and $ \theta\s \sigma \cdot \partial\s \theta^\dagger= \eta\s \eta^\dagger\, \PP + \eta\s \tilde{\eta}^\dagger\, \PP_\perp + \tilde{\eta}\s \eta^\dagger\, \PP_\perp^* + \tilde{\eta}\s\tilde{\eta}^\dagger\, \tilde{\PP}$.
Additionally, the two component fermion $u^\alpha$ may be expanded as:  
\begin{align}
\label{eq:fermionExpan}
&u^\alpha =  \tilde{\xi}^\alpha\s u - \xi^\alpha\s \tilde{u} \,,  \quad \text{so that} \quad \xi^\alpha\s u_\alpha = u\, , \quad \tilde{\xi}^\alpha \s u_\alpha = \tilde{u}\,,
\end{align}
which projects out the two helicity states of $u^\alpha$ \cite{Cohen:2018qvn,Cohen:2016dcl}.
With these replacements, \Eq{eq:PhiFulltheta} becomes:
\newpage
 \begin{align}
 \label{eq:PhiFull}
  \bPhi^{\full} & =      \phi - i \left( \eta\s \eta^\dagger \,\PP + \eta\s \tilde{\eta}^\dagger\, \PP_\perp + \tilde{\eta}\s \eta^\dagger \,\PP_\perp^* + \tilde{\eta}\s\tilde{\eta}^\dagger\, \tilde{\PP} \right)\phi + \tilde{\eta}\s\eta\s \tilde{\eta}^\dagger\s \eta^\dagger \big(\PP\s \tilde{\PP} - \PP_\perp^*\s \PP_\perp \big)\s \phi \nonumber\\[5pt] 
 & \hspace{12pt}+ \sqrt{2}\, \big(\eta\s u +  \tilde{\eta}\s\uu\big)  + i\s \sqrt{2}\, \tilde{\eta}\s \eta\s \Bigl[ \eta^\dagger \left(\PP_\perp^*\s u - \PP\s \tilde{u} \right) + \tilde{\eta}^\dagger \left(\tilde{\PP}\s u - \PP_\perp\s \tilde{u} \right)  \Bigr] - 2\, \tilde{\eta}\s\eta\s F  \,,
 \end{align}
 where we have used $\partial_\mu\s u^\alpha (\sigma^\mu)_{\alpha \dot{\alpha}}\, \theta^{\dagger \dot{\alpha}}  
=  \eta^\dagger \left(\PP_\perp^* u - \PP\s \tilde{u} \right) + \tilde{\eta}^\dagger \left(\tilde{\PP}\s u - \PP_\perp\s \tilde{u} \right) $.
We emphasize that the above expression is still valid in the full $\mathcal{N} = 1$ superspace.

Now going to collinear superspace by setting $\tilde{\eta} = 0$, the chiral superfield reduces to \Eq{eq:PhiSuperfield}:
\begin{equation}
\label{eq:bPhi_definition}
\bPhi \equiv \bPhi^\full \Big|_{\tilde{\eta} = 0} = \phi + \sqrt{2}\,  \eta\s u + i\s \eta^\dagger\s \eta \, \PP\s \phi \,.
\end{equation}
Note that the constraint $\tilde{\bar{ \D}}\s  \bPhi^{\full} = 0$ no longer makes sense when $\tilde{\eta} = 0$, but $\bar{ \D}\s  \bPhi = 0$ still holds.
This collinear chiral superfield has two bosonic (the complex scalar $\phi$) and two fermionic (the complex Grassmann scalar fermion $u$) degrees of freedom.
It does not include an auxiliary $F$-term field, even though the full expression in \Eq{eq:PhiFull} does.
We will see how to recover these $F$-terms in \Sec{sec:Fterm}.

\subsection{Lagrangians in Collinear Superspace}
\label{eq:Lagrangian}
To build a SUSY Lagrangian, we can follow the canonical strategy of writing it as an explicit total derivative, but now in \emph{collinear} superspace:%
\footnote{For those unfamiliar with expressing superspace Lagrangians in terms of super-derivatives as opposed to integrating over superspace, see \emph{e.g.}~\cite{Bertolini:2013via}.}
\begin{align}
\label{eq:genericL}
\mathcal{L} \sim \D\s \bar{\D} \big[\dots \big] \big|_{0}\,,
\end{align}
where the $|_0$ notation indicates the restriction to $\eta = 0 = \eta^\dag$.
This ensures that the resulting Lagrangian preserves collinear SUSY.
(It may or may not preserve the full $\mathcal{N} = 1$ SUSY depending on RPI-II; see \Sec{subsec:RPIIsuperspace}.)
In \Sec{subsec:massterm}, we will also consider a superpotential-like term of the form:
\begin{align}
\label{eq:genericLalt}
\mathcal{L} \sim \D \big[\dots \big] \big|_{0} + \text{h.c.}\,,
\end{align}
where the $\dots$ represent a fermionic combination of superfields that is chiral up to a total derivative (see \Sec{subsec:singleflavor} below for details).

As an example, the kinetic term for a single chiral superfield is:%
\footnote{\label{footnote:sqrt2rule} The $1/2$ prefactor can be traced to the normalization of the SUSY derivatives $\D$ and $\bar{\D}$ that descends from the factor of 2 in the definition of the supercharges; see \Eq{eq:collinearSUSYalgebra}.  The rule of thumb is that every superderivative comes with a factor of $1/\sqrt{2}$.} 
\begin{align}
\label{eq:SUSYSCETLagrangian}
\mathcal{L}  &= \frac{1}{2}\, \D\s \bar{\D} \bigg[ \bPhi^\dagger \frac{i\s \Box}{\PP}  \bPhi \bigg] \bigg|_{0}   = -\phi^* \Box \phi + i\s u^\dag \frac{\Box}{\PP} u\,.
\end{align}
Using the transformations in \Tabs{table:RPIobjects}{table:RPIfieldsChiral}, it is straightforward to see that this operator is RPI-I and RPI-III invariant.
Despite the appearance of $\PP$ in the denominator, this Lagrangian yields a local theory.
Recall that $\D$ and $\bar{\D}$ are mass dimension $1/2$ objects, such that the operator in the brackets of \Eq{eq:SUSYSCETLagrangian} has mass dimension three as required.%
\footnote{As in \Ref{Cohen:2018qvn}, we perform power counting based on mass dimension, which one can show is equivalent to the SCET power counting $(\tilde{\PP}, \PP, \PP_\perp) \sim Q (\lambda^2, 1, \lambda)$ when RPI-III is taken into account.}

\subsection{The Role of RPI-II}
\label{subsec:RPIIsuperspace}

The collinear SUSY algebra in \Eq{eq:collinearSUSYalgebra} closes and is compatible with RPI-I and RPI-III transformations.
As discussed in \Ref{Cohen:2018qvn}, collinear superspace is not compatible with RPI-II.
The reason is that RPI-II acts as a rotation on the $\bar{n}^\mu$-direction, which in superspace corresponds to a translation of $\tilde{\eta}$.
This is incompatible with the defining constraint of collinear superspace: $\tilde{\eta} =0$.
In the context of chiral multiplets, the constraint $\bar{ \D}\s \bPhi = 0$ is not RPI-II invariant, since the collinear superspace derivative $\D$ has a non-trivial RPI-II shift inherited from $\PP$:  $\D \to \D - \eta^\dagger \big(\rpiiiC\, \PP_\perp + \rpiii\, \PP_\perp^*\big)$.

To ensure Lorentz invariance of the theory, we must enforce RPI-II.
This in turn ensures that the theory respects (at least) $\mathcal{N} = 1$ SUSY, because \Eq{eq:collinearSUSYalgebra} is a graded algebra, and $\mathcal{N} = 1$ SUSY is the smallest graded algebra compatible with Lorentz invariance~\cite{Haag:1974qh}.

While we cannot impose RPI-II directly in collinear superspace, we can impose it at the level of the component action.
As an example, consider the Lagrangian for a free chiral multiplet in \Eq{eq:SUSYSCETLagrangian}.
The scalar term is manifestly invariant under RPI-II.
However, the fermionic mode $u$ shifts as
\begin{align}
\label{eq:uLRPIii}
u \RPIii  u + \rpiii \, \tilde{u}\,.
\end{align}
Since $\tilde{u}$ does not appear in \Eq{eq:SUSYSCETLagrangian}, the fermion kinetic term is not invariant under this (linear) RPI-II transformation.
This is not surprising, since RPI-II deforms the theory away from the collinear slice chosen in \Eq{eq:nAndnBar}.

As discussed in \Ref{Cohen:2018qvn}, \Eq{eq:SUSYSCETLagrangian} does respect an alternative realization of RPI-II:
\begin{align}
\label{eq:uNLRPIii}
u  \RPIii u + \rpiii \, \frac{\PP_\perp^*}{\PP}  u\,.
\end{align}
From the bottom up, this can be derived by writing down a candidate RPI-II transformation law consistent with RPI-I and RPI-III and checking that it satisfies the RPI algebra.
From the top down, this can be derived by integrating out $\tilde{u}$ from the full Lorentz-invariant theory.
In this paper, we will restore $\tilde{u}$ to the Lagrangian such that we can use the linear RPI-II transformation in \Eq{eq:uLRPIii} throughout.

\subsection{Abelian Gauge Transformations}
\label{subsec:gaugetheoryReview}

The last topic we review is Abelian gauge fields in collinear superspace.
Starting from a four-component Abelian gauge field $A_\mu$, we can project onto a complex scalar: 
\begin{align}
\alc =\frac{1}{\sqrt{2}}\, \xi^\alpha\s (\sigma \cdot A )_{\alpha \dot{\alpha}}\, \tilde \xi^{\dagger \dot{\alpha}}\,, \qquad \alc^* = \frac{1}{\sqrt{2}}\, \tilde{\xi}^\alpha\s (\sigma \cdot A )_{\alpha \dot{\alpha}} \,\xi^{\dagger \dot{\alpha}}\,.
\end{align}
In LCG, $\alc$ contains the two propagating degrees of freedom, \emph{i.e.}, those that are transverse to the light cone.
These can then be packaged together with the corresponding propagating gaugino $\lambda \equiv \xi^\alpha\s \lambda_\alpha$ into a collinear chiral superfield with $\bar{\D}\s \bPhialc  = 0$: 
\begin{align}
\bPhialc = \alc^* - \sqrt{2} \, i\s \eta\s \lambda^\dagger + i \eta^\dagger\s \eta \, \PP\s \alc^*\,.
\label{eq:bPhialc}
\end{align}
Note that \Eq{eq:bPhialc} contains the \emph{conjugate} degrees of freedom $\alc^*$ and $\lambda^\dagger$ (and the anti-chiral superfield will similarly contain $\alc$ and $\lambda$).
The justification for this seemingly unusual form will be given in \Sec{subsec:TopDownSuperfieldsGauge}.

In the present work, we are interested in ``rediscovering" the non-propagating degrees of freedom, so it is useful to identify the two remaining degrees of freedom in $A_\mu$:
\begin{align}
\bar n \cdot A = \xi^\alpha\s (\sigma \cdot A)_{\alpha \dot{\alpha}}\, \xi^{\dot{\alpha} \dagger}, \qquad n\cdot A = \tilde \xi^\alpha \s(\sigma \cdot A)_{\alpha \dot{\alpha}}\, \tilde \xi^{\dagger \dot{\alpha}} \,.
\label{eq:nAandnBarA}
\end{align}
In this way, the full $A^\mu$ field can be decomposed as: 
\begin{align}
\left(\sigma \cdot A \right)_{\alpha \dot{\alpha}} = \xi_\alpha\, \xi^{\dagger}_{\dot{\alpha}}\, n\cdot A + \tilde{\xi}_\alpha\, \tilde{\xi}^\dagger_{\dot{\alpha}}\, \bar{n}\cdot A + \sqrt{2} \, \xi_\alpha\, \tilde{\xi}^\dagger_{\dot{\alpha}}\, \alc^* + \sqrt{2} \, \tilde{\xi}_\alpha\, \xi^{\dagger}_{\dot{\alpha}}\, \alc\,.
\end{align}
Throughout this paper, we work in LCG where \cite{Leibbrandt:1987qv}:
\begin{align}
\bar{n}\cdot A = 0\,.
\label{eq:LCG}
\end{align}
This removes one of the degrees of freedom in \Eq{eq:nAandnBarA}, leaving $n\cdot A$ as a non-propagating component of the gauge field.
In \Ref{Cohen:2018qvn}, we integrated out $n\cdot A$ to write a down a gauge kinetic term that only depended on $\bPhialc$.
In \Sec{sec:SailingForTheD}, we will package $n\cdot A$ into a real supermultiplet $\bD$ in order to write down gauge interactions for matter fields.

As discussed in \Ref{Cohen:2018qvn}, there is a residual gauge redundancy even after imposing LCG.
Starting from the standard gauge transformation with local (real) gauge transformation parameter $\omega$: 
\begin{align}
\label{eq:A4vectorGauge}
A_\mu \gauge A_\mu + \partial_\mu \omega \, ,
\end{align}
the restriction in \Eq{eq:LCG} requires $\PP \omega = 0$.
Remarkably, this residual gauge transformation can be parametrized by a superfield $\RCA$ with the unusual property of being both chiral and real (and therefore anti-chiral):
\begin{align}
\D \RCA = 0\, , \quad \bar{\D} \RCA = 0\,, \quad \RCA = \RCA^\dagger \qquad \Longrightarrow \qquad \RCA = \omega\,, \quad \PP \s\omega = 0\,.
\label{eq:RCA}
\end{align}
While the form of \Eq{eq:RCA} may seem strange, we note that $\RCA$ is still a non-trivial collinear SUSY multiplet; although $\PP\s \omega = 0$, the derivative projections along the other directions ($\PP_\perp\s \omega$ and $\tilde{\PP}\s \omega$) are non-vanishing \cite{Cohen:2018qvn}.
This gauge transformation acts on the gauge chiral superfield as
\begin{equation}
\label{eq:Phialcgauge}
\bPhialc \gauge \bPhiA + \s\PP_\perp^*\s \RCA \,,
\end{equation}
which preserves the chirality condition $\bar{\D} \bPhialc = 0$.

\section{Setting Sail for the \emph{F}-term}
\addtocontents{toc}{\protect\vspace{2.5pt}}
\label{sec:Fterm}

The chiral superfield in \Eq{eq:PhiSuperfield} includes only propagating degrees of freedom, which was sufficient in \Ref{Cohen:2018qvn} to write the Lagrangian for free chiral matter.
Here, we are interested in constructing Lagrangians for SUSY theories with non-zero $F$-terms, and therefore new ingredients are required.
In this section, we explain how to encode the degrees of freedom that are ``missing'' from the collinear chiral superfield.
This will allow us to write Lagrangians that involve Yukawa couplings such as the Wess-Zumino model, without appealing to external currents as was done in~\Ref{Cohen:2016dcl}.

The key ingredient is an auxiliary chiral superfield $\bU$, defined in \Eq{eq:Usuperfield} below, containing the non-propagating degrees of freedom:  the opposite helicity fermion $\uu \equiv \tilde{\xi}^\alpha u_\alpha$ and the auxiliary scalar $F$-term.
The unique form of $\bU$ will be derived from the bottom up by relying on consistent SUSY and RPI transformation properties of the components.
We also motivate the exotic constrained superfield $\bC$ by appealing to its RPI-II transformation properties.
Finally, we present a top-down derivation of $\bU$ and $\bC$ by applying $\tilde{\D}$ operators to the full $\mathcal{N}=1$ chiral superfield before reducing to collinear superspace.

\subsection{Component RPI and SUSY Transformations}
\label{subsec:RPIuanduu}

\begin{table}[t]
\renewcommand{\arraystretch}{1.8}
\setlength{\arrayrulewidth}{.3mm}
\centering
\small
\setlength{\tabcolsep}{0.45 em}
\begin{tabular}{ |c || c | c | c | }
    \hline
    Component &  RPI-I &  RPI-II &   RPI-III   \\ \hline  \hline
   $\phi  $ &    $\phi$  &   $\phi$ &    $ \phi$    \\  
   $ u $ &   $u$ &   $ u + \rpiii\, \uu$ &   $  e^{-\rpiiii/2}\, u $   \\ 
   $ \uu $ &   $\uu  +\rpii\, u$ &   $\uu$ &   $  e^{\rpiiii/2}\, \uu $    \\ 
    $F  $ &    $F$  &   $F$ &    $ F$    \\
    \hline   
\end{tabular}
\caption{
The RPI transformations for the chiral matter degrees of freedom.
Here, $u \equiv \xi^\alpha u_\alpha$ is the component of the fermion along the light cone and $\uu \equiv \tilde{\xi}^\alpha u_\alpha$ has the opposite helicity.
With $\uu$ in the spectrum, RPP-II is linearly realized.
}
\label{table:RPIfieldsChiral}
\end{table}

As argued in \Sec{subsec:RPIIsuperspace}, we need to incorporate the opposite helicity fermion $\uu$ in order to have a linear realization of RPI-II.
In \Tab{table:RPIfieldsChiral}, however, we see that $\uu$ transforms non-trivially under RPI-I, so if $\uu$ is going to be a component of a SUSY multiplet, then that whole multiplet must have non-trivial RPI-I transformation properties.
This is in contrast to the $u$ field, which is inert under RPI-I, such that $u$ can be the highest component of the RPI-I invariant $\bPhi$ superfield.
In \Sec{subsec:exoticconstrained}, we will build an exotic multiplet $\bC$ which has $\uu$ directly as the lowest component, but from the bottom-up perspective, it is more convenient to start with combinations of fields that are RPI-I invariant.

Another challenge of working directly with $\uu$ is that collinear SUSY necessarily mixes $\uu$ with the propagating degrees of freedom.
This is easiest to see starting from the full $\mathcal{N}= 1$ SUSY transformations for the standard chiral components, but writing the SUSY transformation parameter as $\zeta_\alpha = \xi_\alpha\s \epsilon + \tilde{\xi}_\alpha\s \tilde{\epsilon}$ and setting $\tilde{\epsilon} = 0$ to isolate the collinear supersymmetry transformation:
\begin{align}
\label{eq:collinearsusyauxcomp}
\delta \phi &= \sqrt{2}\, \epsilon\, u \,, \notag \\[3pt] 
\delta u &= - i\s \sqrt{2}\,  \epsilon^\dagger\, \PP \phi  \, ,\notag \\[3pt]
\delta \uu &= - i\s \sqrt{2}\,   \epsilon^\dagger\, \PP_\perp^* \phi - \sqrt{2}\, \epsilon\, F \, , \notag \\[3pt]
 \delta F &=   i \s\sqrt{2}\,  \epsilon^\dagger (\PP \uu - \PP_\perp^* u)   \, .
\end{align}
As expected given the consistency of $\bPhi$, the $\phi$ and $u$ fields rotate into each other under collinear SUSY.
By contrast, the $\uu$ field mixes into the auxiliary $F$ term (as desired) and also mixes into the propagating $\phi$ field.

Satisfyingly, both of the above complications can be solved by the introduction of the field combination
\begin{equation}
\label{eq:Utildedefinition}
\tilde{U} \equiv \uu - \frac{\PP_\perp^*}{\PP}\,u\,.
\end{equation}
This object is invariant under RPI-I, and it closes with $F$ under collinear SUSY: 
\begin{align}
\label{eq:collinearsusyauxcomp_restrict}
\delta \tilde{U} &= - \sqrt{2}\, \epsilon\, F \, , \notag \\[3pt] 
 \delta F &=   i\s \sqrt{2}\,  \epsilon^\dagger\,\PP \tilde{U} \, .
\end{align}
Note that $\tilde{U} = 0$ yields the equation of motion for free chiral fermions.
When we discuss charged matter in \Sec{sec:Landfall}, we will have to contend with the fact that \Eq{eq:Utildedefinition} does not transform homogeneously under gauge transformations, even after restricting to LCG.
Nevertheless, we find that $\tilde{U}$ is more convenient to work with than $\uu$, and (as we show in \Sec{subsec:covariantchiralaux}) it is possible to construct a covariant version of $\tilde{U}$ by effectively replacing $\PP_\perp^*$ with a gauge covariant version $\nabla_\perp^*$.

\subsection{The Auxiliary Chiral Superfield}
\label{subsec:auxchiral}

Armed with the RPI and collinear SUSY transformations of the component fields, we can now construct the auxiliary chiral superfield which contains the $\uu$ and $F$ fields: 
\begin{align}
\label{eq:Usuperfield}
\bU \big(x^\mu, \eta, \eta^\dagger\big)  &\equiv  \tilde{U} -\sqrt{2}\, \eta\, F + i\s  \eta^\dagger \eta \, \PP \tilde{U} \, ,
\end{align}
where $\tilde{U}$ is defined in \Eq{eq:Utildedefinition}.  
Assuming that the auxiliary field $F$ appears in the highest component of a chiral multiplet, the form of $\bU$ is uniquely fixed by requiring closure under collinear SUSY and homogeneity under both RPI and mass dimension.

Unlike a standard chiral multiplet, $\bU$ is fermionic with mass dimension $3/2$ and RPI-III charge $+1/2$ (instead of bosonic with mass dimension $1$ and RPI-III charge $0$).
These scalings are consistent with the highest component being the auxiliary $F$-term, which has mass dimension $2$ and no RPI-III charge.
By comparing the SUSY transformation properties of $\tilde{U}$ and $F$ in \Eq{eq:collinearsusyauxcomp_restrict} to those of $\phi$ and $u$ in \Eq{eq:collinearsusyauxcomp}, one can conclude that $\bU$ is indeed a valid collinear chiral multiplet satisfying $\bar{\D} \bU = 0$.

For completeness, we write the conjugate multiplet: 
\begin{align}
\bU^\dag  \big(x^\mu, \eta, \eta^\dagger\big) =  \tilde{U}^\dagger - \sqrt{2}\, \eta^\dagger\, F^* - i \s \eta^\dagger \eta \, \PP \tilde{U}^\dagger\,,  \quad\qquad \tilde{U}^\dagger = \uu^\dagger - \frac{\PP_\perp}{\PP}\,u^\dagger\, ,
\end{align}
which is an anti-chiral superfield satisfying $\D \bU^\dag = 0$. 
Since both $\tilde{U} = 0$ and $F = 0$ for a free chiral multiplet, we can consistently write the superfield expression $\bU = 0$, which is why the $\bU$ field did not need to appear in \Ref{Cohen:2018qvn}.

\begin{table}[t]
\renewcommand{\arraystretch}{2}
\setlength{\arrayrulewidth}{.3mm}
\centering
\small
\setlength{\tabcolsep}{0.5 em}
\begin{tabular}{ |c || c | c | c | c | c |}
    \hline
    Superfield &  Construction & Constraint & RPI-I &   RPI-III  & Mass Dim. \\
    \hline \hline
  $ \bPhi $ & $\bPhi^{\full}|_{\tilde{\eta} = 0}$ & $\bar{\D} \bPhi = 0$ &  $\bPhi$ &   $  \bPhi $     &  $1$  \\
    $\bC $ & $\frac{1}{\sqrt{2}}\, \tilde{\D} \bPhi^{\full}|_{\tilde{\eta} = 0}$ & $\bar{\D} \bC = -i\s\sqrt{2}\, \PP_\perp^* \bPhi$ &    $\bC+   \frac{\rpii}{\sqrt{2}} \, \D \bPhi$  &  $e^{\kappa_{\rm III} /2}\, \bC $     &  $3/2$   \\ 
    $\bPhiF^\dagger $ & $ \frac{1}{\sqrt{2}}\, \D \bC$ & $\D \bPhiF^\dagger =0$ &    $\bPhiF^\dagger $  &  $\bPhiF^\dagger $     &  $2$  \\ 
     $\bU $ & $\frac{ i}{\sqrt{2}}\,\frac{\bar{\D}}{ \PP}\,  \bPhiF^\dagger $ & $\bar{\D} \bU = 0$ &    $\bU$  &  $e^{\kappa_{\rm III}/2} \,\bU$     &  $3/2$  \\[4pt]
       \hline 
\end{tabular}
\caption{
Collinear superfields for chiral matter, along with their top-down construction, constraints, and symmetry properties.
Note that RPI-II is omitted since it is only well-defined at the level of components.
The RPI-I and RPI-III charges of a superfield are inherited from its lowest component.
}
\label{table:RPIsuperfieldsChiral}
\end{table}

\subsection{An ``Almost Chiral'' Superfield}
\label{subsec:exoticconstrained}
Though $\bU$ is RPI-I invariant, it has complicated transformation properties under RPI-II owing to the $(\PP_\perp^* / \PP)\, u$ term in $\tilde{U}$.
This complicates the construction of RPI-II invariant Lagrangians, especially for superpotential terms in \Sec{subsec:massterm}.

To address this issue, we introduce an exotic constrained superfield $\bC$, build from the following combination of $\bPhi$ and $\bU$:\footnote{\footnotesize{Since the symbol $\bU$ was already taken, we chose the notation $\bC$ because it looks like $\boldsymbol{U}$ on its side.}}
\begin{align}
\bC (x^\mu, \eta) &\equiv  \, \bU + \frac{1}{\sqrt{2}}\, \frac{\PP_\perp^*}{\PP}\, \D \bPhi \nonumber\\[3pt]
& =    \uu - \sqrt{2}\, \eta\, F - i\s\sqrt{2}\, \eta^\dagger\, \PP_\perp^* \phi - i\s \eta^\dagger \eta\, \big( 2\, \PP_\perp^* u - \PP \uu \big) \,.
\label{eq:Cfield_def}
\end{align}
This superfield has the opposite helicity fermion $\tilde{u}$ as its lowest component, which is RPI-II invariant, though the multiplet as a whole does transform non-trivially under RPI-II.

Like $\bU$, the $\bC$ superfield is a fermionic multiplet with mass dimension $3/2$ and RPI-III charge $+1/2$.
Unlike $\bU$, though, $\bC$ has a non-trivial (but still well-defined) RPI-I transformation:
\begin{align}
\label{eq:RPIiofC}
\bC \RPIi \bC+\frac{\rpii}{\sqrt{2}}\, \D \bPhi \,,
\end{align}
which follows from $\PP_\perp^*  \to \PP_\perp^*  + \rpii \, \PP$.

This exotic object $\bC$ is neither chiral nor real.
It does, however, satisfy an interesting constraint.
Acting on \Eq{eq:Cfield_def} with the $\bar{\D}$ derivative:
\begin{equation}
\label{eq:Cconstraint}
\bar{\D} \bC = -i\s\sqrt{2} \,  \PP_\perp^* \bPhi\,,
\end{equation}
where we used $\big\{\bar{\D},\D\big\} = -2\s i\, \PP$.
It is straightforward to check that \Eq{eq:Cconstraint} is compatible with the RPI-I transformation in \Eq{eq:RPIiofC}.
So while $\bC$ is not a chiral multiplet, it is chiral up to a chiral term.
Therefore, we will call this an ``almost chiral'' superfield.
This property of $\bC$ is rather unfamiliar from the perspective of standard $\mathcal{N} = 1$ superspace,%
\footnote{This ``almost chiral'' constraint is perhaps reminiscent of the constraints on the Ferrara-Zumino multiplet~\cite{Ferrara:1974pz}; see discussion in \Ref{Dumitrescu:2011zz}.}
but we can make use of \Eq{eq:Cconstraint} to avoid tedious component manipulations.

\subsection{Top-Down Derivation of Auxiliary Chiral Superfield}
\label{subsec:topdownchiral}
The derivation of $\bU$ in \Sec{subsec:auxchiral} was bottom up, relying only on the symmetries and mass dimensions of the component fields.
As summarized in \Tab{table:RPIsuperfieldsChiral}, a top-down way to derive $\bU$ is by starting with a full $\mathcal{N} = 1$ chiral multiplet, and then taking collinear superspace derivatives before setting $\tilde{\eta} = 0$.
Along the way, we will rediscover the exotic constrained superfield $\bC$ from \Sec{subsec:exoticconstrained} as well as an anti-chiral superfield $\bPhiF^\dagger$ whose lowest component is the $F$ auxiliary field. 
Recall from \Eq{eq:bPhi_definition} that just setting $\tilde{\eta} = 0$, we have $\bPhi = \bPhi^\full |_{\tilde{\eta} = 0}$.

We begin by exploring what happens when we act with $\tilde{\D}$ on a full chiral superfield $\bPhi^\full$, before reducing to collinear superspace by setting $\tilde{\eta} = 0$:
\begin{equation}
\bC \big(x^\mu, \eta, \eta^\dagger\big) \equiv \frac{1}{\sqrt{2}} \left( \tilde{\D} \bPhi^{\full} \right) \bigg|_{\tilde{\eta} = 0}  =  \uu - \sqrt{2}\, \eta F - i\s\sqrt{2} \, \eta^\dagger\, \PP_\perp^* \phi - i\s  \eta^\dag \eta \big( 2\, \PP_\perp^* u - \PP \uu \big)\,.
\end{equation}
This is precisely the exotic multiplet $\bC$ introduced in \Eq{eq:Cfield_def}.
Because it is constructed using $\tilde{\D}$, it has non-trivial RPI-I transformation properties, as shown in \Eq{eq:RPIiofC}.
We can derive the constraint in \Eq{eq:Cconstraint} from the top down via:
\begin{align}
\label{eq:Cconstraint_topdown}
\bar{\D} \left( \frac{1}{\sqrt{2}}\, \tilde{\D} \bPhi^{\full} \right)= -i\s\sqrt{2} \,  \PP_\perp^* \bPhi^{\full}-  \frac{1}{\sqrt{2}}\, \tilde{\D} \Big(\bar{\D} \bPhi^\full  \Big) &  \quad  \colleta  \quad \bar{\D} \bC = -i\s\sqrt{2} \,  \PP_\perp^* \bPhi \,,
\end{align}
where we used $\big\{\bar{\D}, \tilde{\D} \big\} = - 2\s i\, \PP_\perp^*$ and the chirality constraint $(\bar{\D} \bPhi^{\full})|_{\tilde{\eta} =0} = \bar{\D} \bPhi = 0$.

Next, we can construct a multiplet derived from $\bC$ that is RPI-I invariant and has mass dimension 2: 
\begin{align}
\label{eq:Fmultiplet}
\bPhiF^\dagger \equiv 
 \frac{1}{2} \left(  \D \tilde{\D} \bPhi^\full \right) \bigg|_{\tilde{\eta} = 0} = \frac{1}{\sqrt{2}}\, \D \bC 
 =   - F  -  i\s \sqrt{2}\, \eta^\dagger\, \PP \tilde{U} +  i\s \eta^\dagger \eta\, \PP F  \,.
\end{align}
This multiplet has the auxiliary field $F$ as the lowest component, but we have written it as $\bPhiF^\dagger$ because it satisfies the anti-chirality constraint $\D \bPhiF^\dagger = 0$.

Finally, to derive $\bU$, we act with another superderivative:
\begin{align}
\bU \big(x^\mu, \eta, \eta^\dagger\big) \equiv 
\frac{i}{\sqrt{2}}\s \frac{\bar{\D}}{ \PP}\, \bPhiF^\dagger  = \frac{ i}{2\,  \PP }\left(  \bar{\D} \D \tilde{\D} \bPhi^{\rm full} \right) \bigg|_{\tilde{\eta} = 0}  =  \tilde{U} -\sqrt{2}\, \eta\, F + i\s  \eta^\dagger \eta \, \PP \tilde{U} \,,
\end{align}
which reproduces the component expression in \Eq{eq:Usuperfield}.
Note that $\bar{\D}^2 = 0$ provides insight into why $\bU$ is an chiral superfield, \emph{i.e.}, that it satisfies $\bar{\D} \bU = 0$.
The sequence of steps from $\bPhi^\full$ to $\bU$ is summarized in \Tab{table:RPIsuperfieldsChiral}.

Of course, these fields are not independent.
We can relate $\bU$ to $\bC$ by applying the chiral projector from \Eq{eq:chiralprojector}: 
\begin{equation}
\label{eq:UasCchiralprojector}
\bU = \frac{i\s\bar{\D} \D}{2\, \PP}\, \bC\,.
\end{equation}
Similarly, we can relate $\bPhiF^\dagger$ to $\bU$ through a superspace derivative: 
\begin{equation}
\label{eq:UCFrelations}
\D \bU = \D \bC =  \sqrt{2}\, \bPhiF^\dagger\,.
\end{equation}

We now have both a bottom-up and a top-down way of understanding how to package the auxiliary $F$-term into a collinear superspace multiplet.
While $\bU$ is the most natural object to construct from bottom-up considerations, $\bC$ is more natural from the top-down perspective, and both will be convenient for constructing RPI-II invariant actions.
The field $\bPhiF^\dagger$ plays only a minor role in this paper, though it could in principle be used as a building block since it contains the same component fields.

\newpage
\section{Casting Anchor: The Wess-Zumino Model}
\addtocontents{toc}{\protect\vspace{2.5pt}}
\label{sec:WZ}

We now have the ingredients necessary to rediscover the Wess-Zumino model in the language of collinear superspace.
Recall that the obstruction encountered in~\Ref{Cohen:2016dcl} was due to the fact that the Yukawa interaction necessarily involves flipping the helicity of the fermion, but this opposite-helicity degree of freedom, $\uu$, was already eliminated using the light-cone fermion equations of motion.
In~\Ref{Cohen:2016dcl}, this problem was addressed by introducing an external current to encode the desired interaction.
In this section, we demonstrate a new approach that utilizes the $\bU$ and $\bC$ superfields.

We begin with a bottom-up construction.
Using the transformations given in \Tabs{table:RPIobjects}{table:RPIfieldsChiral}, we can construct Lagrangians that are invariant under both RPI and collinear SUSY, yielding consistent theories involving $\bPhi$, $\bU$, and $\bC$ expressed in collinear superspace.
As already emphasized, RPI-II must be checked at the component level, and here we will see how it leads to a non-trivial relation between the kinetic term for $\bPhi$ and $\bU$, as well as an alternative way of understanding the structure of the superpotential via $\bC$.  
We also provide a top-down derivation of the Wess-Zumino model, in order to show the explicit connection to the full $\mathcal{N} = 1$ theory.

\subsection{The Kinetic Term}
\label{sec:WZKinTerm}

Just as there was a unique kinetic term for the collinear chiral superfield $\bPhi$ in \Eq{eq:SUSYSCETLagrangian}, there is a unique ``kinetic term'' for the auxiliary chiral superfield $\bU$:%
\footnote{We will use the standard terminology of calling the bi-linear term for the auxiliary field a ``kinetic term'' even though it may not propagate.}
\begin{align}
\label{eq:UkinL}
\mathcal{L}  \supset   \frac{n_K}{2}\, \D \bar{\D} \bigg[ \bU^\dagger \bU \bigg] \bigg|_{0}   \,\,, 
\end{align}
where $n_K$ is a normalization factor that will be determined in what follows.
This is the only mass-dimension four bi-linear term for $\bU$ allowed by RPI-I, RPI-III, and collinear SUSY.
RPI-I is trivially realized since $\bU$ is invariant, and RPI-III is satisfied since $\D \bar{\D}$ has charge $-1$ which compensates the $+1$  charge of $\bU^\dagger \bU$.
Expanding this term in components yields 
\begin{align}
\label{eq:UkinLcomp}
\frac{1}{n_K}\, \mathcal{L} \supset F^* F + i  \left( \uu ^\dagger\,   \PP \uu - u^\dagger\, \PP_\perp \uu -  \uu^\dagger\, \PP_\perp^* u + u^\dagger\, \tilde{\PP}    u \right) - i\s u^\dagger \frac{\Box}{\PP} u \,.
\end{align}

Next, we note that the $\bPhi$ kinetic term in \Eq{eq:SUSYSCETLagrangian} is \emph{not} invariant under RPI-II, as defined by \Eq{eq:uLRPIii}.
Specifically, the fermion kinetic term of \Eq{eq:SUSYSCETLagrangian} transforms as:
\begin{align}
\label{eq:noninvariantfermionkinetic}
i\s u^\dag \frac{\Box}{\PP} u \RPIii  i\s u^\dag \frac{\Box}{\PP} u +  \rpiiiC\, i\s \tilde{u}^\dag \frac{\Box}{\PP} u + \rpiii\, i\s u^\dag \frac{\Box}{\PP} \tilde{u} + \mathcal{O}\big(\kappa_{\text{II}}^2\big)\,.
\end{align}
This is in contrast to the case with the alternative realization of RPI-II in \Eq{eq:uNLRPIii}, where $\tilde{u}$ is integrated out using its equations of motion and the $\bPhi$ kinetic term is RPI-II invariant.
Here, we are working with a theory where $\tilde{u}$ appears explicitly, and as such we must include it in the RPI transformations as given in \Tab{table:RPIfieldsChiral}.

When combining the $\bPhi$ and $\bU$ kinetic terms together with $n_K = 1$, though, the problematic term in \Eq{eq:noninvariantfermionkinetic} drops out, leading to%
\footnote{To our knowledge, this way of expressing the Lagrangian as a contribution from a propagating and a non-propagating superfield was first written down in Eq.~(6.6) of \Ref{Siegel:1981ec} by relying on top-down arguments.}  
\begin{align}
\mathcal{L} &= \frac{1}{2}\, \D \bar{\D} \bigg[ \bPhi^\dagger \frac{i\s \Box}{\PP}  \bPhi +  \bU^\dagger \bU \bigg] \bigg|_{0}  \notag\\[5pt]
&= -\phi^* \Box \phi +   F^* F + i  \left( \uu ^\dagger\,   \PP \uu - u^\dagger\, \PP_\perp \uu -  \uu^\dagger\, \PP_\perp^* u + u^\dagger\, \tilde{\PP}    u \right)\,.
\label{eq:PhiUTildeKinTerm}
\end{align}
This last term in parenthesis is just the Lorentz-invariant fermion kinetic term expanded in light-cone coordinates and is therefore manifestly RPI-II invariant.
In essence, RPI-II has forced a non-trivial relation between the $\bPhi$ and $\bU$ fields that is only satisfied for $n_K = 1$; this is to be expected since the connection between the degrees of freedom contained within these two separate superfields must be enforced by Lorentz invariance.
Thus, by imposing RPI-II from the bottom up, we have recovered the full Lorentz-invariant kinetic term for a chiral superfield.
In \Sec{subsec:fulltheorymatchingKahler}, we will present a top-down argument for the form of \Eq{eq:UkinL}.

\subsection{The Majorana Mass Term}
\label{subsec:massterm}

As mentioned above, the new superfield object $\bC$ is useful for building a superpotential.%
\footnote{One might wonder why it is preferable to use $\bC$ instead of $\bU$ for building the superpotential.   In fact, it is possible to write the Majorana mass term as $\mathcal{L} \supset (\D/\sqrt{2}) \big[m\,\bU \bPhi\big]\big|_0 + \text{h.c.}$, which yields the same expression as \Eq{eq:massterm} below after making the replacement $\tilde{U} u \to \uu u$ in the component expression.  This last manipulation is valid because of the Grassmann nature of $u$, since up to total derivatives we have:
\begin{equation}
u \left(\frac{\PP_\perp^*}{  \PP}\, u \right) = \left(\frac{\PP_\perp^*}{  \PP}\, u \right) u = - u \left(\frac{\PP_\perp^*}{  \PP}\, u \right) = 0\,,
\end{equation}
where we used integration by parts for the first manipulation, and anti-commutation of Grassmann variables for the second.
Note that this special structure will not hold for the more general interactions considered in \Secs{subsec:singleflavor}{sec:MultiFlavorWZ} below, which is why we formulate what follows in terms of $\bC$ since it automatically has $\uu$ as its lowest component.
}
To demonstrate this concretely, we take a single flavor of chiral superfield $\bPhi$ along with its corresponding $\bC$, and show how to construct the simplest superpotential, namely a SUSY Majorana mass term.
A fermion mass involves both helicities, so we need a bilinear term involving both fields.
The unique mass-dimension four term that respects collinear SUSY, RPI-I, and RPI-III is 
\begin{align}
\mathcal{L}  \supset   \frac{\D}{\sqrt{2}}  \bigg[m \, \bC  \bPhi \bigg] \bigg|_{0} + \text{h.c.} = m \, \uu\s u   -  m\, F\s \phi + \text{h.c.}  \, ,
\label{eq:massterm}
\end{align}
where the mass parameter $m$ is dimension one.
Note that the expression in square brackets is fermionic chiral multiplet up to a total derivative (see \Sec{subsec:singleflavor} for a discussion of this property), as required by \Eq{eq:genericLalt}, since $\bC$ is a fermionic almost-chiral superfield.  

As desired, \Eq{eq:massterm} contains the component mass terms of a single field Wess-Zumino model.
Invariance under RPI-I and RPI-III follows because $\bPhi$ and $\bC$ do not transform under RPI-I, and the RPI-III transformation of $\bC$ cancels against that of $\D$. 
Using the component expression in \Eq{eq:massterm}, we can check that RPI-II is also satisfied:
\begin{align}
\label{eq:RPIiiofuu}
& \uu\s u \RPIii \uu\s u + \rpiii \,\uu^2 = \uu\s u \, , \quad \text{since} \quad \uu^2 = 0  \, ,
\end{align}
so RPI-II does not play a role in constraining the form of the mass operator.

\subsection{Single-Flavor Interactions}
\label{subsec:singleflavor}

Relying on the same logic that led to \Eq{eq:massterm} allows us to construct more general interactions.
Consider the following interaction term: 
\begin{align}
\mathcal{L}  \supset \frac{1}{\sqrt{2}}\, \D  \bigg[ \bC\, h'(\bPhi) \bigg] \bigg|_{0} + \text{h.c.} =  \, h''(\phi) \, u \s \uu  -  \, F \, h'(\phi) + \text{h.c.}   \, ,
\label{eq:singleflavorsuperpotential}
\end{align}
where with malice aforethought we have written $h'(\phi)$ as the derivative of a holomorphic function.
Note that this does not actually enforce any constraints on the form of $h'$ since any sufficiently smooth function can be written as the derivative of another.

Remarkably, \Eq{eq:singleflavorsuperpotential} yields a valid collinear superspace Lagrangian.
Though $\bC$ is not chiral, we can use \Eq{eq:Cconstraint} to show that the term in square brackets is indeed chiral up to a total derivative:  %
\begin{equation}
\label{eq:chiral_check_single_field}
\bar{\D} \big(\bC\, h'(\bPhi) \big) = h'(\bPhi)\, \bar{\D} \bC  =  -\sqrt{2}\s i\,  h'(\bPhi)\, \PP_\perp^* \bPhi = -\sqrt{2}\s i\,  \PP_\perp^*\, h(\bPhi)\,,
\end{equation}
where in the last step we used the reverse chain rule.
Similarly, though $\bC$ is not RPI-I invariant, we can use \Eq{eq:RPIiofC} and a reverse chain rule trick to show that \Eq{eq:singleflavorsuperpotential} is RPI-I invariant up to total derivatives:
\bea
\hspace{-20pt} \D  \bigg[ \bC\, h'(\bPhi) \bigg] \bigg|_{0} \RPIi  &&  \D  \bigg[ \bC\, h'(\bPhi) \bigg] \bigg|_{0}+ \frac{\rpii}{\sqrt{2}}\,  \D  \bigg[ (\D \bPhi )\,h'(\bPhi) \bigg] \bigg|_{0}\nonumber \\[5pt] 
 && =  \D \bigg[ \bC\, h'(\bPhi) \bigg] \bigg|_{0}+ \frac{\rpii}{\sqrt{2}} \,  \D  \bigg[\D h(\bPhi) \bigg] \bigg|_{0}  =  \D \bigg[ \bC\, h'(\bPhi) \bigg] \bigg|_{0}\,.
\eea
Finally, the component expression in \Eq{eq:singleflavorsuperpotential} is manifestly RPI-II invariant by inspection, since it only involves RPI-II invariant objects.

In \Sec{sec:MultiFlavorWZ}, we will see how RPI-II enforces the expected structure of a multi-flavor superpotential, such that we can identify
\begin{equation}
h'(\phi) \equiv W^{\s\prime}(\phi) = \frac{\partial W}{\partial \phi},
\end{equation}
where $W(\phi)$ is the standard superpotential. 
In the single field case, this imposes no restrictions on the form of $h'(\phi)$ in \Eq{eq:singleflavorsuperpotential}.

Putting all of the above ingredients together, we have the complete Lagrangian for a single-field Wess-Zumino model:%
\footnote{Following footnote~\ref{footnote:sqrt2rule}, we see the rule that every $\D$ or $\bar{\D}$ comes with a $1/\sqrt{2}$ to yield the correct normalization.}
\begin{align}
\label{eq:LKPlusMandY}
\mathcal{L} =  \frac{1}{2}\, \D \bar{\D} \bigg[ \bPhi^\dagger \frac{i\s \Box}{\PP} \bPhi + \bU^\dagger \bU \bigg] \bigg|_{0} + \frac{1}{\sqrt{2}}\, \D \bigg[ \bC\, W^{\s\prime}(\bPhi) \bigg] \bigg|_{0} + \text{h.c.}\,\,.
\end{align}
In the renormalizable case, we can write
\begin{equation}
W(\phi) = f\, \phi + \frac{1}{2}\, m\, \phi^2 + \frac{1}{3}\, y\, \phi^3\,,
\end{equation}
where $f$ is a source term for the $F$-term auxiliary field (relevant for SUSY breaking), $m$ is the Majorana mass term, and $y$ is the Yukawa coupling.
In summary, we have achieved a collinear superspace formalism that lets us incorporate the non-propagating degrees of freedom $\uu$ and $F$.

\subsection{Equations of Motion}
\label{subsec:matterEOM}

To connect to the discussion in \Refs{Cohen:2016dcl,Cohen:2018qvn}, it is instructive to integrate out the auxiliary field $\bU$.
Starting from \Eq{eq:LKPlusMandY} and using \Eq{eq:Cfield_def}, we can compute the equations of motion for the non-propagating degrees of freedom directly in superspace: 
\begin{align}
\frac{\delta \mathcal{L} }{\delta \bU^\dagger} = 0 \quad \Longrightarrow \quad  i \s \PP  \bU  + \frac{\bar{\D}}{\sqrt{2}}\,  W^{\s\prime}(\bPhi^\dagger)  = 0\, ,
\label{eq:IntOutUTildeGeneral}
\end{align}
where we have used \Eq{eq:Cfield_def} and  $\bar{\D} \D  \bU = -2\s i \,\PP \bU$, which holds since $\bU$ is chiral.
Plugging this back into \Eq{eq:LKPlusMandY}, we find 
\begin{align}
\label{eq:LKPlusMandYintegratedout}
\mathcal{L} =  \frac{1}{2}\, \D \bar{\D} \bigg[ \bPhi^\dagger \frac{i\s \Box}{\PP} \bPhi +  i\, W^{\prime}\big(\bPhi^\dagger\big) \frac{1}{\PP} W^{\prime}\big(\bPhi\big) \bigg] \bigg|_{0}
+ \frac{1}{2}\, \D \bigg[  \bPhi\,  \frac{\PP_\perp^*}{\PP}\, \D W^{\s\prime}(\bPhi)  \bigg] \bigg|_{0}  + \text{h.c.}\,.
\end{align}
Noting that $W^{\prime}(\phi)$ is mass dimension 2, this is a valid collinear SUSY Lagrangian.

Given the analysis of \Ref{Cohen:2018qvn}, \Eq{eq:LKPlusMandYintegratedout} is a rather surprising expression.
There, we started from a local theory and integrated out the non-propagating $\uu$ and $F$ fields to derive an apparently non-local Lagrangian expressed only in terms of propagating degrees of freedom.
Since it was derived by integrating out $\bU$ from a theory that respects RPI-II, \Eq{eq:LKPlusMandYintegratedout} must also respect RPI-II.  
However, the RPI-II transformations must be revisited in the theory where $\uu$ has been integrated out, in the same spirit as going from \Eq{eq:uNLRPIii} to \Eq{eq:uLRPIii}.

Taking the lowest component of \Eq{eq:IntOutUTildeGeneral}, we find the equation of motion for the $\uu$ field:
\begin{align}
\frac{\delta \mathcal{L} }{\delta \uu^\dagger} = 0 \quad \Longrightarrow \quad \uu = \frac{\PP_\perp^*}{\PP}\, u + i\s   \frac{1}{\PP}\, W^{\s\prime\s\prime}(\phi^*)\, u^\dagger \,.
\label{eq:IntOutUTildeGenerallowest}
\end{align}
It is straightforward to check that \Eq{eq:LKPlusMandYintegratedout} respects a nonlinear RPI-II transformation of $u$ that involves both $u$ and $u^\dagger$:
\begin{equation}
u  \RPIii u + \rpiii \, \left(\frac{\PP_\perp^*}{\PP}\, u + i \s  \frac{1}{\PP}\, W^{\s\prime\s\prime}(\phi^*)\, u^\dagger \right)\,,
\end{equation}
where we simply inserted the $\uu$ equation of motion into \Eq{eq:uLRPIii}.
In this way, RPI-II is non-trivially modified by the presence of interactions arising from the off-shell degrees of freedom.

It is instructive to consider a few special cases.
When $W = f\, \phi$, SUSY is spontaneously broken, and the vacuum energy must be non-zero.
This shows up in \Eq{eq:LKPlusMandYintegratedout} by the fact that $W^{\s\prime}(\bPhi)$ is just a constant, so $(1/\PP) W^{\s\prime}(\bPhi)$ is not well-defined.
Thus, to have a healthy collinear SUSY theory with spontaneous SUSY breaking, the $\bU$ field must remain in the spectrum in order for $F$ to get an expectation value.

When $W = \frac{1}{2}\, m\, \phi^2$, we return to the theory in \Eq{eq:massterm}.
Even though the mass term flips the helicity, it is possible to achieve the same physics with just the $u$ field, with the helicity flip encoded in a modified propagator:
\begin{align}
\label{eq:LKPlusMandYintegratedoutJustmass}
\mathcal{L} =  \frac{1}{2} \,\D \bar{\D} \bigg[ \bPhi^\dagger\, \frac{i\, (\Box + m^2)}{\PP}\, \bPhi \bigg] \bigg|_{0},
\end{align}
and a modified RPI-II transformation $u \to u + \rpiii \, \left(\frac{\PP_\perp^*}{\PP}\, u + i \s  \frac{1}{\PP}\, m\, u^\dagger \right)$.
That said, it is often more convenient to keep the $\uu$ field in the spectrum for practical calculations.

Finally, when $W = \frac{1}{3}\, y\, \phi^3$, \Eq{eq:LKPlusMandYintegratedout} only has quartic interactions, and not the cubic interactions expected of a Yukawa theory.
As discussed in \Ref{Cohen:2016dcl}, once $\uu$ has been integrated out, external sources are needed to reproduce the helicity-flipping Yukawa interaction.  
By keeping the $\bU$ field in the spectrum, we can incorporate the off-shell modes $\uu$ and $F$ in collinear superspace and avoid the need for external currents.

\subsection{Multiple Flavors}
\label{sec:MultiFlavorWZ}

Extending the analysis of \Sec{subsec:massterm} to the case of multiple flavors $\bPhi_j$ is straightforward and illuminating.
After imposing RPI and collinear SUSY, the most general renormalizable theory we can write takes the form
\begin{align}
\label{eq:WZmultiflavor}
\mathcal{L}_\text{WZ} = \frac{1}{2}\, \D \bar{\D} \bigg[\bPhi_j^\dagger\, \frac{i\s \Box}{\PP} \, \bPhi_j+  \bU^\dagger_j  \bU_j\bigg] \bigg|_{0} + \frac{1}{\sqrt{2}} \left( \D \bigg[ \bC_j \,  W_j(\bPhi) \bigg] \bigg|_{0} + \text{h.c.} \right)\,,
\end{align}
where a sum over $j$ is implied.
Without loss of generality, we are using $\bC_j$ instead of $\bU_j$ to make it easier to enforce RPI-II invariance.
At the moment, $W_j(\bPhi)$ is just a holomorphic function of the collinear superfields labeled by $j$ (and not involving any derivatives).
Here, we have rotated and rescaled the fields in order to make the kinetic terms diagonal and canonically normalized.

Because $\bC$ is neither chiral nor RPI-I invariant, we must impose constraints on $W_j(\bPhi)$ for this to be a valid collinear superspace Lagrangian.
Repeating the exercise in \Eq{eq:chiral_check_single_field}, we have
\begin{equation}
\label{eq:multiple_flavor_chiral_check}
\bar{\D} \big(\bC_j\, W_j(\bPhi) \big) = W_j(\bPhi)\, \bar{\D} \bC_j  =  -i\s\sqrt{2} \,  W_j(\bPhi)\, \PP_\perp^* \bPhi_j\,.
\end{equation}
In general, this is not a total derivative, unless%
\footnote{For the remainder of this paper, we use the standard notation of subscripts on functions corresponding to derivatives.}
\begin{equation}
\label{eq:Wj_as_derivative}
W_j(\phi) \equiv \frac{\partial W}{\partial \phi_j}\,,
\end{equation}
where $W$ is now the familiar superpotential.
One can check that \Eq{eq:Wj_as_derivative} is necessary and sufficient for \Eq{eq:WZmultiflavor} to be RPI-I invariant (up to a total derivative) as well.

For RPI-II, the kinetic terms are manifestly RPI-II invariant following the same logic as \Sec{sec:WZKinTerm}.
To check whether the interaction terms are RPI-II invariant, we have to write down the component expression.
The terms with non-trivial RPI-II transformation properties are:
\begin{equation}
\mathcal{L}_\text{WZ} \supset \uu_j \,\frac{\partial W_j}{\partial \phi_k}\, u_k  \RPIii  \uu_j \,\frac{\partial W_j}{\partial \phi_k}\, u_k + \rpiii \, \uu_j\,\frac{\partial W_j}{\partial \phi_k}\, \uu_k\,.
\end{equation}
The only way for this expression to be RPI-II invariant is if $\partial W_j/\partial \phi_k$ is symmetric in $j$ and $k$, since $\uu_j$ are anti-commuting fields.
Indeed, this is case if we impose \Eq{eq:Wj_as_derivative}.

This is a novel way to understand the required structure of the superpotential.
In the standard $\mathcal{N} = 1$ superspace language, the superpotential has to be holomorphic to respect SUSY, and the fact that interactions arise from derivatives of the superpotential then follow from the component expansion.
In collinear superspace, $W_j(\phi)$ has to be holomorphic in order to respect collinear SUSY, but if we used $\bU_j \, W_j(\bPhi)$ instead of $\bC_j  \, W_j(\bPhi)$, the form of $W_j$ would naively look to be unconstrained.
The fact that we have to use $\bC_j$ then follows from RPI-II invariance, which in turn requires $W_j$ to be the derivative of the superpotential by chirality and RPI-I.
Of course, these two perspectives are related, which we will see in the top-down derivation below.

As an example of \Eq{eq:WZmultiflavor}, consider the superpotential coupling between three chiral superfields $W = \lambda \,\phi_1\, \phi_2\, \phi_3$.
In collinear superspace, the interaction terms take the form
\begin{align}
\mathcal{L}_\text{WZ}  \supset  \frac{1}{\sqrt{2}} \lambda \, \D \bigg[  \bC_1\,   \bPhi_2\, \bPhi_3  + \bC_2 \,  \bPhi_1\, \bPhi_2 + \bC_3 \,  \bPhi_1\,\bPhi_2\bigg] \bigg|_{0} + \text{h.c.}\,,
\end{align}
where one can check explicitly that the same coupling $\lambda$ must multiply all three terms in order to preserve RPI-II.

\subsection{Top-Down Derivation of Superpotential}
\label{subsec:fulltheorymatchingSuperpotential}

Using the strategy of \Sec{subsec:topdownchiral}, we can understand the structure of the superpotential interactions from the top down.
The full $\mathcal{N} = 1$ superpotential term, written in the notation of \Sec{eq:lightcone}, is
\begin{equation}
\mathcal{L}_W =  \frac{1}{2}\, \D \tilde{\D}  \bigg[ W\big(\bPhi^\full\big) \bigg] \bigg|_{\theta = 0 = \theta^\dagger} + \text{h.c.}\,\,.
\end{equation}
Here, we used the fact that
\begin{equation}
\int \text{d}^2 \theta\, \big[\dots \big] = -\frac{1}{4}\, \mathcal{D}^\full_\alpha\s \mathcal{D}^{\full \,\alpha}\s \big[\dots \big] = \frac{1}{2} \,\D \tilde{\D}  \big[\dots \big] \,,
\end{equation}
up to total derivatives. Carrying out the $\tilde{\D}$ derivative, setting $\tilde{\eta} = 0$ to go to collinear superspace, and using $\bC = \frac{1}{\sqrt{2}} (\tilde{\D} \bPhi^\full)|_{\tilde{\eta} = 0}$, we have
\begin{equation}
\label{eq:CversionW}
\mathcal{L}_W = \frac{1}{\sqrt{2}}\, \D \bigg[  \bC_j\, W_j(\bPhi) \bigg] \bigg|_{0} + \text{h.c.} \,,
\end{equation}
where again a sum over $j$ is implied and we using the standard notation $W_j = \partial W / \partial \phi_j$. 
This is precisely the form of \Eq{eq:WZmultiflavor} obtained from the bottom up.

While the exotic $\bC$ field might have seemed strange from the bottom-up perspective, \Eq{eq:CversionW} emphasizes why it was indeed necessary.
Of course, $\bC$ is just a function of $\bPhi$ and $\bU$ by \Eq{eq:Cfield_def}, so we can write \Eq{eq:CversionW} using $\bPhi$ and $\bU$ alone:
\begin{align}
\label{eq:UversionW}
\mathcal{L}_W &=\frac{1}{\sqrt{2}} \,\D \bigg[ \bU_j\, W_j(\bPhi) - \frac{1}{\sqrt{2}}\, \frac{\PP_\perp^*}{\PP} \big(\D \bPhi_j \big)\, W_j(\bPhi)  \bigg] \bigg|_{0} + \text{h.c.}\,\,.
\end{align}
This form, however, offers no additional insight into the structure of the problem.
The first term is manifestly chiral and RPI-I invariant for any $W_j(\phi)$, but not RPI-II invariant.
The second term is neither chiral nor RPI-I invariant unless $W_j(\phi) \equiv \partial W / \partial \phi_j$, nor is it RPI-II invariant.
Only this combination exhibits the desired symmetries, which is why it makes sense to package them together via $\bC_j$.

\subsection{Top-Down Derivation of K\"ahler Potential}
\label{subsec:fulltheorymatchingKahler}
In \Sec{sec:WZKinTerm}, we only wrote down the minimal kinetic term and did not include more general K\"ahler potential interactions.
The reason is that RPI-II imposes rather non-trivial constraints on K\"ahler interactions, which is easiest to see from the top down.
The full $\mathcal{N} = 1$ K\"ahler potential term is:
\begin{equation}
\mathcal{L}_K = \frac{1}{4} \, \bar{\D} \D \bar{\tilde{\D}} \tilde{\D} \bigg[ K\big(\bPhi^\full, \bPhi^{\full \, \dagger}\big) \bigg] \bigg|_{\theta\s =\s 0\s =\s \theta^\dagger}\,,
\end{equation}
where $K(\phi, \phi^*)$ is real function.
We will use the notation $K_j \equiv \partial K / \partial \phi_j$ and $K_{\bar{k}} \equiv \partial K / \partial \phi_{k}^*$ for derivatives with respect to fields.

Analogous to \Sec{subsec:fulltheorymatchingSuperpotential}, we can carry out the $\tilde{\D}$ and $\bar{\tilde{\D}}$ derivatives, set $\tilde{\eta} = 0$, and use the fields defined in \Sec{subsec:topdownchiral}:
\begin{equation}
\label{eq:Kahler}
\mathcal{L}_K = \frac{1}{2}\, \bar{\D} \D \bigg[ K_{j \bar k}\big(\bPhi, \bPhi^{\dagger}\big)\, \bC_j\,  \bC^\dagger_{\bar{k}}  + i \s K_j\big(\bPhi, \bPhi^{\dagger}\big)\, \tilde{\PP} \bPhi_j  \bigg] \bigg|_{0}\,,
\end{equation}
where we used $\bar{\tilde{\D}} \tilde{\D} \bPhi^\full_k= -2\s i\, \tilde{\PP} \bPhi^\full_k$.
To symmetrize the second term and make this expression manifestly real, one can use the fact that $\tilde{\PP} K = K_j \, \tilde{\PP} \phi_j +  K_{\bar{k}} \, \tilde{\PP} \phi_{\bar k}^*$ is a total derivative.

We have not found a rewriting of \Eq{eq:Kahler} that makes RPI-I invariance obviously manifest.  
Like in \Eq{eq:UversionW}, it is possible to use \Eq{eq:Cfield_def} to replace $\bC$ with $\bU$ and $\bPhi$, but we find that there are always terms involving $\PP_\perp$ or $\tilde{\PP}$ that remain, which obscures RPI-I.  
For this reason, it is challenging to find valid K\"ahler potential interactions from the bottom up.

Since it is non-obvious, we provide the details of the check of that  \Eq{eq:Kahler} is RPI-I invariant, critically including the shift of $\bC$ in \Eq{eq:RPIiofC}.  First, note that 
\begin{align}
\D K_{\bar{k}} = K_{j\bar{k}}\, \D \bPhi_j \qquad \text{and} \qquad \bar{\D} K_{j} = K_{j\bar{k}}\, \bar{\D} \bPhi^\dag_{\bar{k}}\,.
\label{eq:Kidentity}
\end{align}
Then
\begin{align}
\mathcal{L}_K & \RPIi  \frac{1}{2}\,\bar{\D} \D \bigg[ K_{j\bar{k}} \left(\bC_j + \frac{\rpii}{\sqrt{2}} \,\D \bPhi_j \right)  \left(\bC^\dagger_{\bar k} + \frac{\rpiiC}{\sqrt{2}}\, \bar{\D} \bPhi_{\bar k}^\dagger\right) +  i\s K_j  \Big(\tilde{\PP}  + \rpiiC\, \PP_\perp^* + \rpii \,\PP_\perp \Big) \bPhi_j  \bigg] \bigg|_{0}  \nonumber \\[5pt]
&=\mathcal{L}_K +  \frac{1}{2}\,\bar{\D} \D \bigg[\rpii \bigg(\frac{1}{\sqrt{2}}\,K_{j\bar{k}} \big(\D\bPhi_j\big)\, \bC^\dag_{\bar k} + i\s K_j\, \PP_\perp \bPhi_j \bigg) \notag\\[2pt]
&\hspace{100pt}+ \rpiiC\,  \bigg(\frac{1}{\sqrt{2}}\,K_{j\bar{k}}\, \big(\bar{\D}\bPhi^\dag_{\bar k}\big)\, \bC_{j} + i\s K_j\, \PP^*_\perp \bPhi_j \bigg)  \bigg] \bigg|_{0}\notag \\[5pt]
&= \mathcal{L}_K +  \frac{1}{2}\,\bar{\D} \D \bigg[\rpii \bigg( \frac{1}{\sqrt{2}}\, \big(\D K_{\bar{k}}\big)\, \bC^\dag_{\bar k} + i\s K_j \,\PP_\perp \bPhi_j \bigg) + \rpiiC\,  \bigg(\frac{1}{\sqrt{2}}\, \big( \bar{\D} K_{j}\big) \, \bC_{j} + i\s K_j\, \PP^*_\perp \bPhi_j \bigg) \bigg] \bigg|_{0} \notag\\[5pt]
&= \mathcal{L}_K +  \frac{1}{2}\,\bar{\D} \D \bigg[\rpii \Big(K_{\bar{k}} \,\big(i\s \PP_\perp \bPhi^\dag_{\bar k}\big)  + i\s K_j \,\PP_\perp \bPhi_j \Big)  \bigg] \bigg|_{0}  \nonumber\\[5pt] 
&= \mathcal{L}_K +  \frac{1}{2}\,\bar{\D} \D \bigg[\rpii \, i \s \PP_\perp K( \bPhi_{\bar{k}}^\dagger, \bPhi_j)  \bigg] \bigg|_{0} =  \mathcal{L}_K \, ,
\end{align}
where in the second line we only kept terms to leading order in $\rpii$, in the third line we used \Eq{eq:Kidentity}, in the fourth line we used integration by parts on $\D$ and $\bar{\D}$ (noting that there is no minus sign because $\bC$ is fermionic) along with the almost-chiral constraint in \Eq{eq:Cconstraint}, and in the final line we used the reverse chain rule.
The expression in parentheses on the final line is a total derivative, $i\s \PP_\perp K$, demonstrating the RPI-I invariance of \Eq{eq:Kahler}.

In the case of the canonical K\"ahler potential $K = \phi^*_j \,\phi_j$, it is straightforward to show that 
\begin{equation}
\label{eq:minimalKahler}
\mathcal{L}_{\rm canonical} = \frac{1}{2}\,\bar{\D} \D \bigg[ \bC_j \, \bC^\dagger_j +  i\s \bPhi_j^{\dagger}\, \tilde{\PP} \bPhi_j  \bigg] \bigg|_{0} = \frac{1}{2}\, \bar{\D} \D \bigg[  \bPhi_j^\dagger\, \frac{i\s \Box}{\PP}\, \bPhi_j +  \bU_j^\dagger \bU_j\bigg] \bigg|_{0}\,,
\end{equation}
where in addition to \Eq{eq:Cfield_def}, we used integration by parts and the fact that $\tilde{\PP}  = (\Box+\PP_\perp^* \PP_\perp)/\PP$.
This reproduces the kinetic terms of \Eq{eq:PhiUTildeKinTerm}, with the normalization of the $\bPhi$ and $\bU$ kinetic terms linked, demonstrating the connection between the full Lorentz-invariant theory and the collinear superspace realization of the same physics.

\section{Setting Sail  for the \emph{D}-term}
\addtocontents{toc}{\protect\vspace{2.5pt}}
\label{sec:SailingForTheD}

Having seen how to construct theories out of the building blocks that explicitly separate the propagating from the non-propagating degrees of freedom for a chiral multiplet in collinear superspace, we will now apply the same strategy to the theory of an Abelian gauge multiplet.
In addition to the propagating degrees of freedom $\alc$ and $\lambda$ contained in $\bPhialc$ (see \Eq{eq:bPhialc}), our goal is to write down an object ---  the auxiliary real superfield $\bD$ below in \Eq{eq:bDbottomup} --- that encodes the gauge potential $n \cdot A$, the opposite-helicity fermion $\tilde{\lambda} = \tilde{\xi}^\alpha \lambda_\alpha$, and the scalar auxiliary $D$-term.
One crucial new aspect of the gauge theory case (as compared to the auxiliary fermion field $\tilde{U}$ discussed in \Sec{sec:Fterm}) is that the auxiliary field $n \cdot A$ transforms non-trivially under RPI-I, implying that $\bD$ must as well.
Nevertheless, we will show how to take appropriate superderivatives of $\bD$ to define the RPI-I invariant object $\bU_\lambda$, which will be convenient for constructing the gauge kinetic term.

In this section, we first emphasize the importance of working in LCG and WZG in collinear superspace.
We then take a bottom-up perspective to argue that the form of $\bD$ is fixed by RPI-I, RPI-III, and collinear SUSY.
This section concludes with a top-down derivation of $\bPhialc$ and $\bD$ starting from a full $\mathcal{N} = 1$ vector superfield.

\subsection{The Importance of Gauge Fixing}
\label{eq:gaugefixing}

To make SUSY manifest, SUSY gauge theories have additional gauge redundancies beyond those of non-SUSY gauge theories.
Taking an Abelian $U(1)$ gauge symmetry for simplicity, recall that the real gauge transformation parameter $\omega$ is promoted to a full chiral superfield $\RCA^{\rm full}$, where the full superscript anticipates that this will be related to $\RCA$ from \Eq{eq:RCA} in collinear superspace.
An $\mathcal{N} = 1$ vector superfield transforms as 
\begin{align}
\label{eq:Vfullgauge}
\bV^{\rm full} \gauge \bV^{\rm full} + \frac{i}{2} \left(\RCA^{\rm full} - \RCA^{ \full\, \dagger}\right) \, ,
\end{align}
which encodes the transformation of the Abelian gauge field $A_\mu \to A_\mu + \partial_\mu \omega$, along with additional redundancies associated with the modes of  $\RCA^{\rm full}$.

For practical calculations, it is convenient to work in WZG, where the additional SUSY redundancies are fixed by setting various components of $\bV^{\rm full}$ equal to zero.
We start from the full $\mathcal{N} = 1$ vector superfield that satisfies $\bV^{\full} = \bV^{\full\, \dagger}$:  
\begin{align}
 \bV^{\text{full}} =\,\,& C + i\s \Big(\eta \xi^\alpha + \tilde{\eta} \tilde{\xi}^\alpha\Big)\chi_\alpha - i\s \Big(\eta^\dagger \xi^{\dagger}_{\dot{\alpha}} + \tilde{\eta}^\dagger \tilde{\xi}^\dagger_{\dot{\alpha}}\Big)\s \chi^{\dagger \dot{\alpha}} + i\s \eta \tilde{\eta}(M + i\s N) - i\s \tilde{\eta}^\dagger \eta^\dagger (M - i\s N)\nonumber \\[5pt] 
&+  \left(\eta \eta^\dagger\, \bar{n}\cdot A + \sqrt{2} \, \eta \tilde{\eta}^\dagger \alc + \sqrt{2}\, \tilde{\eta} \eta^\dagger \alc^* + \tilde{\eta} \tilde{\eta}^\dagger\, n \cdot A  \right)  \nonumber\\[5pt] 
&- 2\s i\,  \eta \tilde{\eta}\Big(\xi^{\dagger}_{\dot{\alpha}}\s \eta^\dagger + \tilde{\xi}^{\dagger}_{\dot{\alpha}}\s \tilde{\eta}^\dagger\Big)\! \left[ \lambda^{\dagger \dot{\alpha}} + \frac{i}{2}\, (\bar{\sigma} \cdot \partial \chi)^{\dot{\alpha}}   \right]+2\s i\,  \tilde{\eta}^\dagger \eta^\dagger  \big(\xi^\alpha \eta + \tilde{\xi}^\alpha \tilde{\eta}\big)\! \left[  \lambda_\alpha + \frac{i}{2} \big(\sigma \cdot \partial \chi^\dagger\big)_\alpha \right]\notag\\
& - 2\,  \eta \tilde{\eta} \tilde{\eta}^\dagger \eta^\dagger \left( D + \frac{1}{2}\, \Box C \right)\,, 
\label{eq:VfullNotWZ}
\end{align}
where for later convenience, we have used  \Eq{eq:theta} to replace $\theta_\alpha$ with $\eta$ and $\tilde{\eta}$. 
Here, $A^\mu$ is the gauge field, $\lambda_\alpha$ and its conjugate are the mass dimension $3/2$ gauginos, and $D$ is the auxiliary $D$-term.
In WZG, the mass dimension $1/2$ fermion $\chi_\alpha$ and the real scalars $C$, $M$, and $N$ are fixed to zero using the additional redundancy encoded in \Eq{eq:Vfullgauge}, so that the only remaining component is the collinear SUSY gauge transformation superfield $\RCA^{\rm full}|_{\rm WZG} = \omega$, where $\omega$ is real.

In collinear superspace, it is convenient to work in LCG with $\bar{n} \cdot A = 0$.
One compelling reason is that $\alc \to \alc + \rpii \, \bar{n}\cdot A$ under RPI-I, so working in LCG makes $\alc$ manifestly RPI-I invariant.
At first glance, WZG and LCG seem to be independent choices, but we now show that maintaining manifest collinear SUSY requires imposing both of these conditions simultaneously.

Taking a top-down perspective (see further discussion in \Sec{subsec:TopDownSuperfieldsGauge}), consider the following collinear chiral superfield by acting on \Eq{eq:VfullNotWZ} with collinear superspace derivatives and setting $\tilde{\eta} = 0$:
\begin{align}
\label{eq:PhibarnA}
\bPhi_{\bar{n}\cdot A} = \Big(\bar \D \D \bV^{\text{full}}\Big)\Big|_{\tilde{\eta} = 0}  = \big( \bar{n}\cdot A - i\s \PP C \big) + 2\, \eta\, \xi^\alpha\s \PP \chi_\alpha + i\s \eta^\dagger \eta \, \PP \big( \bar{n}\cdot A - i\s \PP C \big)\,.
\end{align}
This field has RPI-III charge $+1$ and satisfies the collinear chiral constraint $\bar{\D} \bPhi_{\bar{n}\cdot A} = 0$ by construction.
Since both $\bar{n}\cdot A$ and $C$ are real scalars, $\bPhi_{\bar{n}\cdot A}$ has the correct degrees of freedom to be a collinear chiral multiplet.

If we were to fix either WZG or LCG, then $\bPhi_{\bar{n}\cdot A}$ would no longer be a valid collinear chiral multiplet since it would be missing various components.
However, simultaneously imposing both WZG and LCG yields $\bPhi_{\bar{n}\cdot A} = 0$, which is a valid constraint that preserves collinear SUSY.%
\footnote{Strictly speaking, this condition does not require setting $M = N = 0$, though one can make a similar argument by inspecting $\Big(\D \tilde{\D} \bV^{\text{full}}\Big)\Big|_{\tilde{\eta} = 0}$.}
For this reason, we work in both WZG and LCG for the remainder of this paper.

As already discussed in \Sec{subsec:gaugetheoryReview}, going to LCG further restricts the form of $\RCA^{\rm full}$ to be
\begin{equation}
\RCA^{\rm full}_{\rm WZG, LCG} = \omega \qquad \text{with} \qquad \PP\s \omega = 0\,.
\end{equation}
Going to collinear superspace with $\tilde{\eta} = 0$, this corresponds to the real, chiral, anti-chiral superfield $\RCA$ from \Eq{eq:RCA}:
\begin{equation}
\RCA^{\rm full}_{\rm WZG, LCG} \big|_{\tilde{\eta} = 0} \equiv \RCA \,.
\end{equation}
More explicitly, acting $\bar \D \D$ on the gauge transformation in \Eq{eq:Vfullgauge} we have 
\begin{align}
\bPhi_{\bar{n}\cdot A} \gauge \bPhi_{\bar{n}\cdot A}  +  \PP\s \RCA^{ \rm full}\big|_{\tilde{\eta} = 0} \,,
\end{align}
where we can parametrize
\begin{align}
\RCA^{ \rm full}\Big|_{\tilde{\eta} = 0} = (\omega + i \s \tau) + \sqrt{2}\, \eta\s \psi_\omega + i\s \eta^\dagger \eta \, \PP (\omega + i\s  \tau)\,,
\label{eq:RCAfull}
\end{align}
with $\omega$ and $\tau$ being real.
Correlating with \Eq{eq:PhibarnA}, we can use $\tau$ to set $C$ to zero, $\psi_\omega$ to set $\chi$ to zero, and LCG is compatible with $\omega \neq 0$ satisfying $\PP\s \omega = 0$.

\subsection{Component SUSY, RPI, and Gauge Transformations} 
\label{subsec:DcollSUSYtrans}

Constructing the analog of \Eq{eq:collinearsusyauxcomp} for gauge superfields, we can check that collinear SUSY closes on the component fields in WZG and LCG. 
Acting with the collinear supersymmetry charges $\Q$ and $\bar{\Q}$ (or equivalently, performing the full $\mathcal{N} = 1$ SUSY transformation with SUSY transformation parameter $\zeta_\alpha = \xi_\alpha\s \epsilon$) yields:  
\begin{align}
\delta \alc  &=   \sqrt{2}\s i\,    \lambda\s  \epsilon^\dagger    \,, \notag \\[3pt]
\delta \alc^* &=    - \sqrt{2}\s i\,  \epsilon\s \lambda^\dagger  \,, \notag\\[3pt]
\delta (n\cdot A) &=    2\s i \Big( \tilde{\lambda}\s \epsilon^\dagger - \epsilon\s \tilde{\lambda}^\dagger  \Big ) \,,\notag \\[3pt]
\delta \lambda &=  \sqrt{2}\,  \epsilon \s  \PP \alc  \,, \notag \\[3pt]
\delta \tilde{\lambda} &=   -  i\s \epsilon\s  D - \frac{1}{2}\, \epsilon\s \PP (n \cdot A)  + \frac{1}{\sqrt{2}}\,\epsilon\s \big( \PP_\perp \alc^* -  \PP_\perp^* \alc \big)  \,, \notag\\[3pt]
\delta D & =  \epsilon\s  \PP \tilde{\lambda}^\dagger -  \epsilon\s  \PP_\perp \lambda^\dagger  -  \epsilon^\dagger\s   \PP \tilde{\lambda}+ \epsilon^\dagger \s  \PP^*_\perp \lambda \,,
\label{eq:SUSYTransformGaugeComponents} 
\end{align}
where the two gaugino helicities are $\lambda = \xi^\alpha \lambda_\alpha$ and $\tilde{\lambda} = \tilde{\xi}^\alpha \lambda_\alpha$.
The LCG choice $\bar{n} \cdot A = 0$ is compatible with collinear supersymmetry since $\delta (\bar{n}\cdot A) = 0$.
Note that $\big(\epsilon^\dagger \lambda\big)^\dagger = \lambda^\dagger \epsilon = - \epsilon\s \lambda^\dagger$, so that for instance $n \cdot A$ transforms into a real field as it should.  

In WZG and LCG, the scalar field $\alc$ and fermion $\lambda$ transform exactly like the $\phi$ and $u$ fields in \Eq{eq:collinearsusyauxcomp}, which explains why it is consistent for them to be packaged into the chiral multiplet $\bPhialc$ in \Eq{eq:bPhialc}.
Building a multiplet containing the non-propagating $n \cdot A$, $\tilde{\lambda}$, and $D$ fields is more complicated, since the collinear SUSY transformations rotate them into the $\alc$ and $\lambda$ fields.
However, this issue can be circumvented by noting that the combination 
\begin{equation}
D_\alc \equiv  D  +  \frac{i}{\sqrt{2}} \left(     \PP_\perp \alc^* - \PP_\perp^* \alc \right)
\label{eq:Dalccombination}
\end{equation}
allows us to identify three fields that close under collinear SUSY (in LCG): 
\begin{align}
\delta (n\cdot A) &= 2\s i\, \Big( \tilde{\lambda}\s \epsilon^\dagger - \epsilon\s \tilde{\lambda}^\dagger    \Big) \,,\notag \\[5pt]
\delta \tilde{\lambda} &= - i\s \epsilon\s D_\alc - \frac{1}{2}\, \epsilon\s \PP (n\cdot A) \,, \notag\\[5pt]
\delta D_\alc &= \epsilon \s \PP \tilde{\lambda}^\dagger-  \epsilon^\dagger \s  \PP \tilde{\lambda}  \, ,
\label{eq:SUSYTransformGaugeComponentsAuxiliary} 
\end{align}
where to derive these transformations, we simply applied \Eq{eq:SUSYTransformGaugeComponents}  to \Eq{eq:Dalccombination}.
These will be the three components of the auxiliary real superfield $\bD$.

\begin{table}[t]
\renewcommand{\arraystretch}{1.8}
\setlength{\arrayrulewidth}{.3mm}
\small
\setlength{\tabcolsep}{0.45 em}
\hspace{-5pt}
\begin{tabular}{ |c || c | c | c  | c|}
    \hline
    Component &  RPI-I &  RPI-II &   RPI-III& Res.~Gauge   \\
    \hline 
    \hline
 $\alc  $ &  $\alc+\frac{\rpiiC}{\sqrt{2}} \,\bar n \cdot A$     &   $\alc +\frac{\rpiii}{\sqrt{2}}  \, n \cdot A$ &    $\alc$  & $\alc +  \frac{\PP_\perp}{\sqrt{2}}  \, \omega $\\
 $\alc^*  $ &  $\alc^*+\frac{\rpii}{\sqrt{2}}  \, \bar n\cdot A$  &   $\alc^* + \frac{\rpiiiC}{\sqrt{2}}\, n\cdot A$ &   $\alc^*$  & $\alc^* + \frac{\PP_\perp^*}{\sqrt{2}}  \, \omega$ \\
 $n\cdot A  $ &    $n \cdot A + \sqrt{2}\,\big(\rpiiC\, \alc^* + \rpii\, \alc\big)$  &   $n\cdot A  $ &    $e^{\rpiiii}\, n\cdot A$  & $n \cdot A +  \tilde{\PP} \omega$ \\ 
 $\bar{n}\cdot A  $ &    $\bar{n}\cdot A$  &   $\bar{n}\cdot A + \sqrt{2}\,\big(\rpiii\, \alc^* + \rpiiiC\, \alc\big)$ &    $e^{-\rpiiii} \,\bar{n}\cdot A$  & $\bar{n}\cdot A$ \\ 
 \hline
       $\lambda  $ &  $\lambda $     &   $ \lambda + \rpiii \,\tilde{\lambda}$ &    $e^{-\rpiiii/2}\, \lambda$  & $\lambda$\\
 $\tilde{\lambda} $ &  $\tilde{\lambda}  + \rpii \,\lambda$     &   $\tilde{\lambda}$ &    $e^{\rpiiii/2}\, \tilde{\lambda}$  & $\tilde{\lambda} $\\
 \hline
 $D  $ &    $D$  &   $D$ &    $D$  & $D$ \\ \hline 
\end{tabular}
\caption{
Transformation properties for the Abelian gauge theory degrees of freedom.
Throughout, we work in LCG with $\bar{n}\cdot A = 0$, such that $\alc$ and $\alc^*$ are RPI-I invariant.
Like in the chiral matter case, RPI-II is linearly realized with the inclusion of the $\tilde{\lambda}$ and $n\cdot A$ components.
}
\label{table:RPIfieldsGauge}
\end{table}

Next, we turn to the RPI transformations of the component fields, which are summarized in \Tab{table:RPIfieldsGauge}.
As we already anticipated and will show explicitly in \Sec{subsec:auxvector}, the auxiliary real superfield $\bD$ must have non-trivial RPI-I transformation properties, so we highlight the RPI-I transformations of the fields in \Eq{eq:SUSYTransformGaugeComponentsAuxiliary}, restricted to LCG: 
\begin{align}
n \cdot A &\RPIi n \cdot A  +\sqrt{2}\, \big( \rpiiC\, \alc^*  + \rpii\, \alc \big) \,, \nonumber \\[4pt]
\tilde{\lambda} &\RPIi \, \tilde{\lambda}  + \rpii\, \lambda \,,  \nonumber \\[4pt]
D_\alc &\RPIi D_\alc + \frac{i}{\sqrt{2}}\,\PP\s \big( \rpiiC \,\alc^* - \rpii\, \alc \big) \, .
\label{eq:RPIiVectorComponents}
\end{align}
Under RPI-III, $n \cdot A$ has charge +1, $\tilde{\lambda}$ has charge $+1/2$, and $D_\alc$ has RPI-III charge $0$; using these facts, we will see below that $\bD$ has homogenous RPI-III scaling.

In parallel with \Sec{subsec:RPIuanduu}, it is convenient to identify objects that are invariant under RPI-I.
For the fermionic component $\tilde{\lambda}$, this unsurprisingly implies the same combination as in \Eq{eq:Utildedefinition} above, involving the propagating gaugino helicity $\lambda$:
\begin{align}
\tilde{U}_\lambda \equiv  \tilde{\lambda}  - \frac{ \PP_\perp^*}{\PP} \lambda \,.
\label{eq:DefUtildelambda}
\end{align}
For the bosonic components, $D$ is already RPI-I invariant by itself, as is
\begin{align}
\label{eq:partialA}
\partial_\mu A^\mu = \PP (n \cdot A) - \sqrt{2} \,\big( \PP_\perp \alc^* + \PP^*_\perp \alc \big)\,.
\end{align}
These RPI-I invariant combinations will appear in the $\bU_\lambda$ field defined below in \Sec{subsec:auxvector_chiral}.
Note that $\tilde{U}_\lambda$ has RPI-III charge $1/2$ and $\partial_\mu A^\mu$ is RPI-III invariant.

The (Abelian) residual gauge transformation of the components are also given in \Tab{table:RPIfieldsGauge}.
Specifically, for the non-propagating degrees of freedom, we find   
\begin{align}
n \cdot A &\gauge  n \cdot A + \tilde{ \PP} \omega \,, \nonumber \\[3pt]
\tilde{\lambda} &\gauge  \tilde{\lambda} \,, \nonumber \\[3pt]
D_\alc &\gauge D_\alc \,,
\label{eq:fullloretnzgaugetrans}
\end{align}
where we have used the fact that $\PP \omega = 0$ in LCG to simplify the last line.

\subsection{The Auxiliary Real Superfield}
\label{subsec:auxvector}

With the collinear SUSY and RPI transformations in place and fixing to WZG and LCG, we can write down a real collinear superfield that contains the non-propagating gauge degrees of freedom:
\begin{align}
\label{eq:bDbottomup}
\bD  &  =  n \cdot A +  2\s i\,  \eta\s \tilde{\lambda}^{\dagger}  + 2\s  i\,   \eta^\dagger\s \tilde{\lambda} + 2\, \eta^\dagger \eta\, D_\alc \,,
\end{align}
where $D_\alc$ is defined in \Eq{eq:Dalccombination}.
This field has mass dimension $1$, RPI-III charge $+1$, and it satisfies the reality condition $\bD = \bD^\dagger$ because $\big(i\s \eta\s \tilde{\lambda}^\dagger +  i\s \eta^\dagger\s \tilde{\lambda} \big)^\dagger = - i\s \tilde{\lambda}\s \eta^\dagger - i \s\tilde{\lambda}^\dagger\s \eta = i\s \eta^\dagger\s \tilde{\lambda} + i\s \eta \tilde{\lambda}^\dagger$.
One can perform a collinear SUSY transformation of this superfield by taking $\eta \to \eta + \epsilon$ to verify that the components transform as advertised in \Eq{eq:SUSYTransformGaugeComponentsAuxiliary}.
Assuming that $n \cdot A$ is in the lowest component of $\bD$, the form of \Eq{eq:bDbottomup} is completely fixed.

As we saw in \Eq{eq:RPIiVectorComponents}, the components of $\bD$ have non-trivial RPI-I transformations.
To maintain collinear SUSY, $\bD$ must transform into some combination of superfields that also satisfies the reality condition, which it does:
\begin{align}
\label{eq:VdRPIi}
\bD \RPIi \bD + \sqrt{2}\, \Big( \rpiiC \, \bPhialc + \rpii\, \bPhialc^\dagger \Big) \,.
\end{align}
This can be derived by invoking the RPI-I transformations of the components.
Invariance under \Eq{eq:VdRPIi} will be an important tool for building collinear superspace operators for gauge theories, especially when we introduce charged matter in \Sec{sec:Landfall}.
As emphasized in \Sec{subsec:RPIIsuperspace}, RPI-II can only be imposed at the level of the component action.

Finally, we can lift the gauge transformation in \Eq{eq:fullloretnzgaugetrans} to collinear superspace: 
\begin{align}
\bD \gauge \bD + \tilde{\PP}\s \RCA\,,
\label{eq:Dgauge}
\end{align}
where $\RCA$ is the real, chiral, anti-chiral superfield defined in \Eq{eq:RCA}.
As we will see in \Sec{subsec:gaugecovderiv}, $\bD$ allows one to lift the gauge covariant derivative along the $n \cdot A$ direction to collinear superspace \textemdash\ a necessary ingredient for constructing collinear superspace Lagrangians for charged chiral matter.

\subsection{The Auxiliary Gauge Chiral Superfield}
\label{subsec:auxvector_chiral}

While $\bD$ will be advantageous later for constructing gauge-covariant derivatives, it is not convenient for building the gauge kinetic term because of its non-trivial RPI-I and gauge transformation properties; see \Eqs{eq:VdRPIi}{eq:Dgauge}.
We can easily construct an RPI-I and gauge invariant combination, though, by taking appropriate covariant derivatives.

First note that $\D \bD$ is gauge invariant (for an Abelian theory), since $\RCA$ in \Eq{eq:Dgauge} is anti-chiral.
It is not, however, RPI-I invariant:
\begin{equation}
\D \bD \RPIi \D \bD +  \sqrt{2}\, \rpiiC\, \D \bPhialc \,.
\end{equation}
To make further progress, we can take a $\bar{\D}$ derivative
\begin{equation}
\bar{\D} \D \bD \RPIi \bar{\D} \D \bD - 2\s \sqrt{2}\s i\, \rpiiC\,  \PP \bPhialc \,,
\end{equation}
where we used the anti-commutation relation $\big\{\D, \bar{\D} \big\} = - 2\s i\, \PP$.
While this also shifts under RPI-I, we can use the fact that  $\PP_\perp  \to \PP_\perp +\rpiiC \, \PP$ to construct an RPI-I invariant combination with $\bPhiA$:  
\begin{align}
\bPhiD &\equiv  \frac{i}{2}\, \bar{\D} \D \bD -  \sqrt{2}\, \PP_\perp \bPhiA \nonumber \\[5pt]
&= \biggl[ \frac{1}{2}\, \partial^\mu A_\mu- i D   \bigg] - 2\s i\, \eta \, \PP \tilde{U}^\dagger_\lambda + i\s \eta^\dagger \eta \, \PP  \bigg[  \frac{1}{2}\, \partial^\mu A_\mu- i D   \bigg]\,,
\label{eq:phiD_def}
\end{align}
where $\tilde{U}_\lambda$ is defined in \Eq{eq:DefUtildelambda}, and we have simplified this expression using the fact that (in LCG)
\bea
\label{eq:DAtoD}
 -  i D   - \frac{1}{2} \partial^\mu A_\mu   = -i D_\alc -   \frac{1}{2} \PP\s (n \cdot A) +   \sqrt{2} \,  \PP_\perp^*  \alc  \,.
\eea
%
This field is a chiral multiplet $\big(\bar{\D} \bPhi_D = 0\big)$, whose lowest component involves the auxiliary $D$ term, so it plays a similar role to $\bPhiF$ in  \Eq{eq:Fmultiplet}.
It is RPI-III invariant, with mass dimension $2$, and it has the following gauge transformation property:
\begin{equation}
\bPhiD \gauge \bPhiD -     \PP_\perp \PP^*_\perp \RCA \, .
\end{equation}

Following this analogy with $\bPhiF$ further, we can define a fermionic chiral multiplet in the same spirit as $\bU$:
\begin{align}
\bU_\lambda \equiv -\frac{i}{2}\, \frac{\bar{\D}}{\PP}\s \bPhiD^\dagger &= -\frac{i}{2}\, \bar{\D} \left(\bD - \sqrt{2} \, \frac{\PP_\perp^*}{\PP} \s\bPhialc^\dagger \right) \nonumber \\[4pt]
&= \tilde{U}_\lambda - \eta \left( i D + \frac{1}{2}\, \partial^\mu A_\mu \right) + i\s \eta^\dagger \eta\, \PP \tilde{U}_\lambda  \,,
\label{eq:Ulambda_def}
\end{align}
where in the second step we used $\bPhiD^\dagger = \frac{i}{2}\, \D \bar{\D} \bD - \sqrt{2}\, \PP_\perp^* \bPhialc^\dagger $.%
 \footnote{The potentially confusing sign of the first term in $\bPhiD^\dagger$ results from the fact that $\big(\bar{\D} \D \big)^\dagger = - \D \bar{\D}$.}
This $\bU_\lambda$ multiplet is also invariant under RPI-I and it is now gauge invariant (in the Abelian case).
It is analogous to $\bU$ in \Eq{eq:Usuperfield}, with mass dimension $3/2$ and RPI-III charge +1/2.
Note that both $\bPhiD$ and $\bU_\lambda$ contain the same auxiliary field content as $\bD$, but they also depend on the propagating degrees of freedom in $\bPhiA$, which is necessary to make them RPI-I invariant.

\subsection{Top-Down Derivation of Gauge Superfields}
\label{subsec:TopDownSuperfieldsGauge}

\begin{table}[t]
\renewcommand{\arraystretch}{1.8}
\setlength{\arrayrulewidth}{.3mm}
\footnotesize
\setlength{\tabcolsep}{0.35 em}
\hspace{-14pt} 
\begin{tabular}{ |c || c | c | c | c | c | c | c|}
    \hline
    Superfield &  Construction & Constraint & RPI-I &    RPI-III  & Res.~Gauge  & Mass Dim.  \\
    \hline \hline
  $\bPhialc$  & $\frac{1}{\sqrt{2}}\, \bar{\D} \tilde{\D} \bV^\full_\text{WZG}\Big|_{\tilde{\eta}=0}$ & $\bar{\D} \bPhialc = 0$ &  $\bPhialc$   &  $\bPhialc$   &  $\bPhialc +\frac{\PP_\perp^*}{\sqrt{2}}\, \RCA $  & 1   \\ 
$\bD $ & $\tilde{\bar \D} \tilde{\D}\bV^\full_\text{WZG}\bigg|_{\tilde{\eta} = 0} $ & $\bD^\dag = \bD$ &    $\begin{array}{c}\bD+\\[-10pt]  \rpiiC \bPhialc + \rpii\, \bPhialc^\dag \end{array}$  &  $e^{\rpiiii}\, \bD$  &  $\bD + \tilde{\PP}\, \RCA$ & 1  \\ 
$\bPhi_D$ & $ \frac{i}{2}\, \bar{\D} \D \bD-  \PP_\perp \bPhiA $ & $\bar{\D}\bPhi_D =0$ & $\bPhi_D$ & $\bPhi_D$ & $\bPhiD -  |\PP_\perp|^2\, \RCA$ & 2\\ 
$\bU_\lambda$  & $\frac{i}{\sqrt{2}}\, \frac{\bar{\D}}{\PP} \bPhiD^\dagger$ &$\bar{\D}\bU_\lambda =0$ & $\bU_\lambda$  & $e^{\rpiiii/2}\, \bU_\lambda$ &$\bU_\lambda$ & 3/2 \\
$\bC_\lambda$ & $\bU_\lambda  -\frac{1}{\sqrt{2}}\, \frac{\PP_\perp^*}{\PP} \D \bPhiA$ &$\bar{\D}\bC_\lambda =-\sqrt{2}\s i\, \PP_\perp^* \bPhiA$ & $\bC_\lambda+ \frac{1}{\sqrt{2}}\, \rpii\, \D \bPhialc$ & $e^{\rpiiii/2}\, \bC_\lambda $ & $\bC_\lambda$ & 3/2\\
\hline
\end{tabular}
\caption{
Collinear superfields for the Abelian gauge theory, along with their top-down construction and transformation properties.
Note that we are working in both WZG and LCG, in which $\bV^\full_\text{WZG}|_{\tilde{\eta}=0} = 0$, $\bar \D \D \bV_\text{WZG}^{\text{full}}|_{\tilde{\eta} =0} = 0$, and $\D \tilde{\D} \bV_\text{WZG}^{\text{full}}|_{\tilde{\eta} =0} = 0$.
}
\label{table:RPIsuperfieldsGauge}
\end{table}

The form of $\bD$ in \Eq{eq:bDbottomup} was dictated by bottom-up considerations, as the most general superfield we could construct with $n \cdot A$ as the lowest component, given the constraints imposed by RPI-I, RPI-III, collinear SUSY, and gauge transformations.
Note that we could have also determined $\bU_\lambda$ in \Eq{eq:Ulambda_def} in an analogous way by requiring $\tilde{U}_\lambda$ as the lowest component.
We can gain further insight into the structure of $\bD$ by performing a top-down derivation starting from the full $\mathcal{N} = 1$ vector multiplet.

As in \Sec{subsec:topdownchiral}, we act on the full superfield with superspace derivatives and then set $\tilde{\eta} = 0$ to go to collinear superspace.
We already saw an example of this type when deriving \Eq{eq:PhibarnA}, where we argued that $\bPhi_{\bar{n}\cdot A} = \bar \D \D \bV^{\text{full}}|_{\tilde{\eta} = 0}$ could be consistently set to zero in WZG and LCG.
The construction, constraints, and transformations of the superfields derived in what follows are summarized in \Tab{table:RPIsuperfieldsGauge}.

The full $\mathcal{N} = 1$ vector superfield $\bV^{\rm full} = \bV^{\rm full\,\dagger}$ in WZG is
\begin{align}
\bV^\full_\text{WZG} &=    \left(\eta \eta^\dagger\, \bar{n}\cdot A + \sqrt{2} \, \eta \tilde{\eta}^\dagger\, \alc + \sqrt{2}\, \tilde{\eta} \eta^\dagger\, \alc^* + \tilde{\eta} \tilde{\eta}^\dagger\, n \cdot A  \right)  \nonumber\\[4pt] 
&\hspace{15pt}+ 2\s i\,  \eta \tilde{\eta}\Big(\eta^\dagger \s \lambda^\dagger  +  \tilde{\eta}^\dagger\s \tilde{\lambda}^\dagger\Big) +2\s i\,  \tilde{\eta}^\dagger \eta^\dagger  \big( \eta\s \lambda +  \tilde{\eta}\s \tilde{\lambda}\big)  - 2 \, \eta \tilde{\eta} \tilde{\eta}^\dagger \eta^\dagger\s D \,.
\end{align}
Note that $\big(\bV^\full_\text{WZG}\big)^3 = 0$, which will be useful for the top-down derivation of the charged matter kinetic term in \Sec{subsec:topdownchargedkinetic}.
Furthermore, in LCG with $\bar{n}\cdot A = 0$ and moving to collinear superspace, we have
\begin{equation}
\label{eq:WZG_LCG_lowestV_simplification}
\bV^\full_\text{WZG}\Big|_{\tilde{\eta}=0} = 0\,.
\end{equation}

We can derive the gauge chiral superfield $\bPhialc$ in \Eq{eq:bPhialc} by acting on $\bV^\full_\text{WZG}$ with two collinear superspace derivatives: \
\begin{align}
\label{eq:Phialc}
\bPhialc = \frac{1}{\sqrt{2}}\, \bar{\D} \tilde{\D} \bV^\full_\text{WZG}\Big|_{\tilde{\eta}=0} &= \alc^* + i\s \sqrt{2} \, \eta \lambda^\dagger + i\s \eta^\dagger \eta \, \PP \alc^* \,, \notag\\[6pt]
\bPhialc^\dagger = - \frac{1}{\sqrt{2}}\, \D \tilde{\bar{\D}} \bV^\full_\text{WZG}\Big|_{\tilde{\eta}=0} &= \alc + i\s \sqrt{2}\,  \eta^\dagger \lambda - i \s \eta^\dagger \eta \, \PP \alc \,.
\end{align}
Here, we have made use of $\big(\D \bar{\D}\big)^\dagger = - \bar{\D} \D$, and the factor of $1/\sqrt{2}$ is introduced to ensure canonically normalized kinetic terms in \Sec{sec:abeliangaugekinetic}.
Since $\bar{\D} \bPhialc = 0$ and $\D \bPhialc^\dagger = 0$ by construction, these respectively correspond to chiral and anti-chiral superfields under collinear SUSY.
This analysis clarifies why $\alc$ enters into the anti-chiral superfield while the conjugate field $\alc^*$ field enters into the chiral superfield, since in analogy with the $\bPhiF^\dagger$ field in \Eq{eq:Fmultiplet}, the $\tilde{\D}$ derivative isolates the complex conjugate representation.

Acting with a different pair of collinear superspace derivatives, we can identify a superfield whose lowest component is $n \cdot A$:
\begin{align}
\label{eq:bD}
\bD    \equiv \left( \tilde{\bar \D} \tilde{\D} \bV^\full_\text{WZG}\right)\!\bigg|_{\tilde{\eta} = 0} = - \left( \tilde{\D} \tilde{ \bar{ \D} } \bV^\full_\text{WZG}\right)\!\bigg|_{\tilde{\eta} = 0} = n \cdot A +  2\s i\,  \eta\s \tilde{\lambda}^{\dagger}  + 2\s  i\,   \eta^\dagger \s \tilde{\lambda} + 2\, \eta^\dagger \eta\s D_\alc \,,
\end{align}
where $D_\alc$ is defined in \Eq{eq:Dalccombination}, and in the second step we made use of the anti-commutation relations in \Eq{eq:DanticommAll} along with the simplification in \Eq{eq:WZG_LCG_lowestV_simplification}.
Note that $\bD = \bD^\dagger$, which is necessary since the lowest component $n \cdot A$ is a real degree of freedom.
This is precisely the form of $\bD$ we found using bottom-up logic in \Eq{eq:bDbottomup}.

By acting with more supercovariant derivatives, we could construct $\bPhiD$ and $\bU_\lambda$, though the derivation is the same as in \Sec{subsec:auxvector_chiral} since it does not require $\tilde{\bar{\D}}$ or $\tilde{\D}$.
In the bottom-up derivation, we did not encounter the analog of the $\bC$ field introduced above, though we can define one in analogy with \Eq{eq:Cfield_def}:
\begin{align}
\bC_\lambda &= \bU_\lambda  + \frac{1}{\sqrt{2}}\s \frac{\PP_\perp^*}{\PP}\s \D \bPhiA \,,
\label{eq:Clambda}
\end{align}
such that $\bC_\lambda$ obeys the almost-chiral constraint $\bar{\D} \bC_\lambda = - i\s \sqrt{2}\, \PP_\perp \bPhialc$.
This $\bC_\lambda$ field does not play as critical a role in the gauge theory case as it did above for chiral matter (where it was the natural object to use for building a superpotential), although it will appear briefly in \Eq{eq:FIterm} below when we discuss $D$-term SUSY breaking.
It is satisfying that chiral matter and gauge fields have exactly parallel structures in collinear superspace (with LCG and WZG); see further discussion in \Sec{subsec:parallels}.

For the manipulations in \Sec{subsec:topdownchargedkinetic}, we will encounter expressions with a single $\tilde{\D}$ acting on $\bV^\full_\text{WZG}$, which can be simplified using the following relation:
\begin{equation}
\tilde{\D} \bV^\full_\text{WZG}\Big|_{\tilde{\eta}=0} = \frac{i\s \D}{2\, \PP}\, \bPhi_\alc \,.
\end{equation}
This can be verified by acting by $\bar{\D}$ on both sides of this expression and recognizing that the right-hand side involves the chiral projector from \Eq{eq:chiralprojector}.

Finally, we can perform a gauge transformation on $\bV_\text{WZG}^\text{full}$ to derive the transformation properties of $\bPhialc$ and $\bD$: 
\begin{align}
\bPhialc  = \frac{1}{\sqrt{2}}\, \bar{\D} \tilde{\D}  \bV^{\rm full} \Big|_{\tilde{\eta} = 0} & \gauge \bPhialc +  \frac{i}{2\, \sqrt{2}}\, \bar{\D} \tilde{\D}  \RCA^{\rm full} \Big|_{\tilde{\eta} = 0} = \bPhialc + \frac{1}{\sqrt{2}}\,\PP_\perp^* \RCA \nonumber \,, \\[6pt]
\bD = \tilde{\bar{\D}} \tilde{\D} \bV_\text{WZG}^{\rm full} \Big|_{\tilde{\eta} = 0}  & \gauge \bD  + \frac{i}{2}\, \tilde{\bar{\D}} \tilde{\D} \Big( \RCA^{\rm full} - \RCA^{\dagger \rm \, full} \Big) \Big|_{\tilde{\eta} = 0} = \bD + \tilde{\PP} \RCA \, ,
\end{align}
where we used the anti-commutation relations in \Eq{eq:DanticommAll} and the fact that the full theory anti-chiral field satisfies $\tilde{\D} \RCA^{\dagger \rm full} |_{\tilde{\eta} = 0} = 0$. 
Without choosing a gauge, the only constraint on $\RCA \equiv \RCA^{\rm full} \big|_{\tilde{\eta} = 0}$ is that it is chiral, but it is also real in LCG and WZG, as discussed already in \Sec{eq:gaugefixing}.
These transformations agree with \Eqs{eq:Phialcgauge}{eq:Dgauge}, and furthermore will be used in \Sec{subsec:gaugecovderiv} to lift $\PP_\perp^*$ and $\tilde{\PP}$ into the gauge-covariant derivative $\nabla_\perp^*$ and $\tilde{\nabla}$.

\section{Casting Anchor: Gauge Theories}
\addtocontents{toc}{\protect\vspace{2.5pt}}
\label{sec:AbelianGaugeTheory}

We now have all the necessary ingredients to construct gauge theory Lagrangians in collinear superspace utilizing $\bPhialc$, $\bD$, and $\bU_\lambda$.
In the Abelian case, the bottom-up derivation of the kinetic term exactly parallels that of \Sec{sec:WZKinTerm}, since $\bPhialc$ and $\bU_\lambda$ have exactly the same transformation properties as $\bPhi$ and $\bU$ for a free chiral multiplet.
From a top-down perspective, this might seem a bit surprising, so we will show how it can be derived from the standard $\mathcal{N} = 1$ treatment involving a holomorphic gauge kinetic term.
The non-Abelian case is conceptually straightforward but algebraically tedious, since it requires the use of gauge-covariant derivatives, so we present only select aspects of its construction in \App{sec:nonAbelianGauge}.

\subsection{The Abelian Gauge Kinetic Term}
\label{sec:abeliangaugekinetic}

When working with only propagating degrees of freedom, \emph{i.e.},~the components of the chiral multiplet $\bPhialc$,  the form of the gauge kinetic term is fully determined by RPI-I, RPI-III, and gauge invariance.
As shown in \Ref{Cohen:2018qvn}, this unique form is given by:
\begin{equation}
\mathcal{L}  = \frac{1}{2}\, \D \bar{\D} \bigg[ \bPhialc^\dagger \frac{i\s \Box}{\PP}  \bPhialc \bigg] \bigg|_{0}\,,
\end{equation}
which has the same structure as for free chiral matter in \Eq{eq:SUSYSCETLagrangian}.
Even though $\bPhialc \to \bPhiA + \s\PP_\perp^*\s \RCA/\sqrt{2}$ under a gauge transformation, this kinetic term is gauge invariant because $\RCA$ is both chiral and anti-chiral.

Constructing a kinetic term directly for the auxiliary superfield $\bD$ is more complicated because of its non-trivial RPI-I transformation properties.
For that reason, it is more convenient to work with the RPI-I invariant chiral auxiliary field $\bU_\lambda$, which contains the same component fields as $\bD$.
Imposing RPI-I, RPI-III, and gauge invariance, the unique bilinear kinetic term we can construct is:
\begin{equation}
\mathcal{L} \supset \frac{n_V}{2}\, \D \bar{\D} \bigg[ \bU_\lambda^\dagger \bU_\lambda \bigg] \bigg|_{0} \, .
\end{equation}
This is the same structure that we already encountered in \Sec{sec:WZKinTerm}, and analogously, we have to set $n_V = 1$ in order to achieve RPI-II invariance of the component action.
The final gauge kinetic term is 
\begin{align}
\mathcal{L}  &= \frac{1}{2} \D \bar{\D} \bigg[ \bPhialc^\dagger \frac{i\s \Box}{\PP}  \bPhialc + \bU_\lambda^\dagger \bU_\lambda \bigg] \bigg|_{0}  \nonumber \\[4pt]
& = -\alc^* \Box \alc + i  \left( \tilde{\lambda}^\dagger\,   \PP \tilde{\lambda} - \lambda^\dagger\, \PP_\perp \tilde{\lambda} -  \tilde{\lambda}^\dagger\, \PP_\perp^* \lambda + \lambda^\dagger\, \tilde{\PP}    \lambda \right) + \frac{1}{2} D^2 + \frac{1}{8} (\partial^\mu A_\mu)^2\,,
\label{eq:LagrangianD}
\end{align}
exactly parallel to \Eq{eq:PhiUTildeKinTerm}.

In the absence of matter fields, $\bD$ only appears inside of $\bU_\lambda$, so we can integrate it out via the equations of motion for $\bU_\lambda$:
\begin{equation}
\label{eq:integrateoutUlambda}
 \frac{\delta \mathcal{L}}{\delta \bU_\lambda^\dagger} = 0 \quad \Longrightarrow \quad  \bU_\lambda = 0 \,, \quad  \text{so that}  \quad \bD = \sqrt{2} \left(\frac{\PP_\perp^*}{\PP} \bPhialc^\dagger + \frac{\PP_\perp}{\PP} \bPhialc \right) \,.
\end{equation}
This implies, using \Eq{eq:partialA} to set $\partial^\mu A_\mu = 0$, that
\begin{equation}
\tilde{\lambda} = \frac{\PP_\perp^*}{\PP} \lambda \,, \qquad
D = 0 \,, \qquad
n\cdot A = \frac{\sqrt{2} }{\PP}\, \big(\PP_\perp^* \alc + \PP_\perp \alc^*\big) \,,
\end{equation}
which are the expected light-cone equations of motion.

\subsection{Parallels with Chiral Matter}
\label{subsec:parallels}

How can it be that chiral matter and Abelian gauge fields have the same kinetic structure in collinear superspace, despite having different spin structures?
We can argue why this makes sense in three steps.
First, after projecting with the spinors $\xi_\alpha$ and $\tilde{\xi}_\alpha$, all bosons look alike, as do all fermions, so the valid multiplets of collinear SUSY only depend on their RPI-I and RPI-III transformation properties.
In LCG, the two polarizations of a gauge field can be packaged into an RPI-I invariant complex scalar $\alc$ with RPI-III charge 0, which is the same as the complex scalar field $\phi$ in a matter multiplet.
Collinear SUSY then tells you that $\alc$ must be part of a chiral multiplet with a fermion of RPI-III charge $-1/2$, namely, the propagating gaugino $\lambda$ (the analog of $u$ for chiral matter). 
Second, RPI-II provides a unique way to relate the spin-1/2 fermion helicities residing in different collinear SUSY multiplets.
Just as $u$ and $\tilde{u}$ are related by RPI-II for chiral matter, $\lambda$ must be related to $\tilde{\lambda}$.
Appealing to the first point, $\tilde{\lambda}$ must then be part of a chiral multiplet $\bU_\lambda$ with another complex scalar $i\s D + \frac{1}{2}\, \partial_\mu A^\mu $ (the analog of $F$ for chiral matter). 
Finally, we already saw in \Sec{sec:WZKinTerm} that the kinetic terms for chiral multiplets were unique, and therefore gauge fields must also have the same kinetic structure in collinear superspace.
Note that this argument relied crucially on working in LCG, since that is the gauge where $\alc$ is RPI-I invariant.

One might wonder if there are further parallels between chiral matter and gauge fields.
For example, can we write a ``superpotential'' for gauge fields?
Because $\bPhialc$ and $\bC_\lambda$ have the same transformation properties as $\bPhi$ and $\bC$, one is allowed to write superpotential term of the form:
\begin{equation}
\label{eq:FIterm}
\mathcal{L}  \supset \D  \bigg[ \frac{i}{2}\, d_\text{FI} \, \bC_\lambda \bigg] \bigg|_{0} + \text{h.c.} = \D  \bigg[\frac{i}{2}\, d_\text{FI} \, \bU_\lambda \bigg] \bigg|_{0} + \text{h.c.} = d_\text{FI}\, D  \,,
\end{equation}
where we used \Eqs{eq:Clambda}{eq:UCFrelations} in the first manipulation and in the final expression we have used $\partial^\mu A_\mu = 0$.
This is just the Fayet-Iliopoulos term~\cite{Fayet:1974jb} for spontaneous SUSY breaking.
Comparing to the superpotential $W = f \phi$ in \Sec{subsec:matterEOM}, it is fascinating that $F$-term and $D$-term SUSY breaking have completely parallel forms in collinear superspace.

Note that a mass term involving $\bC_\lambda\s \bPhialc$ is forbidden by the gauge transformation properties of $\bPhialc$.
Similarly, most non-trivial ``K\"ahler potential'' terms are forbidden by gauge invariance.
Of course, if the gauge symmetry is spontaneously broken, then such interactions would be allowed, as long as one works in unitary gauge or is careful to keep track of the explicit Goldstone fields.

\subsection{Top-Down Derivation of Gauge Kinetic Term}

In standard $\mathcal{N} = 1$ superspace, the gauge kinetic term arises from the holomorphic operator 
\begin{align}
\label{eq:LfullTheory}
\mathcal{L}  \supset   \frac{1}{2}\, \D \tilde{\D}  \bigg[ \bW^{\text{full}\,\alpha} \s \bW^\text{full}_\alpha  \bigg] \bigg|_{\theta = 0 = \theta^\dagger} + \text{h.c.}\,\,, 
\end{align}
where $\bW_\alpha^\text{full} = -\frac{i}{4}\, \bar{\mathcal{D}}^\text{full}_{\dot{\alpha}}\s \bar{\mathcal{D}}^{\text{full}\,\dot{\alpha}}\s \mathcal{D}^\text{full}_\alpha\, \bV_{\text{WZ}}^{\full}$.
By following similar logic to \Secs{subsec:fulltheorymatchingSuperpotential}{subsec:fulltheorymatchingKahler}, we can derive \Eq{eq:LagrangianD} from the top down by first carrying out the $\tilde{\D}$ derivative and then setting $\tilde{\eta} = 0$.
As discussed in \Sec{eq:lightcone}, we suppress the ``full" labels when no confusions would arise.

First, using \Eqs{eq:DFullDecomposed}{eq:DsquaredSimplification}, we can rewrite the chiral gauge superfield in WZG as
\begin{align}
\label{eq:Wdecom}
\bW^\text{full}_\alpha =- \frac{i}{2}\s  \bar{\D} \bar{\tilde{\D}}   \left(  \tilde{\xi}_\alpha\s \D- \xi_\alpha\s \tilde{\D} \right) \bV_{\text{WZG}}^{\full} \,.
\end{align}
Plugging this into \Eq{eq:LfullTheory}, pulling out an overall $\bar{\D}^\text{full}$ derivative, and dropping total derivatives, we find 
\begin{equation}
\mathcal{L}  \supset   \frac{1}{2}\, \D \bar{\D} \tilde{\D}  \bigg[ \Big(\bar{\D} \bar{\tilde{\D}} \D \bV_{\text{WZG}}^{\full}\Big) \Big(  \bar{\tilde{\D}} \tilde{\D}\bV_{\text{WZG}}^{\full}  \Big)  \bigg] \bigg|_{\theta = 0 = \theta^\dagger} + \text{h.c.}\,\,.
\end{equation}
It is straightforward to simplify this expression, although the algebra is a bit tedious.
Carrying out the overall $\tilde{\D}^\text{full}$ derivative, using integration by parts with $\bar{\D}^\text{full}$, repeatedly applying the anti-commutation relation  in \Eq{eq:DanticommAll}, using the definitions of $\bPhialc$ in \Eq{eq:Phialc} and $\bD$ in \Eq{eq:bD}, noting that after taking $\tilde{\eta} = 0$ and fixing to WZG/LCG, one may set  $\bar{\D} \D \bV_{\text{WZ}}^{\full} = 0$ and $\bar{\D} \tilde{\bar{\D}} \bV_{\text{WZ}}^{\full} = 0$, and including the Hermitian conjugate terms, we find
\bea
\hspace{-22pt}\mathcal{L}  \supset\!\frac{1}{4} \s\D \bar{\D} \bigg[\! - \D \bD\, \bar{\D} \bD +    2\s i \, \bPhialc^\dagger\, \tilde{\PP}  \bPhialc  - 2\s i\s \sqrt{2}\, \big( \PP_\perp  \bPhialc \big) \bD + 2\s i\s\sqrt{2}  \,\big( \PP_\perp^*  \bPhialc^\dagger \big) \bD \bigg] \bigg|_{0}\,.
\eea
Next, inserting the chiral projector from \Eq{eq:chiralprojector} and using the definition of $\tilde{\PP} = \frac{\Box}{\PP} + \PP_\perp^* \frac{1}{\PP} \PP_\perp$, we find
\begin{equation}
\label{eq:topdownabelianL}
\mathcal{L}  \supset   \frac{1}{2} \s\D \bar{\D}  \bigg[ \bPhialc^\dagger \,\frac{i\s \Box}{\PP}\, \bPhialc  -\frac{1}{4}\, \D \left(\bD - \sqrt{2} \, \frac{\PP_\perp}{\PP} \s\bPhialc \right) \bar{\D} \left(\bD - \sqrt{2} \, \frac{\PP_\perp^*}{\PP} \s\bPhialc^\dagger  \right)  \bigg] \bigg|_{0}\,.
\end{equation}
Following \Eq{eq:Ulambda_def}, we recognize the second expression as $\bU_\lambda^\dagger \bU_\lambda$, so we indeed recover the expected Lagrangian from \Eq{eq:LagrangianD}.
This shows how the holomorphic kinetic term in standard $\mathcal{N} = 1$ superspace becomes a real kinetic term in collinear superspace.

\section{Landfall: Interacting Charged Matter}
\addtocontents{toc}{\protect\vspace{2.5pt}}
\label{sec:Landfall}

Now that we understand how to construct collinear superspace Lagrangians for gauge fields, in this section we show how to couple them to charged chiral matter.
To simplify the presentation and minimize the algebra, we work only in the case of a single Abelian $U(1)$ gauge theory, though analogous manipulations hold for multiple (non-)Abelian theories, up to the same kinds of complications discussed in \App{sec:nonAbelianGauge}.
The key new ingredients relative to \Ref{Cohen:2018qvn} are gauge-covariant derivatives, alluded to already in \Sec{subsec:auxvector}.
As with the previous derivations in \Secs{sec:WZ}{sec:AbelianGaugeTheory}, we start by presenting the constructions using bottom-up reasoning, and then show how the same structures can be obtained from the top down.

\subsection{Gauge-Covariant Derivatives}
\label{subsec:gaugecovderiv}

\begin{table}[t]
\renewcommand{\arraystretch}{1.8}
\setlength{\arrayrulewidth}{.3mm}
\centering
\small
\setlength{\tabcolsep}{0.35 em}
 \begin{tabular}{ |c || c | c | c | c | c|}
 \hline
     Component & RPI-I &  RPI-II &  RPI-III  & Gauge    \\ \hline \hline
     $\phi_M$ & $\phi_M$  & $\phi_M$  & $\phi_M$ &  $e^{\s i\s q\s g\s \omega} \s \phi_M $ \\ 
     $u_M$ & $u_M$ & $ u_M + \rpiii \,\tilde{u}_M$  &  $ e^{-\rpiiii/2}\,u_M $& $e^{\s i\s q\s g\s \omega}\s  u_M$ \\ 
     $\tilde{u}_M$ & $\tilde{u}_M + \rpii \, \tilde{u}_M$ & $ \tilde{u}_M$  &  $ e^{\rpiiii/2}\,\tilde{u}_M $& $e^{\s i\s q\s g\s \omega}\s  \tilde{u}_M$ \\ 
     $F_M$ & $F_M$ & $ F_M $  &  $ F_M $& $e^{\s i\s q\s g\s \omega}\s  F_M$ \\ \hline
\end{tabular}
\caption{The transformations for the component charged matter fields. 
}
\label{Table:transComp}
\end{table}

We introduce a chiral matter multiplet $\bM$ with $U(1)$ charge $q$, as\footnote{Here, we use the $\bM$ notation (instead of denoting a charged chiral multiplet superfield using $\bPhi$) to minimize possible confusions with the gauge field $\bPhiA$.}
\begin{align}
\bM = \phi_M + \sqrt{2}\, \eta \, u_M + i\s \eta^\dagger \eta \, \PP \phi_M \,.
\end{align}
The gauge transformation of $\bM$ is
\begin{equation}
\label{eq:matter_transformation}
\bM \gauge e^{i\s q\s g \s\RCA} \bM\,,
\end{equation}
where $g$ is the gauge coupling, and there is the factor of $i$ in the exponent because $\RCA$ is real. 
The gauge and RPI transformation properties of the $\phi_M$ and $u_M$ components are summarized in \Tab{Table:transComp}.

Together with $\bPhialc$, we can use $\bD$ to define gauge-covariant derivatives acting on charged multiplets.
Recalling that $\PP\s \RCA = 0$ in LCG, we do not need to covariantize $\PP$.
We do need gauge-covariant versions of $\tilde{\PP}$, $\PP_\perp$, and $\PP_\perp^*$, though, for which it is straightforward to check that these act as expected for gauge-covariant derivatives:
\begin{alignat}{2}
\tilde{\nabla} \bM &\equiv \left(\tilde{\PP} - i\s q\s g\s \bD \right) \bM &&\gauge e^{i \s q\s g\s \RCA} \big(\tilde{\nabla} \bM\big)\,, \notag\\[4pt]
\nabla_\perp \bM &\equiv \left(\PP_\perp - i\s q\s g\s \sqrt{2}\s \bPhiA^\dagger \right) \bM &&\gauge e^{i \s q\s g\s \RCA} \big(\nabla_\perp \bM\big) \,,\notag\\[4pt]  
\nabla_\perp^*\bM &\equiv \left(\PP_\perp^* - i\s q\s g\s  \sqrt{2}\s \bPhiA \right) \bM &&\gauge e^{i \s q\s g\s \RCA} \big(\nabla_\perp^* \bM\big) \,.  \label{eq:nabla_def}
\end{alignat}
In \Sec{subsec:kinetictermchiralcharged}, we will covariantize the $\bM$ kinetic term by making the replacement 
\begin{equation}
\label{eq:BoxReplaceM}
\frac{\Box}{\PP} \quad \Rightarrow \quad \tilde{\nabla} - \nabla_\perp\s \frac{1}{\PP}\s \nabla_\perp^*\,.
\end{equation}
Unlike for standard $\mathcal{N} = 1$ superspace, these gauge-covariant derivatives are essential in collinear superspace because of the explicit space-time derivatives in \Eq{eq:SUSYSCETLagrangian}.

One important aspect of these gauge-covariant derivatives is that they \emph{do not} generically preserve superfield chirality.
If $\bM$ is a chiral multiplet satisfying $\bar{\D} \bM = 0$, then $\nabla_\perp^* \bM$ is a chiral multiplet because $\bPhiA$ is chiral.
However, $\tilde{\nabla} \bM$ and $\nabla_\perp \bM$ are not chiral multiplets, since they involve the real multiplet $\bD$ and the anti-chiral multiplet $\bPhiA^\dagger$, respectively.
This implies that when manipulating these objects, one has to be careful to keep track of tricky facts like $\bar{\D} \big(\tilde{\nabla} \bM\big) \neq 0$.

\subsection{The Covariant Chiral Auxiliary Superfield}
\label{subsec:covariantchiralaux}

Next, we address the auxiliary chiral multiplet $\bU_M$:
\begin{align}
\hspace{-15pt} \bU_M = \tilde{U}_M - \sqrt{2}\, \eta \, F_M +i\s\eta^\dagger \eta \, \PP \tilde{U}_M\,,  \qquad\text{with}\qquad \tilde{U}_M = \tilde{u}_M - \frac{\PP_\perp^*}{\PP}\s u_M\,,
\end{align}
whose analog is given in \Eq{eq:Usuperfield}.
In contrast to $\bM$, the object $\bU_M$ has complicated gauge transformation properties, owing to the ordinary derivative $\PP_\perp^*$ appearing its fermionic component.
We emphasize that $\bU_M$ is still a valid chiral superfield, but since it does not transform covariantly under gauge transformations, it is tedious to construct its gauge-invariant kinetic term.

\newpage

This issue motivates the construction of a covariantized $\bU$ superfield, which can be accomplishing by involving the gauge field $\bPhiA$:  
\begin{align}
\bU_M^{\rm cov} &\equiv \bU_M - i\s q\s g\, \frac{i\s \bar{\D} \D}{2\, \PP} \bigg(\bM\, \frac{\D}{\PP}\, \bPhiA  \bigg)\notag \\[6pt]
& =  \bigg[ \uu_M - \frac{1}{\PP}\nabla_\perp^* u_M  + \sqrt{2}\, q\s g\, \frac{1}{\PP}  \bigg( \big( \PP\s \phi_M \big) \frac{1}{\PP}\s \lambda^\dagger  \bigg)  \bigg] +\sqrt{2}\, \eta \bigg[ F_M -  \sqrt{2}\,q\s g\, u_M\s \frac{1}{\PP}\s \lambda^\dagger \bigg] \notag \\[4pt]
& \hspace{17pt}+ i\s \eta^\dagger \eta\, \PP \bigg[ \uu_M - \frac{1}{\PP} \nabla_\perp^* u_M  + \sqrt{2}\, q\s g \, \frac{1}{\PP} \bigg( \big( \PP\s \phi_M \big) \frac{1}{\PP}\s \lambda^\dagger   \bigg)  \bigg] \,,
 \label{eq:UMcovariant}
\end{align}
where the component covariant derivative $\nabla_\perp^* u_M = \big( \PP_\perp^* - i\s \sqrt{2}\,q\s g\, \alc^* \big) u_M$ is being used here to make it clear that this object is gauge covariant.
Note that the order of derivatives is crucial here.  
We are unaware of a simple bottom-up argument that yields this expression beyond simply checking the components directly; the top-down logic that leads to this expression is presented below in \Eq{eq:topdownUMcovariant}.

There are a number of ways to cross check the form of $\bU_M^{\rm cov}$.
First, using the component expression in \Eq{eq:UMcovariant}, it is straightforward to check that the whole multiplet transforms covariantly:
\begin{equation}
\bU_M^{\rm cov}  \gauge e^{i\s q\s g\s \RCA}\, \bU_M^{\rm cov}\,.
\end{equation}
Second, the additional gaugino terms in \Eq{eq:UMcovariant} ensure that the components of $\bU_M^{\rm cov}$ transform like a chiral multiplet, as expected by the presence of the chiral projector.
Finally, this multiplet inherits the RPI-I invariance of $\bPhiA$ and $\bM$.
These reasons back up the assertion that $\bU_M^{\rm cov}$ is a valid covariant chiral auxiliary multiplet.

Similarly, we can build a covariant version of $\bC_M$ following \Eq{eq:Cfield_def}: 
\begin{equation}
\label{eq:CcovDefinition}
\bC^{\rm cov}_M = \bU^{\rm cov}_M  + \frac{1}{\sqrt{2}} \frac{1}{\PP}\s \nabla_\perp^*\s \D \bM \,.
\end{equation}
This superfield transforms as
\begin{equation}
\bC^{\rm cov}_M \gauge e^{i\s q\s g\s \RCA}\s \bC^{\rm cov}_M\,,
\label{eq:CcovGaugeTrans}
\end{equation}
and it satisfies a covariant almost-chiral constraint equation, analogous to \Eq{eq:Cconstraint}:
\begin{equation}
\bar{\D} \bC^{\rm cov}_M = -i \s \sqrt{2}\s \nabla_\perp^*\s \bM\,.
\label{eq:CcovConstraint}
\end{equation}
In deriving this expression, it is useful to recall that $\nabla_\perp^*$ depends on the chiral multiplet $\bPhiA$.

\subsection{The Kinetic Term and Superpotential}
\label{subsec:kinetictermchiralcharged}

With these ingredients, we can lift the kinetic term for uncharged chiral matter in \Eq{eq:PhiUTildeKinTerm} to the charged scenario,  including the non-propagating degrees of freedom.
Using \Eq{eq:BoxReplaceM} to replace $\Box$ with its covariant version yields
\begin{equation}
\label{eq:chargedchiralkineticterm}
\mathcal{L} = \frac{1}{2}\, \D \bar{\D} \bigg[ \bM^\dagger\, i \left(\tilde{\nabla} - \nabla_\perp\s \frac{1}{\PP}\s \nabla_\perp^* \right)  \bM + \bU_M^{\rm cov \dagger} \bU_M^{\rm cov} \bigg] \bigg|_{0} \,\,.
\end{equation}

The generalization to multiple matter fields is straightforward.
In addition to having a copy of \Eq{eq:chargedchiralkineticterm} for each $\bM$/$\bU_M^{\rm cov}$ pair, we can also write down a gauge-invariant superpotential
\begin{equation}
\label{eq:charged_superpotential}
\mathcal{L}\supset  \frac{1}{\sqrt{2}} \, \D \bigg[ \bC_{M,j}^{\rm cov} \,W_j(\bM) \bigg] \bigg|_{0} + \text{h.c.}\,.
\end{equation}
As in \Sec{sec:MultiFlavorWZ}, in order for the term in square brackets to be chiral and for the action to be RPI-I invariant, it is crucial that $W_j \equiv \partial W / \partial \phi_{M,j}$.
Compared to the analogous manipulation in \Eq{eq:multiple_flavor_chiral_check}, the only difference is the replacement of  $\PP_\perp^*$ with $\nabla_\perp^*$.
This imposes the additional restriction that $W$ must be gauge invariant, such that $\nabla_\perp^* W  = \PP_\perp^* W$ is a total derivative.

\subsection{Top-Down Derivation of Covariant Superfields}

\begin{table}[t]
\renewcommand{\arraystretch}{2}
\setlength{\arrayrulewidth}{.3mm}
\small
\setlength{\tabcolsep}{0.35 em}
\hspace{-10pt}
 \begin{tabular}{ |c || c | c | c | c | c | c|}
    \hline
    Superfield &  Construction & Constraint &  Gauge    \\ \hline \hline
 $\bM $ & $\bM^{\full}\big|_{\tilde{\eta} = 0}$  & $\bar{\D} \bM = 0$ &    $e^{i\s q\s g\s \RCA} \bM$     \\[4pt]  
    $\bC_M $ & $\frac{1}{\sqrt{2}}\, \tilde{\D} \bM^{\full}|_{\tilde{\eta} = 0}$ & $\bar{\D} \bC_M = -i\s\sqrt{2}\, \PP_\perp^* \bM$ &   $e^{i\s q\s g\s \RCA}\s \big( \bC_M + i \s q \s g \s \bM \s \RCA_C  \big)$   \\ [4pt]  
 $ \bC^{\rm cov}_M $ & $\bC_M - i \s q\s g \left( \bM \s \frac{\D}{\PP} \bPhiA \right)$  & $\bar{\D} \bC^{\rm cov}_M = -i\s\sqrt{2}\,\nabla_\perp^* \bM$  &  $e^{i\s q\s g\s \RCA}\s \bC^{\rm cov}_M$     \\[4pt]  
 $\bU^{\rm cov}_M $ & $\frac{i\s \bar{\D} \D}{2\,\PP}\, \bC^{\rm cov}_M = \bU_M + i\s q\s g\s \frac{i\s \bar{\D} \D}{2\, \PP} \left(\bM \frac{\D}{\PP} \bPhiA  \right)$  & $\bar{\D} \bU^{\rm cov}_M = 0$  &  $e^{i\s q\s g \s\RCA} \bU_M^{\rm cov}$     \\[4pt] \hline 
 \end{tabular}
 \caption{Collinear charged matter superfields, along with their top-down construction and gauge transformation properties.
 The RPI transformations are the same as the analogous ones in \Tab{table:RPIsuperfieldsChiral}.
}
\label{Table:transSuperfields}
\end{table}

To better understand the necessity for the covariant multiplets $\bU_M^{\rm cov}$ and $\bC_M^{\rm cov}$, it is instructive to repeat the top-down derivation in \Sec{subsec:topdownchiral}, but now accounting for gauge transformations.
Again, we restrict ourselves to the Abelian case to minimize the algebra.
A summary of the various fields we construct and their gauge transformations are given in \Tab{Table:transSuperfields}.

We start with the gauge transformation for a full matter field in WZG:
\begin{equation}
\bM^\full \gauge e^{i\s q\s g\s \RCA_{\rm WZG}^\full}\s \bM^\full\,.
\end{equation}
Restricting to $\tilde{\eta} = 0$ yields the expected gauge transformation for $\bM$ in \Eq{eq:matter_transformation}.
Taking a $\tilde{\D}$ derivative before setting $\tilde{\eta} = 0$ yields
\begin{align}
\frac{1}{\sqrt{2}} \big(\tilde{\D}\bM^\full \big) \Big|_{\tilde{\eta} = 0} \equiv \bC_M \gauge e^{i\s q\s g\s \RCA}\s \big( \bC_M + i \s q \s g \s \bM \s \RCA_C  \big)\,,
\end{align}
where
\begin{align}
\RCA_C \equiv \frac{1}{\sqrt{2}} \Big(\tilde{\D} \RCA_{\rm WZG}^\full \Big)\Big|_{\tilde{\eta} = 0}\,.
\end{align}
Thus, there is a contribution to the gauge transformation from the new object $\RCA_C$.
We emphasize that in LCG, $\RCA_C$ is not independent from $\RCA$; from the component expressions, one can show that
\begin{equation}
\RCA_C \equiv \frac{1}{\sqrt{2}}\bigg(\frac{\PP_\perp^*}{\PP}\, \D\s \RCA^{\rm full}_{\rm WZG} \bigg) \bigg|_{\tilde{\eta} = 0} = -i \s \sqrt{2}\s \eta \, \PP_\perp^* \omega\,.
\end{equation}
This implies the following relations:%
\footnote{Because of this first relation, the gauge transformation analog of $\bPhi_F$ in \Eq{eq:Fmultiplet} is zero when working in WZG and LCG.}
\begin{equation}
\D\s \RCA_C = 0\,, \qquad \bar{\D}\s \RCA_C = -i \s \sqrt{2} \s \PP_\perp^* \RCA\,.
\end{equation}

While it is possible to work with $\RCA_C$, we find it much more convenient to directly covariantize $\bC_M$:
\begin{equation}
\label{eq:CcovDefinition_topdown}
\bC^{\rm cov}_M = \bC_M - i \s q\s g\s \bigg( \bM \s \frac{\D}{\PP} \bPhiA \bigg)\,,
\end{equation}
where we are using the fact that $\bPhiA \to \bPhiA + \s \PP_\perp^*\s \RCA / \sqrt{2}$. 
Although it is not obvious, \Eq{eq:CcovDefinition_topdown} has exactly the same components as \Eq{eq:CcovDefinition}.
The gauge transformation for $\bC^{\rm cov}_M$ is given in \Eq{eq:CcovGaugeTrans}, and it depends on $\RCA$, as opposed to $\RCA_C$.

We can define $\bU_M^{\rm cov}$ in analogy with \Eq{eq:UasCchiralprojector}:
\begin{equation}
\label{eq:topdownUMcovariant}
\bU^{\rm cov}_M = \frac{i\s \bar{\D} \D}{2\,\PP}\, \bC^{\rm cov}_M = \bU_M - i\s q\s g\s \frac{i\s \bar{\D} \D}{2\, \PP} \bigg(\bM \s\frac{\D}{\PP} \s \bPhiA \bigg)\,,
\end{equation}
which is identical to \Eq{eq:UMcovariant}.
In parallel to \Sec{sec:Fterm}, from the bottom-up perspective it is more natural to define $\bC^{\rm cov}_M$ from $\bU^{\rm cov}_M$ using \Eq{eq:CcovDefinition}, while from the top down it is more natural to define $\bU^{\rm cov}_M$ from $\bC^{\rm cov}_M$ using \Eq{eq:topdownUMcovariant}.

\subsection{Top-Down Derivation of Charged Matter Lagrangian}
\label{subsec:topdownchargedkinetic}
We now turn to deriving the charged chiral matter Lagrangian in \Eq{eq:chargedchiralkineticterm} from the top down.
The full $\mathcal{N} = 1$ kinetic term is
\begin{equation}
\mathcal{L} = \frac{1}{4}\, \bar{\D} \D \bar{\tilde{\D}} \tilde{\D} \bigg[ \bM^{\full \dagger}\, e^{q\s g\s \bV^\full}\, \bM^\full \bigg] \bigg|_{\theta = 0 = \theta^\dagger}\,.
\end{equation}
As always, we are working in WZG, in which case $\big(\bV^\full\big)^3 = 0$, so that we can expand 
\begin{align}
e^{q\s g\s \bV^\full} = 1 + q\s g\, \bV^\full + \frac{1}{2}\,q^2\s g^2\, \big(\bV^\full\big)^2\,.
\end{align}
Carrying out the $\bar{\tilde{\D}}$ and $\tilde{\D}$ derivatives and using the definitions $\bC^{\rm cov}_M$ from \Eq{eq:CcovDefinition}, $\bPhiA$ from \Eq{eq:Phialc}, and $\bD$ from \Eq{eq:bD}, we find
\begin{equation}
\mathcal{L} = \frac{1}{4}\, \bar{\D} \D \bigg[ \bC_M^{\rm cov \dagger} \bC^{\rm cov}_M + i\s \bM^\dagger\s \tilde{\nabla} \bM \bigg] \bigg|_{0}\,.
\end{equation}
Using \Eq{eq:topdownUMcovariant} and performing integration by parts, we can rewrite this as
\begin{equation}
\mathcal{L} = \frac{1}{2}\, \D \bar{\D} \bigg[ \bM^\dagger\, i \left(\tilde{\nabla} - \nabla_\perp\s \frac{1}{\PP}\s \nabla_\perp^* \right)  \bM + \bU_M^{\rm cov \dagger} \bU_M^{\rm cov} \bigg] \bigg|_{0} \,,
\end{equation}
in agreement with \Eq{eq:chargedchiralkineticterm}.

For multiple fields interacting via a superpotential, the initial manipulation is identical to \Sec{subsec:fulltheorymatchingSuperpotential}, leading to
\begin{align}
\mathcal{L} \supset \frac{1}{2}\, \D \tilde{\D}  \bigg[ W\big(\bM^\full\big) \bigg] \bigg|_{\theta = 0 = \theta^\dagger} + \text{h.c.} \,\, \Longrightarrow \,\, \frac{1}{2}\, \D \bigg[  \bC_{M,j}\, W_j(\bM) \bigg] \bigg|_{0} + \text{h.c.} \, \,.
\end{align}
Naively, this differs from \Eq{eq:charged_superpotential} by the replacement of $\bC^{\rm cov}_{M,j}$ with $\bC_{M,j}$. 
Substituting in \Eq{eq:CcovDefinition_topdown}, the difference is
\begin{equation}
\label{eq:Ccov_mismatch}
\delta \mathcal{L} = \frac{1}{\sqrt{2}} \bigg[ i g \Big(\sum_j q_j \, \bM_j \, W_j(\bM) \Big) \frac{\D}{\PP} \bPhiA \bigg] \bigg|_{0} + \text{h.c.} \,,
\end{equation}
where we have make the sum over $j$ explicit.
For  generic choices of $W$, this expression is non-zero, but if $W$ is gauge invariant, then  $W(\phi_M) = W(e^{iqg \omega} \phi_M)$.
Taylor expanding this for small $\omega$ implies
\begin{equation}
\sum_j q_i \, \bM_j \, W_j(\bM) = 0 \,,
\end{equation}
so the mismatch term in \Eq{eq:Ccov_mismatch} vanishes, and we recover \Eq{eq:charged_superpotential} as expected.

\subsection{The Component Charged Matter Kinetic Term}

For completeness, we provide the component expression for the charged matter kinetic term by expanding out the superspace expression of \Eq{eq:chargedchiralkineticterm}.
First, we analyze the term that is bi-linear in $\bM$:
\begin{align}
\mathcal{L}_{\bM} = \, & \frac{1}{2}\, \D \bar{\D} \bigg[ \bM^\dagger\, i\s \bigg(\tilde{\nabla} - \nabla_\perp\s \frac{1}{\PP}\s \nabla_\perp^* \bigg)  \bM   \bigg] \bigg|_{0} \nonumber \\[6pt] 
=\, & - \phi_M^*\s \Big( \tilde{\nabla} \PP - \nabla_\perp \nabla_\perp^* \Big)\s \phi_M    + i \s u_M^\dagger \bigg(\tilde{\nabla}-  \nabla_\perp \frac{1}{\PP} \nabla_\perp^* \bigg)\s u_M  \nonumber \\[6pt] 
&   -  q \s g\, \phi_M^*\,\lambda\, \frac{1}{\PP}\s\nabla_\perp^*\s u_M    - q\s g\,   u_M^\dagger\,  \nabla_\perp\s \frac{1}{\PP}\s \big(\lambda^\dagger\, \phi_M\big)     -  \sqrt{2}\, q\s g\,  \phi_M^*\, \tilde{\lambda}\, u_M \nonumber \\[6pt] 
&  - i\s q\s g\, \phi_M^*  \bigg(\! -  i \s D   - \frac{1}{2} \,\partial^\mu A_\mu \bigg)\s \phi_M - \sqrt{2}\, q^2\s g^2\, \phi_M^*\,\lambda\, \frac{1}{\PP}\s \big( \lambda^\dagger\, \phi_M\big)\, ,
\label{eq:CompChargedMatterL1}
\end{align}
where the component covariant derivatives are as expected, \emph{e.g.} $\tilde{\nabla} \phi_M = \big( \tilde{\PP} - i\s q\s g\, n\cdot A \big) \,\phi_M$ and  $\nabla_\perp \phi_M = \big( \PP_\perp - i\s q\s g\, \sqrt{2} \alc \big) \phi_M$. 
Importantly, note that the kinetic terms generated in \Eq{eq:CompChargedMatterL1} are the covariantization of \Eq{eq:SUSYSCETLagrangian} when the scalar and fermion both carry a $U(1)$ charge $q$. 
Note that in simplifying the interactions in the second and third lines of \Eq{eq:CompChargedMatterL1}, we have used \Eq{eq:DAtoD} to combine terms.

Next, consider the term that is bi-linear in $\bU_M^{\rm cov}$.  Expanding it out in components yields
\begin{align}
\mathcal{L}_{\bU_M^{\rm cov}} =\, & \frac{1}{2}\, \D \bar{\D} \bigg[  \bU_M^{\rm cov \dagger} \bU_M^{\rm cov} \bigg] \bigg|_{0} \notag \\[6pt]
=\, & i \left( \uu_M^\dagger\, \PP\s \uu_M - u_M^\dag \nabla_\perp \uu_M -  \tilde{u}_M^\dagger \nabla_\perp^* u_M  +   u_M^\dagger \nabla_\perp  \frac{1}{\PP}  \nabla_\perp^* u_M   \right) +  F_M^* F_M \notag \\[6pt]
&+\bigg(- i \s \sqrt{2} \, q\s g\, \bigg[ \uu_M^\dagger - \frac{1}{\PP} \nabla_\perp u_M^\dagger     \bigg]  \bigg[   \big( \PP\s \phi_M \big) \frac{1}{\PP}\s \lambda^\dagger  \bigg]   -   \sqrt{2}\, q\s g\, F_M^*\s  u_M\s \left( \frac{1}{\PP}\s \lambda^\dagger \right)  + \text{h.c.}\bigg) \notag\\[6pt] 
& -2\s  i\,  q^2\s g^2 \bigg[  \frac{1}{\PP}  \bigg( \big( \PP\s \phi_M^\dagger \big) \frac{1}{\PP}\s \lambda \bigg)  \bigg]  \bigg[ \big( \PP\s \phi_M \big) \frac{1}{\PP}\s \lambda^\dagger   \bigg] + 2\, q^2\s g^2  \bigg[ u_M^\dagger \s \frac{1}{\PP}\s \lambda \bigg] \bigg[ u_M\s \frac{1}{\PP}\s \lambda^\dagger \bigg] \,.
\label{eq:UUcompL}
\end{align}
The total component kinetic term is simply given by $\mathcal{L} = \mathcal{L}_{\bM} + \mathcal{L}_{\bU_M^{\rm cov}}$.
One non-trivial cross check is that the terms that are bi-linear in the matter scalar, fermion, and auxiliary fields (including covariant derivatives) reduce to the full theory covariantized kinetic term (recall that covariantizing $\Box \rightarrow  \tilde{\nabla} \PP - \nabla_\perp \nabla_\perp^*$ in collinear superspace):
\begin{align}
\mathcal{L} = \, &- \phi_M^*\s \Big( \tilde{\nabla} \PP - \nabla_\perp \nabla_\perp^* \Big)\s \phi_M  \notag\\
&+i \left( u_M^\dagger \tilde{\nabla}\s u_M+ \uu_M^\dagger\, \PP\s \uu_M - u_M^\dag \nabla_\perp \uu_M -  \tilde{u}_M^\dagger \nabla_\perp^* u_M   \right) +  F_M^* F_M + \dots\,,
\end{align}
where this can be easily derived by combining the second lines of \Eqs{eq:CompChargedMatterL1}{eq:UUcompL}.
This Lagrangian encodes the $\mathcal{N} = 1$ dynamics of charged chiral matter coupled to an Abelian gauge field expressed on the light-cone, where WZG and LCG has been assumed, but the non-propagating degrees of freedom are made manifest.

\section{Future Horizons}
\label{sec:Outlook}
In this paper, we expanded the effective field theory rules for collinear superspace~\cite{Cohen:2018qvn} to include auxiliary multiplets.
First, we demonstrated that constructing theories of interacting chiral matter such as the Wess-Zumino model required extending the superfield content beyond $ \bPhi$, to include the fermionic multiplet $\bU$ and its almost-chiral counterpart $\bC$.
We then showed that one can express Abelian gauge theories, including the non-propagating modes, by expressing the superspace Lagrangian in terms of $\bPhialc$, $\bD$, and $\bU_\lambda$.
Finally, we showed how to write the Lagrangian for charged chiral matter $\bM$ coupled to an Abelian supermultiplet, which required covariantizing the derivatives $\tilde{\nabla}$,  $\nabla_\perp$, and $\nabla_\perp^*$, and the auxiliary superfields $\bC^{\rm cov}_M$ and $\bU^{\rm cov}_\lambda$.
(A similar strategy is required for the non-Abelian extension, as we briefly present in \App{sec:nonAbelianGauge}).
We have now come full circle with respect to \Refs{Cohen:2016jzp,Cohen:2016dcl}, by showing that collinear superspace can describe not just free chiral matter and Yang-Mills theories, but the full range of interactions that are consistent with $\mathcal{N} = 1$ SUSY.

The structure of $\mathcal{N} = 1$ SUSY is already well known, of course, but (re)constructing it while obscuring half of the supercharges has yielded new insights.
In particular, our work provides a perspective on the question:  how much symmetry must be made manifest in the action to ensure that physical observables respect those symmetries?
In the case of SUSY theories with $F = 0$ and $D = 0$, \Refs{Cohen:2016jzp,Cohen:2016dcl,Cohen:2018qvn} found that only two (out of four) supercharges and three (out of six) Lorentz generators were required to recover $\mathcal{N} = 1$ SUSY at leading power.
As we have emphasized many times above, RPI-II invariance emerges as an accidental symmetry for these special theories in that it does not impose any constraints on the marginal or relevant interactions.
For contrast, when working with the more general SUSY theories considered in this paper, RPI-II invariance was essential, leading to numerous constraints on the component field expressions that were not manifest in collinear superspace directly.
In addition to setting the normalization of various kinetic terms, RPI-II invariance (along with chirality and RPI-I) enforced that chiral interactions be described via the superpotential form $W_j(\phi) \equiv \partial W / \partial \phi_j$.
Of course, these constraints are familiar from the standard $\mathcal{N} = 1$ superspace formalism where Lorentz invariance is manifest, but it is satisfying to see them emerge from both bottom-up and top-down considerations.

There are a number of novel ingredients that emerged from these studies that do not have obvious counterparts in the standard $\mathcal{N} = 1$ literature.
First, the auxiliary multiplets $\bU$ and $\bU_\lambda$ are needed to house the $F$- and $D$-term components, and they are fully non-propagating and therefore able to be integrated out (in the absence of mass terms).
Second, the superfield $\bC$ was found to be useful for constructing an RPI-II invariant superpotential, and it satisfies an exotic constraint $\bar{\D} \bC = -i\s \sqrt{2} \s \PP_\perp^* \bPhi$  that renders it ``almost chiral.'' 
Third, the gauge transformation parameter $\RCA$ is simultaneously chiral, anti-chiral, and real, yet it is not a constant in collinear superspace such that it can be used to constrain the form of gauge interactions.
Fourth, the gauge covariant derivatives $\tilde{\nabla}$, $\nabla_\perp$, and $\nabla_\perp^*$, which are rarely encountered in standard $\mathcal{N} = 1$ constructions, are essential in collinear superspace because of the explicit derivatives in the charged matter kinetic term.
We emphasize that a number of these features are only possible when working in LCG and WZG, and indeed, only with both of these gauge choices imposed simultaneously is collinear SUSY manifest.

Looking ahead, we suspect that collinear superspace will be particularly illuminating for $\mathcal{N} = 2$ and $\mathcal{N} = 4$ SUSY theories which do not have fully Lorentz-invariant superspace descriptions.
A typical superspace approach for $\mathcal{N} > 1$ is to preserve Lorentz invariance and maintain an $\mathcal{N} = 1$ subgroup, at the expense of obscuring $SU(\mathcal{N})_R$ and many of the supercharges, see \emph{e.g.}~the original treatment of $\mathcal{N} = 4$ SUSY Yang Mills~\cite{Mandelstam:1982cb,Brink:1981nb}.
Using collinear superspace, we could potentially make $SU(\mathcal{N})_R$ manifest along with half of the supercharges, at the expense of obscuring Lorentz invariance.
This is in keeping with the spirit of light-cone superspace~\cite{Siegel:1981ec, Brink:1981nb, Hassoun:1982mr, Mandelstam:1982cb, Brink:1982pd, Brink:1982wv}, with the key difference of keeping track of the RPI structure to enable a fully bottom-up effective field theory treatment.
It would also be interesting to study the connection to recent progress understanding soft theorems (see~\Ref{Strominger:2017zoo} for a review) in the framework of collinear superspace, perhaps making connection with the non-SUSY formalism developed in~\Ref{Larkoski:2014bxa}. 
Additionally, since the collinear superspace building blocks separate propagating degrees of freedom from the non-propagating ones, we anticipate that there are natural connections to be made with the modern on-shell scattering amplitudes program.

It would be interesting to revisit spontaneous SUSY breaking in the collinear superspace language, since the $F$- and $D$-term order parameters now live in auxiliary multiplets, which might lead to novel non-linear realizations of SUSY.
Another avenue of research would be to discover the appropriate building blocks when working with spontaneously broken gauge theories, along with understanding the way in which the super-Higgs mechanism manifests.
It would also be fascinating to construct linearized supergravity in collinear superspace, at minimum to check whether the gravitino/graviton kinetic structure mimics the one for chiral matter, as we already saw was the case for gauge theories.

One of the original motivations to study collinear superspace was to illuminate aspects of Soft Collinear Effective Theory (SCET)~\cite{Bauer:2000yr, Bauer:2001ct, Bauer:2001yt}.
Effective theories in general are defined by a Lagrangian that describes propagation and interactions in the infrared, as well as local operators that source fields.
The framework developed here is directly relevant to writing down SCET theories with $\mathcal{N} = 1$ SUSY.  
We have focused on the rules for constructing the interaction Lagrangians, but the superfields identified here also form fundamental building blocks of SUSY-, RPI-, and gauge-invariant local operators.
Generally, gauge invariance of an operator that carries some charge is accomplished by attaching appropriate Wilson lines.
Maintaining explicit collinear SUSY invariance requires working in WZG and LCG, and this should render the necessary Wilson lines to be simple or even unity.
It would further be interesting to explore if one can derive collinear-soft mode factorization~\cite{Bauer:2002nz}, as well as to extend the analysis to understand the role of Glauber gluons~\cite{Rothstein:2016bsq} in superspace.
Surprising infrared cancellations are known to occur in SUSY gauge theories \cite{Dine:2016rxc}, and a direct formulation of the infrared in superspace may identify the physics responsible.
Resumming a canonical observable such as thrust could also illuminate differences between QCD and SUSY QCD.

With this paper in hand, we now have a chart for navigating collinear superspace.
There are many applications of the formalism that could yield deeper insight into more formal questions related to $\mathcal{N} > 1$ SUSY, and more phenomenological applications via SUSY SCET.
By providing an alternative perspective on the symmetries and fundamental building blocks, collinear superspace opens a new way of thinking about SUSY field theories.

\acknowledgments

We thank Martin Beneke, Cliff Cheung, Howie Haber, and David Pinner for helpful discussions.
TC is supported by the U.S. Department of Energy, under grant number DE-SC0011640 and DE-SC0018191, the Munich Institute for Astro- and Particle Physics (MIAPP) of the DFG cluster of excellence ``Origin and Structure of the Universe,'' and the KITP from the National Science Foundation under Grant No.\ NSF PHY-1748958.
GE is supported by the U.S. Department of Energy, under grant numbers DE-SC0011637 and DE-SC0018191.
JT thanks the Harvard Center for the Fundamental Laws of Nature for hospitality while this work was completed, as well as the Voyages Beyond the Standard Model workshop for transportation and lodging.
JT is supported by the Office of High Energy Physics of the U.S. Department of Energy under grant DE-SC-0012567 and by the Simons Foundation through a Simons Fellowship in Theoretical Physics.

\appendix 
\section{Tacking: Non-Abelian Gauge Theories}
\label{sec:nonAbelianGauge}

In this appendix, we briefly introduce the necessary ingredients to write down pure non-Abelian gauge theory in collinear superspace.
Much of the analysis in \Sec{sec:AbelianGaugeTheory} carries through for non-Abelian gauge theories, albeit with complications arising due to the need for gauge-covariant derivatives, in analog to the considerations discussed in \Sec{sec:Landfall} for building a theory of charged chiral matter. 

We wish to lift the Abelian kinetic term for $\bPhialc$ in \Eq{eq:LagrangianD} to a non-Abelian theory.
This requires replacing spacetime derivatives with the appropriate gauge-covariant versions, \emph{e.g.} $\tilde{\PP} \rightarrow \tilde{\nabla}$.
Let $a$ be a group index and $T^a$ be the associated generator.
Treating the superfields as matrices in color space $\bPhialc \equiv \bPhialc^a\s T^a$ and $\bD \equiv \bD^a T^a$, the final non-Abelian gauge kinetic term is
\begin{align}
\label{eq:LagrangianDnonAbelian}
\mathcal{L}  &=i\s \D \bar{\D} \, \tr \bigg[  \bPhialc^\dagger \left(\tilde{\nabla} - \nabla_\perp \frac{1}{\PP} \nabla_\perp^* \right) \bPhialc + \bU^{\rm cov  \, \dagger}_\lambda \bU^{\rm cov}_\lambda \bigg] \bigg|_{0}\,,
\end{align}
where there is an extra overall factor of $2$ with respect to \Eq{eq:LagrangianD} due to the normalization of the generators: $\tr(T^a\s T^b) = \frac{1}{2}\s \delta^{ab}$.
The detailed definitions of all the objects will be explained in what follows, \emph{e.g.} $\bU^{\rm cov}$ is defined in \Eq{eq:improvedUlambda}.

The gauge transformation parameter $\RCA \equiv \RCA^a\s T^a$ is now a matrix, and the gauge transformations of the superfields are: 
\begin{align}
\bPhialc &\gauge e^{i\s g\s \RCA} \big(\bPhiA + i\s \sqrt{2}\, \PP_\perp^* \big) e^{-i\s g\s \RCA}, \notag\\[4pt]
\bD &\gauge e^{i\s g\s \RCA} \big(\bD + i\s \tilde{\PP} \big) e^{-i\s g\s \RCA}.
\end{align}
In analogy with the discussion in \Sec{subsec:gaugecovderiv}, we define light-cone gauge-covariant derivatives in collinear superspace: 
\begin{alignat}{2}
\tilde{\nabla}\bPhialc & \equiv \tilde{\PP} \bPhialc - i\s  \big[\bD, \bPhialc\big] &&\gauge e^{i\s g\s \RCA} \Big(\tilde{\nabla} \bPhiA - \sqrt{2} \, g \,  \tilde{\PP}\, \PP_\perp^* \Big) e^{-i\s g\s \RCA} \,, \notag\\[5pt]
\nabla_\perp \bPhialc &\equiv \PP_\perp \bPhialc - \frac{i}{\sqrt{2}} \big[\bPhiA^\dagger, \bPhialc\big] &&\gauge e^{i\s g\s \RCA} \Big(\nabla_\perp \bPhiA - \sqrt{2} \, g \,\PP_\perp \PP_\perp^*  \Big) e^{-i\s g\s \RCA} \,, \notag\\[5pt]
\nabla_\perp^* \bPhialc &\equiv  \PP_\perp^* \bPhialc - \frac{i}{\sqrt{2}}\big[\bPhiA , \bPhialc^\dagger \big]  &&\gauge e^{i\s g\s \RCA} \Big(\nabla_\perp^* \bPhiA -\sqrt{2} \,g \,\PP_\perp^* \PP_\perp^* \Big) e^{-i\s g\s \RCA} \,,
\label{eq:nabla_def_nonAbelian}
\end{alignat}
with similar expressions for $\bPhialc^\dagger$.
Then to covariantize the $\bPhialc$ kinetic term from \Eq{eq:LagrangianD}, we make the replacement 
\begin{equation}
\frac{\Box}{\PP} \quad \Rightarrow \quad \tilde{\nabla} - \nabla_\perp \frac{1}{\PP} \nabla_\perp^*\,.
\label{eq:BoxReplace}
\end{equation}
Note that $\nabla_\perp^* \bPhialc$ is still a chiral multiplet (since $\nabla_\perp^*$ itself depends on $\bPhialc$), but $\nabla_\perp \bPhialc$ is neither chiral nor anti-chiral.

The contribution to the Lagrangian in \Eq{eq:LagrangianDnonAbelian} that is bi-linear in $\bPhialc$ is gauge invariant on its own.
While $\tilde{\nabla} \bPhialc$ does not transform covariantly under a gauge transformation, the extra shift in \Eq{eq:nabla_def_nonAbelian} involves a term that is both chiral and anti-chiral, so it only contributes a total derivative to the Lagrangian.
Similar arguments can be made for the term that depends on $\nabla_\perp(1/\PP) \nabla_\perp^*$; a detailed discussion of the gauge invariance of this term was given in \Ref{Cohen:2018qvn}, and so we do not repeat it here.
Taken together, we conclude that the bi-linear in $\bPhialc$ term is gauge invariant.

The kinetic term for $\bU_\lambda \equiv  \bU_\lambda^a\s T^a$ requires more care.
The reason is that $\bU_\lambda$ is defined in \Eq{eq:Ulambda_def} via $\bPhiD$, and from its definition in \Eq{eq:phiD_def}, we see that it only depends on the ordinary $\PP_\perp$ derivative acting on $\bPhialc$ instead of the gauge-covariant derivative $\nabla_\perp$.
Naively, one might try to make the simple replacement $\PP_\perp \bPhialc \Rightarrow \nabla_\perp \bPhialc$ in the definition of $\bPhiD$, but this fails since $\nabla_\perp \bPhialc$ is not a chiral multiplet.
We can, however, covariantize the definition of $\bU_\lambda$ directly, which yields
\begin{equation}
\label{eq:improvedUlambda}
\bU^{\rm cov}_\lambda  = -\frac{i}{2}\, \bar{\D} \bigg(\bD - \sqrt{2} \, g\, \frac{\nabla_\perp^*}{\PP} \bPhialc^\dagger \bigg)\,,
\end{equation}
where we note that $\bar{\D}$ acting on any superfield yields a chiral superfield.
It is easy to see that $\bU^{\rm cov}_\lambda$ does indeed transform covariantly:  
\begin{equation}
\bU^{\rm cov}_\lambda \gauge e^{i\s g\s \RCA}\, \bU^{\rm cov}_\lambda\, e^{-i\s g\s \RCA} \,,
\end{equation}
which follows from noting that the residual term in the transformation of $\bU^{\rm cov}_\lambda$ involves the term $\bar{\D} \big( \tilde{\PP} \RCA + 2\s \frac{\PP_\perp^*}{\PP}\s \RCA  \big)$, which vanishes because $\bar{\D} \RCA = 0$.
Given that the gauge transformation does not include any derivative-dependent shifts, as we had in \Eq{eq:nabla_def_nonAbelian}, it is clear that the term bi-linear in $\bU^{\rm cov}_\lambda$ is gauge invariant on its own.
We conclude that the non-Abelian Lagrangian given in \Eq{eq:LagrangianDnonAbelian} is gauge invariant.

We now provide a cross check that this approach yields the correct Lagrangian.
Noting that $\bD$ is a pure auxiliary superfield, we can integrate it out via its superspace equations of motion.
In the Abelian case, this is equivalent to integrating out $\bU^{\rm cov}_\lambda$,  see \Eq{eq:integrateoutUlambda}.
The situation for the non-Abelian theory is more complicated, since $\bD$ now appears both in $\bU^{\rm cov}_\lambda$ and in $\tilde{\nabla}$. 
Therefore, the most straightforward approach is to integrate out $\bD$ directly:
\begin{equation}
\label{eq:integrateoutVndotA}
 \frac{\delta \mathcal{L}}{\delta \bD} = 0 \qquad \Longrightarrow \qquad \bD = 2\s i\, \bigg(\frac{\nabla_\perp^*}{\PP}\s \bPhialc^\dagger - \frac{\nabla_\perp}{\PP} \s\bPhialc \bigg)  \,.
\end{equation}
This superfield expression encodes the equations of motion for all the non-Abelian gauge superfields, which is clear by expanding this in components:
\begin{equation}
\tilde{\lambda} = \frac{1}{\PP}\s \nabla_\perp^* \lambda \,, \qquad\quad
D =  0\,, \qquad\quad
n\cdot A = \frac{1}{\PP}\s \big(\nabla_\perp^* \alc + \nabla_\perp \alc^*\big) \,,
\end{equation}
where now the covariant derivatives are interpreted as the component expressions.

Using the fact that \Eq{eq:integrateoutVndotA} is valid at the level of superfields, we can plug the solution for $\bD$ back into \Eq{eq:LagrangianDnonAbelian} to derive the collinear superspace Lagrangian for the non-Abelian theory:  
\begin{align}
\label{eq:nonablianLExpanded}
\mathcal{L}
&= \frac{i}{2}\D \bar{\D} \bigg[ \bPhiA^{a \dagger} \frac{\Box}{\PP} \bPhiA^a  \bigg] \bigg|_0 - \frac{i}{2}\s g\s f^{abc}\s \D\s \bar{\D} \bigg[  \left(\bPhiA^{\dagger a}\s \bPhiA^b\right) \frac{\PP_\perp^*}{\PP} \bPhiA^c - \bPhiA^{\dagger a}\s \frac{\PP_\perp}{\PP}\s \left(\bPhiA^{\dagger b}\s \bPhiA^c  \right) \bigg] \bigg|_0 \notag \\[5pt] 
& \qquad - \frac{i}{2} \s g^2\s f^{abc}\s f^{che}\s \D \s \bar{\D} \bigg[  \left(\bPhiA^{\dagger a}\s \bPhiA^b\right) \frac{1}{\PP} \left( \bPhiA^{\dagger h}\s \bPhiA^e \right) \bigg] \bigg|_0  \,\,,
\end{align}
where here we have chosen to express the gauge fields explicitly in the adjoint representation, where the structure constants of the gauge group are $(T^a)_{bc} = - i\s f^{abc}$. 
This is the same Lagrangian derived from the top down in \Refs{Cohen:2016dcl,Cohen:2016jzp} as well as from the bottom up in \Ref{Cohen:2018qvn}.
Recall, though, that these previous derivations did not need to appeal to the non-propagating degrees of freedom in $\bD$.
As we saw in \Sec{sec:Landfall}, the true power of the $\bD$ formalism was for constructed Lagrangians for interacting charged chiral matter, and this power also translates to the non-Abelian case. 

\end{spacing}

\begin{spacing}{1.1}
\addcontentsline{toc}{section}{\protect\numberline{}References}%
\bibliography{OnShellSUSY}
\bibliographystyle{utphys}
\end{spacing}

\end{document}